\documentclass[twocolumn]{aastex61}
\usepackage{floatflt,graphicx}
\usepackage{tabularx}
\usepackage[figuresright]{rotating}
\usepackage{lipsum}
\usepackage{mathtools}
\usepackage{appendix}

\usepackage{rotating}
\usepackage{latexsym}
\usepackage{lineno}

\newcommand{\m}{$\, \mu$m}
\newcommand{\mj}{$\, \mu$Jy}
\newcommand{\sig}{$\, \sigma$}
\newcommand{\arcs}{\arcsec }
\newcommand{\wise}{{\it WISE} }
\renewcommand*{\arraystretch}{0.75}

\accepted{Oct 23, 2019}
\submitjournal{ApJS}

\shorttitle{The Largest \wise Galaxies}
\shortauthors{Jarrett et al.}

\begin{document}

\title{The \wise Extended Source Catalogue (WXSC)\\ I:  The 100 Largest Galaxies}

\correspondingauthor{Thomas Jarrett}
\email{tjarrett007@gmail.com}

\author{T.H. Jarrett}
\affiliation{Department of Astronomy, University of Cape Town, Rondebosch,
 South Africa}

\author{M.E. Cluver}
\affil{Centre for Astrophysics and Supercomputing, Swinburne University of Technology, John Street, Hawthorn 3122, Victoria, Australia}
\affiliation{Department of Physics and Astronomy, University of the Western Cape,
Robert Sobukwe Road, Bellville 7535,South Africa}

\author{M.J.I. Brown }
\affil{School of Physics and Astronomy, Monash University, Clayton 3800, Victoria, Australia}

\author{D.A. Dale}
\affiliation{Department of Physics and Astronomy, University of Wyoming, Laramie, WY 82071, USA}

\author{C.W. Tsai}
\affiliation{Department of Physics and Astronomy, University of California, Los Angeles, Los Angeles, CA, 90095, USA}

\author{F. Masci}
\affiliation{Infrared Processing and Analysis Center, California Institute of Technology, Pasadena, CA 91125, USA}


\begin{abstract}

We present mid-infrared photometry and measured global properties of the 100 largest galaxies in the sky,
including the well-studied Magellanic Clouds, Local Group galaxies M\,31 and M\,33,  the Fornax and Virgo Galaxy Cluster giants, and many of the most spectacular Messier objects (e.g., M\,51 and M\,83).  
This is the first release of a larger catalog of extended sources as imaged in the mid-infrared, called the \wise Extended Source Catalogue (WXSC).
In this study we measure their global attributes, including integrated flux, surface brightness and radial distribution.   The largest of the large are the LMC, SMC and the Andromeda Galaxy, which are also the brightest mid-infrared galaxies in the sky.  We interrogate the large galaxies using 
 \wise colors, which serve as proxies for four general types of galaxies:  bulge-dominated spheroidals, intermediate semi-quiescent disks, star-forming spirals, and AGN-dominated.  The colors reveal a tight ``sequence" that spans 5 magnitudes in W2$-$W3 color, ranging from early to late-types, and low to high star-forming activity;  we fit  
the functional form given by:
${\rm (W1-W2)} = [0.015 \times {\rm e}^{ \frac{{\rm (W2-W3)}}{1.38} }]  -  0.08$.  Departures from this sequence may reveal nuclear, starburst, and merging events.
 Physical properties and luminosity attributes are computed, notably the diameter, aggregate stellar mass and the dust-obscured star formation activity.  To effectively study and compare these galaxy characteristics, we introduce the `pinwheel' diagram which depicts physical properties with respect to the median value observed for \wise galaxies in the local universe.  Utilized with the WXSC, this diagram will delineate between different kinds of galaxies, identifying those with similar star formation and structural properties. Finally, we present the mid-infrared photometry of the 25 brightest globular clusters in the sky, for which many are also the largest and brightest objects orbiting the Milky Way, including  Omega\,Centauri, 47\,Tucanae and a number of famed night-sky targets (e.g., M\,13).

\end{abstract}




\section{Introduction} \label{intro}

In the realm of our galactic neighbours, the most appropriate distance scale is megaparsecs.
At these vast distances, even the nearest of these {\it island universes} appear faint and low surface brightness.
It has been well on 100 years since they were first identified as distinct objects beyond the Milky Way.  Nevertheless, the nearest spiral galaxies have held fascination for centuries -- they were somehow different from the wealth of stars and Galactic nebulae.  The observations of M\,51 by Lord Rosse, in the 1840's, clearly showed spiral arms, the nucleus and a companion galaxy, all of which were singularly unique to astronomers at the time.    

Galaxies such as Andromeda (M\,31), Triangulum (M\,33), the Whirlpool (M\,51), and the Pinwheel (M\,101) have been the focus of observations since as least the times of Lord Rosse, possibly earlier by medieval Persian astronomers\footnote{Notably Abd Al-Rahman Al-Sufi in his "Book of the Fixed Stars", published in 964 AD}.  Their relatively close proximity, less than 10 Mpc, means they are the largest galaxies in angular appearance.   Today, the largest galaxies in the sky are still some of the favourite laboratories to study the internal gas and stellar components, dynamics,  present and historic star formation, as well as used to construct local analogues of high-redshift systems.

There are a variety of galaxy types that are local to the Milky Way, as such they can be studied with physical resolutions that are necessary to connect the star formation processes, e.g. cloud collapse at the pc level, to their environment -- the galaxy ``main sequence" at two orders larger in scale -- to the kpc-scale components such spiral arms, bars, bulges, that represent the `skeleton' of the galaxy. These galaxies are important because they not only offer a window into the workings of these individual galaxies, but can also serve as proxies for galaxies in the early universe.  The most active galaxy building epoch was between redshifts of 1 and 3 (e.g., Hopkins \& Beacom 2006), but even space telescopes cannot access these systems with the detail and fidelity that is (currently only) possible in the local volume, within 10 to 20 Mpc. To begin to understand the galaxy ``ecosystem", you must first study nearby galaxies.

Since the days of Vesto Slipher and Edwin Hubble, nearby galaxies have been carefully studied using a combination of spectroscopy and imaging to decode their nature.  We have a wealth of data on tens of thousands of nearby galaxies, fully articulated in archives such as the NASA Extragalactic Database (NED) and Lyon-Meudon Extragalactic Database (HyperLEDA), mostly from optical observations, both photographic (e.g., RC3) and CCD imaging (e.g., SDSS), but also in the last four decades from UV and infrared whole-sky surveys of galaxies (e.g., {\it GALEX} , {\it IRAS}, 2MASS, {\it WISE}). Completing the electromagnetic spectrum, from the radio to the gamma-ray, the latest surveys hope to provide new and complementary ways of looking at galaxies, unlocking the complex processes of the baryonic cycle:  halo accretion, star formation, feedback and Gyr secular evolution.  

One such survey, the Wide-field Infrared Survey Explorer (\wise; Wright et al. 2010), is well suited to the study of galaxy evolution in general, and to nearby galaxies in particular.  The reason is that \wise has imaged the entire 4$\pi$ sky -- with relatively uniform breadth and depth. The imaging bands are sensitive to both stellar population and ISM processes, which can be used to study the past-to-present star formation history   \citep{Jar13}.  The ALLWISE Catalogue \citep{Cut12} and the upcoming CATWISE catalog \citep{Eis19} are optimized for point sources, both Milky Way stars and high-redshift galaxies.  The {\it WISE} mission did not make provision for systematically and properly measuring resolved sources; hence, these catalogs are either incomplete or poorly constructed for nearby and resolved galaxies.  We estimate that there are at least 2 million resolved galaxies detected in the \wise all-sky imaging, similar to the 2MASS Extended Source Catalogue \citep{Jar00}.  As such, our mission has been to identify and measure resolved galaxies in the {\it WISE} imaging.
{\bf We have carried out a systematic study of nearby galaxies, first introduced in the pilot study by \citet{Jar13}, that endeavours to build dedicated mosaics for all resolved galaxies, extract and characterize their measurable attributes, and produce uniform catalogs.} For those with redshifts or known distances, their physical (global) attributes are derived, including stellar mass and star formation activity. {\bf The over-arching project is called the \wise Extended Source Catalogue (WXSC).} 

The full-sky coverage of \wise means that every galaxy in the sky is imaged (to the sensitivity limits of \wise, see below), including the largest and most extended objects.  Hence, \wise has the unique dual capability of adding ancillary infrared data to any survey or observation, while also the ability to study the largest objects in the sky.  It is this current study that addresses the latter capability.  As a first major release, the largest angular-extent galaxies in the sky lead the way, which is appropriate given their historical and contemporary importance to galaxy studies. Subsequent releases of the WXSC will include well-known samples, such as the NGC, UGC, S$^4$G, as well as new galaxies that may be used to target spectroscopic surveys such as TAIPAN \citep{da16} and SDSS-V \citep{Kol17}.  To date we have measured and cataloged over 70,000 galaxies in the WXSC. 

Inspired by the release of the largest 2MASS galaxies \citep{Jar03}, here we present the angular largest 100 \wise galaxies in the sky, which includes most of the Local Group (LG), and many of the most famous Messier objects (M\,81, M\,51, M\,83, M\,101), as well as `exotic' varieties, the starbursts NGC\,253 and M\,82, and the AGN systems: NGC\,1068 (M\,77) and the Circinus Galaxy.  Some of the WXSC data (mosaics and measurements, including large gaaxies) have already been used in recent studies, notably neutral hydrogen studies of NGC\,253 \citep{Luc15}, M\,83 \citep{Hea16} and M\,33 \citep{Kam17, El19}, 
 multi-wavelength studies of NGC\,6744 \citep{Yew17} and M\,31 \citep{Tom19},  baryonic Tully-Fisher study \citep{Ogle19}, 
 moderate-sized samples (e.g., HIPASS) that study star formation-stellar mass relations \citep{Hall18} and the star-formation-gas connection \citep{Kor18, Park18, Park19}.  A number of studies are currently underway using the WXSC, including a complete census of the Coma and Perseus-Pisces Galaxy Clusters.
  One of the most important applications of the WXSC comes from \citet{Clu17} who derived a new set of star formation rate (SFR) relations using \wise and total infrared luminosities from SINGS \citep{Ken03} and KINGFISH \citep{Ken11} .   We will be using these SFR relations to characterize WXSC galaxies.
 
Here we present a uniform set of mosaics and measurements, and derive global properties, that may be used to study the local volume of galaxies.   We compare the measurements of the 100 Largest with a statistically-significant sample from the local universe, as compiled in the WXSC, to give context and contrast (if any) to the largest angular-sized galaxies.
 For completeness, we also provide measurements of LG galaxies that do not satisfy the angular size requirement.  Finally, measurements for the largest and brightest (Milky Way) globular clusters are also included since the are both large, bright and require similar photometric methods to extract their global fluxes.  We note this current release of the 100 largest galaxies will be directly followed by the periodic release of WXSC galaxies, whose catalog and data products will be fully described in a second paper, hereafter referred to as Paper II.
 

This paper is organized as follows:
Explanation of source measurements and the data are presented in \textsection 2. The 100 largest galaxy sample is introduced in \textsection 3.  Source properties, such as coordinates, size, shape, and photometry are presented in \textsection 4.  We then derive global physical attributes for the sample, and compare with large WXSC samples to assess the nature of the largest galaxies relative to the local universe;  \textsection 5.  In \textsection 6 we introduce a new way to graphically study the physical attributes, which will be used in future released; and finally, \textsection 7 presents the photometric measurements for the brightest globular clusters.

For galaxies without redshift independent distances, the cosmology adopted throughout this paper is H$_0 = 70 $\,km\,s$^{-1}$ Mpc$^{-1}$, $\Omega_M=0.3$ and $\Omega_{\Lambda} = 0.7$.  The conversions between luminosity distance and redshift use the analytic formalism of \citet{Wic10} for a flat, dark energy dominated Universe, assuming standard cosmological values noted above. Length and size comparisons are all carried out within the co-moving reference frame.   All magnitudes are in the Vega system ({\it WISE} photometric calibration described in \citealt{Jar11}).   Photometric colors are indicated using band names; e.g., W1$-$W2 is the [3.4$\, \mu$m]$-$[4.6$\, \mu$m] color. The Vega magnitude-to-flux conversion factors are
309.68, 170.66, 29.05, 7.871 Jy, respectively, for W1, W2, W3, and W4.  We  adopt the new W4 calibration from \citet{Br14b}, in which the central wavelength is 22.8$\, \mu$m (and hence we will refer to W4 as 23$\, \mu$m) and the magnitude-to-flux conversion factor is 7.871 Jy.  
It follows that the conversion from Vega System to the monochromatic AB System conversions are 2.67, 3.32, 5.24 and 6.66 mag.

\section{\wise Imaging \& Source Characterization}

In this section we describe the data, the imaging products, source characterization and selection of the largest galaxies.  All of the data are derived from the \wise mission, notably the level-1 individual frames and the ALLWISE source catalogs \citep{Cut12}.  Mosaics are constructed from the individual-epoch frames, and source characterization commences with both automated pipelines and expert-user interaction.  Measurements are carried out on galaxies, roughly 70,000 to date as part of the \wise Extended Source Catalogue,  located in the local universe, $z <$ 0.3, including the largest known galaxies.  Here we report on the 100 largest in the sky, including four, LMC/SMC and M\,31/M\,33,  of which required special processing due to their extreme angular extent.

\subsection{\wise Mission Details}

\wise is a NASA Medium-Class Explorer mission, launched in December 2009, 
featuring a
40-cm primary-mirror and 
1024$\times$1024 pixel Si:As and HgCdTe solid-hydrogen-cooled arrays that 
simultaneously imaged in four broad-spectral bands \citep{Wri10}.  Simply abbreviated as W1, W2, W3 and W4, the bands are centered on 3.4, 4.6, 12 and 22\m\ in the mid-infrared window,
with a W1 point source sensitivity that reaches $\sim$25\mj\ 
(5\sig ; \citet{Jar17}).  As previously noted, the W4 band has a filter response that is closer to 23\m\ \citep{Br14b}, and hence we refer to the central wavelength of W4 band as 23\m.
This primary `cryogenic' mission fully covered the sky several times (epochs) before the cryogen was exhausted (pre-2011), after which \wise began a new post-cryogen phase in which its two short bands (3.4 and 4.6\m) continued to map the sky with passive thermal cooling, focusing on asteroid science (read all about NEOWISE at  http://wise2.ipac.caltech.edu/docs/release/postcryo/).  


\subsection{Target Galaxies}

Building the \wise Extended Source Catalogue is a multi-phase project, starting with known, cataloged galaxies and extending to galaxies in the local universe that are resolved by {\it WISE}.  The former involves targeting specific galaxy catalogs (or samples), and the latter `blind' detection and characterization to determine the nature of the object: star , galaxy, or otherwise  \citep{Jar17}.   
For the largest galaxies in the sky, we rely on known catalogs, including the optically-based RC3/UGC (de Vaucouleurs et al. 1991; Nilsson, 1973) and S$^4$G (Sheth et al. 2010), and the databases NED and LEDA, to extract large galaxies.   Additionally, since we are building an infrared catalog, specifically using W1 (3.4\m) for the size metric, we include a number of infrared galaxy catalogs (e.g., IRAS Revised Bright Galaxy Sample, and the 2MASS XSC) to construct an initial sample to make sure we collect all of the largest galaxies.  

To date, we have measured over 70,000 nearby galaxies, including discrete Local Group galaxies that are detected by {\it WISE}.  
Source characterization provides the W1 (3.4\m) 1-\sig$_{\rm sky}$ isophotal ($\sim$23 mag arcsec$^{-2}$) size, which is used to identity the 100 largest galaxies.  Most of which are `normal' galaxies, but also includes some physically notable examples, from M\,82 (starburst), to the AGN/Seyferts NGC\,1068 (M\,77) and Circinus Galaxy.
In addition to extragalactic sources, we have measured the brightest (Milky Way) globular clusters since they are some of the largest and brightest objects in the sky, with at least one of them, Omega\,Centauri (NGC\,5139) considered to be a stripped dwarf galaxy (e.g., Noyola, Gebhardt, \& Bergmann 2008).  We describe the characterization process for globulars below, which is similar to that used for the Magellanic Clouds.

\subsection{Mosaic Construction}

Owing to the relatively small primary mirror, the  \wise angular resolution is poor compared to {\it Spitzer}-IRAC, with $\sim$6-8\arcs\ in the short bandpasses and 12\arcs\ in the longest bandpass \citep{Jar12}.
It is therefore important to work with native-resolution mosaic images when measuring extragalactic sources, especially those that are resolved.  Unfortunately the public-release ``Atlas" imaging from the \wise mission does not have native resolution -- they were smoothed to primarily benefit point source detection.  
Hence, we have constructed new mosaics of all galaxies in the WXSC to have native resolution with a field-of-view size, from arc-minutes to degrees, capable of measuring both the target galaxy and its local environment (see \citet{Jar13}).  There is the added benefit of co-adding epochs that were not available during the primary ALLWISE mission, with additional data from NEOWISE \citep{Mai13} in the W1 and W2 bands resulting in deeper, more sensitive photometric imaging.

Mosaics feature re-sampling with  1\arcs\ pixels using a `drizzle' technique developed in the software package ICORE \citep{Mas13} specifically designed for {\it WISE} single-frame images; details of the process and performance can be found in \citet{Jar12}.  All galaxies have this well-sampled (relative to the beam) 1\arcs\ pixel scale except for the LMC, SMC and M\,31:  they are too large to practically accommodate the resulting extreme image sizes.  Details of these exception galaxies and their processing is given below.   Depending on the total coverage (and hence, depth), 
the typical 1-\sig$_{\rm sky}$ surface brightness depths are 23.2, 22.1, 18.4 and 
15.8 mag arcsec$^{-2}$ (Vega mags), respectively for W1, W2, W3 and W4 bands. 

Mosaic construction is relatively straightforward for most galaxies. There are some cases which require extra attention, 
the highest surface brightness objects, specifically, M\,82, NGC\,253, Circinus, NGC\,1068, and NGC\,2070 (see below), as well as some of the globular clusters, all saturated their cores, notably in the longest bands of {\it WISE}.  
 Recovery of the lost information was carried out using a technique developed for {\it Spitzer} IRAC/MIPS image saturation (by the first author of this paper;  
 see also the IRAC instrument handbooks\footnote{https://irsa.ipac.caltech.edu/data/SPITZER/docs/ \\ irac/iracinstrumenthandbook/}
  for more detail), which entails utilizing the PSF and super-resolution image construction \citep{Jar12} to recover the core information. The method was utilized successfully for these cases; nevertheless, the resulting core flux has a large uncertainty, 20- 30\%, which translates to an extra 3-5\% uncertainty for the global flux.

In addition to the images, which include the signal, uncertainty and coverage maps per band, we also construct point spread functions (PSFs) that are used for star removal (see below).

\subsection{Source Characterization}

The goal of source characterization is to extract global parameters and to `reduce' an image (or in this case, four multi-$\lambda$ images) to a set of standard metrics that can be used to study the nature of the target galaxy.  The most important parameters are the integrated flux (or brightness), size and shape (inclination for disk galaxies), surface brightness, and cross-band flux ratios (i.e., colors).  There are many other parameters that are measured and cataloged (e.g., nuclear concentration), but are otherwise not discussed here, details will be given in Paper II.  The basic source characterization pipeline has a heritage that extends from the 2MASS Extended Source Catalogue \citep{Jar00}, also used in the \wise processing pipeline, and various aspects have been discussed in previous \wise work \citep{Jar11, Jar13, Clu14, Jar17, Clu17}.
Below we add some additional detail relevant to large galaxies.

One of the most important first steps is to identify and remove foreground Galactic stars so as not to contaminate the galaxy measurements.  This is particularly important for W1 and W2, which are filled with foreground Milky Way stars;  hence, particular attention is given to the W1 and W2 star-cleaning process.    We identify stars using the ALLWISE Catalog
\citep{Cut12}, which has basic point source characterization, the most important being the fluxes, colors and profile metrics -- notably, the reduced $\chi^2_{\rm reduced}$ which provides an assessment of the point or resolved nature of the object.   We use a combination of the source colors and point-like characteristics to decide if the source should be removed from the image -- that is to say,  if the source is foreground or associated with the target galaxy. 

We have found that sources that are resolved (i.e., galaxies in most cases) have a W1 $\chi^2_{\rm reduced}$ value greater than 2 \citep{Jar11,Clu14,Jar17}.  We use this metric to assess if the source is point-like (i.e., Galactic star, or background, distant galaxy) or resolved.  If resolved, it may be a piece of the target galaxy (e.g., H\,{\sc ii} region), in which case we may not want to remove it, especially if located on a spiral arm or within the disk, or it may be a blend of two or three stars, necessitating its removal from the cleaned image.  An even more powerful discriminant is the W2$-$W3 color,  which is relatively `blue' (low value) for Galactic stars compared to extragalactic sources or star-forming (SF) regions in the target galaxies.  In this way we avoid shredding our target galaxies into multiple pieces, which can be a real challenge for large imaging surveys and automated source extraction.   Moreover, and crucially, we visually inspect every galaxy to make sure that stars have been properly identified and removed, and that the inverse, shredding, has not occurred, in any band.
Stars are removed from the images by PSF subtraction, in the case of faint to moderately bright sources, W1 $\sim$ 15 mag (0.3 mJy), and for bright stars we use a masking process in which pixels are then recovered using the local background.  As noted, visual inspection and human-intervention are used for difficult cases, especially with source crowding and bright nearby stars, and for overall validation.

Once stars have been removed, the next step is to determine the local background value, which is an  iterative -- curve of growth, flux convergence -- process such that the local background is determined in a centered elliptical annulus whose inner and outer radii are safely (30 to 50\%)  beyond the light coming from the target galaxy. 
Stars are excluded from the distribution through subtraction and masking. Using the same method as that developed for \wise point-source processing, the local background is derived from a mean centered about the pixel-value distribution mode of the annulus pixel values (previously cleaned of foreground stars). The RMS uncertainty (\sig$_{\rm sky}$) of this background value is computed from the width of the distribution, as described in the \wise Explanatory Supplement \citep{Cut12}.  For nearly all cases a simple offset is all that is needed to removed the background emission, but in the case of M\,31, a tilted plane was used to account for the slight background light gradient (orthogonal to the Galactic Latitude axis) across the enormous field area, over 25 deg$^2$.

The local background, or sky value, is subtracted from the star-cleaned mosaics, and the primary source characterization is then carried out.  Shape characterization is derived from the 3-\sig$_{\rm sky}$ elliptical isophote, and held fixed for all measurements; i.e., the source is assumed to be elliptical and axi-symmetric.  The maximum extent of the galaxy is determined at the 1-\sig$_{\rm sky}$ elliptical isophote, which then represents the isophotal aperture for integrated flux measurements.
The total light, however, is estimated using both larger apertures (approaching the asymptotic limit) and by first constructing the radial surface brightness distribution, fitting a double S\'{e}rsic function, and extrapolating the extent of the galaxy to three disk scale lengths (based on the S\'{e}rsic  scale lengths).  The resulting `extrapolation' flux represents the ``total" flux, although it is not much more (5 to 10\%) than the isophotal flux, and agrees very well (1 to 2\%) with the asymptotic flux in most cases; details are in Section 4.1. 
Finally,  integrating radially from the center until 50\% of the integrated light is reached, the half-light (or effective) radius and surface brightness are then derived from the total flux.

These are the basic steps that are used to measure resolved galaxies. However, there are four notable exceptions that require further explanation:  M\,31, M\,33, LMC and SMC, given below.


\subsection{Magellanic Clouds}

\begin{figure*}[ht!]
 \includegraphics[width=180mm]{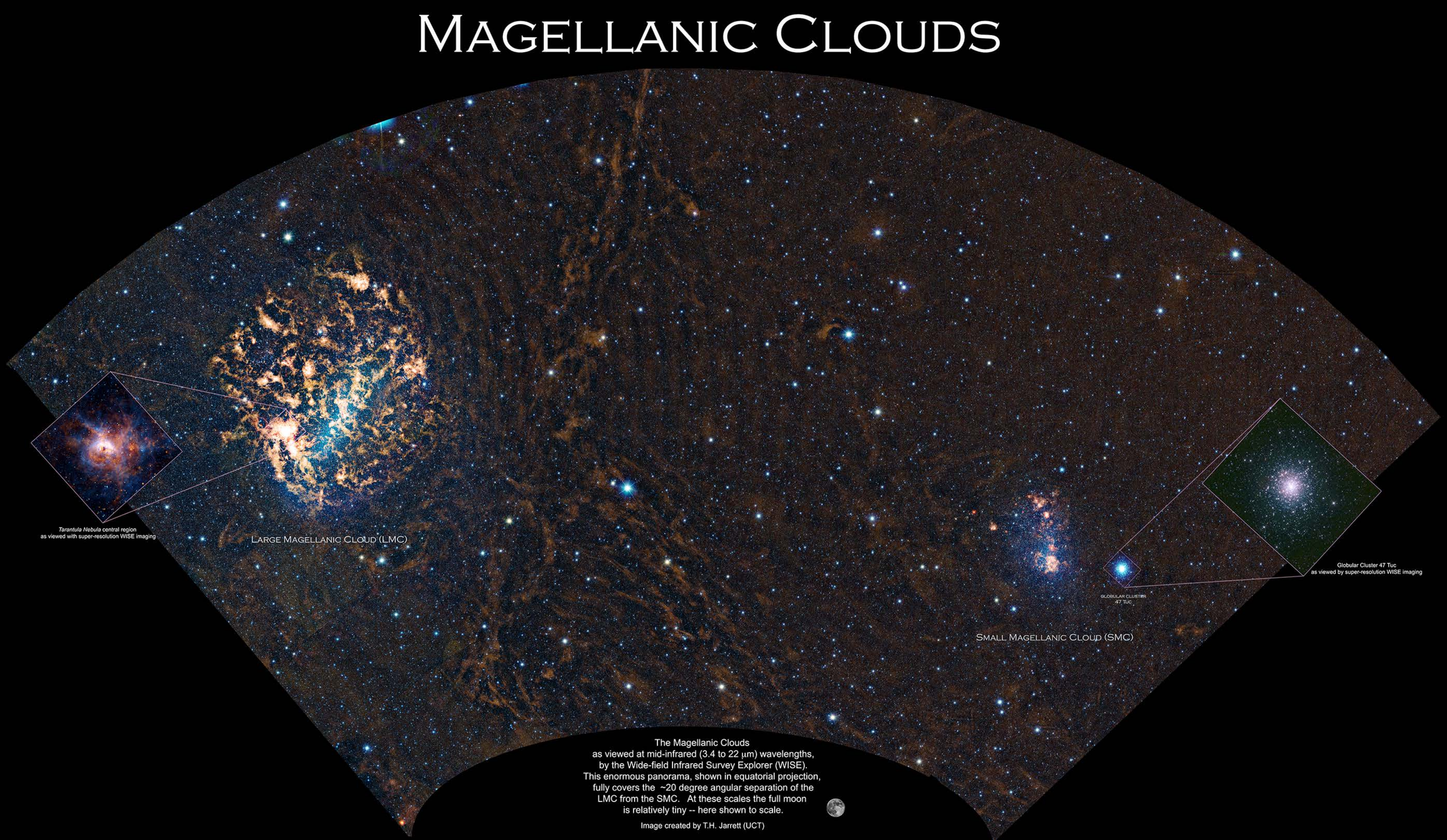}   
\caption{Mid-infrared four-color panorama of the Magellanic Clouds, constructed from \wise imaging. This equatorial projection fully covers the $>$20 degree separation between the LMC and the SMC.  The insets show \wise super-resolution mosaics of the Tarantula Nebula (30 Dor) in the LMC, and the globular cluster 47\,Tuc adjacent to the SMC. Note the long stream of nebulosity running north-south between the clouds, this is real emission likely associated with the tidal interaction between the clouds and the Milky Way.
\label{fig:f1}
}               
\end{figure*}

With the Magellanic Clouds, our challenging aim was to construct a mosaic large enough to encompass both galaxies,  thereby revealing the bridge region between them.  The Large and Small Clouds encompass a region of the sky that is $\sim$35 degrees across, mostly along the equatorial axis.  This is far too large to accommodate 1\arcs\ pixels, and given our primary goal of measuring the global properties of the Clouds, we did not require heavy over-sampling.  Consequently, we chose a pixel scale that was comparable to the resolution of \wise, specifically 8\arcs\ pixels.   

The resulting images (four bands) are 18,501 $\times$ 10,001  pixels, constructed with an equatorial projection (a galactic projection is not all that different), with the LMC to the east, and the SMC to the west.  
Artifacts from bright stars in the images are manually removed, and
the \wise bands are combined to form a four-color panorama of the region; see Figure~\ref{fig:f1}.  The depth of the images is relatively high for \wise because of the proximity to the South Ecliptic Pole; the resulting high coverage ($>$100 epochs on average) is good for optimal sensitivity, but the source confusion in W1 is reaching its peak
\citep[explored in detail by][]{Jar11}.
 So although the longward bands, W3 and W4, have excellent sensitivity in this region, the short bands of W1 and W2 are reaching diminishing returns with confusion noise.

The LMC is graced by the star formation complex 30\,Doradus (aka the Tarantula Nebula),  which is both spectacular, and supremely bright in the mid-infrared, so much so that the central star cluster (NGC\,2070) saturates in all four \wise bands.  The multiple-core saturation was rectified (see Section 2.3), and the restored super-resolution image is shown in Fig.~\ref{fig:f1} with the inset image on the east side.  On the west side, the SMC is shadowed by the magnificent globular cluster 47\,Tucanae, which was also saturated in the core (W1 only) and restored using the PSF and super-resolution; see the inset figure showing 47\,Tuc.

Source characterization for the Magellanic Clouds is carried out on the large mosaics.   
Since both objects are large in angular extent compared to the field of Galactic stars, it is not practical to identify stars within the Clouds that may or may not belong to the Milky Way.  This is also the case for M\,31, M\,33 and, worst case, globular clusters since member stars look just like foreground Galactic stars.   Hence, we employ a technique developed for measuring globular clusters in 2MASS \citep{Jar03}, whereby the mean flux of the foreground Galactic stars is measured and subtracted from the target galaxy.   The assumption is that the mean flux (per pixel) of the starlight as measured in a patch of sky (annulus centered on the target galaxy) is the same as that within the galaxy.  This is statistically robust for faint stars, but clearly there can be fluctuations from bright stars.  To minimize this deleterious effect, we identify bright Milky Way stars, contained in the ALLWISE Catalogue, from both the target and the sky patch and remove them before statistical measurements are taken.

In the infrared, the LMC displays the well-known stellar bar (appearing blue and elliptical in Fig. 1),  but also appears strangely square or ``blocky" in the ISM bands.  This is not an artifact of the \wise imaging, it is real and  can also be seen in the large {\it Spitzer}-SAGE mosaics of the LMC \citep{Meix06} and in {\it IRAS} 60\m\ dust maps.   To measure the global flux of the LMC, we have chosen to use a circular aperture (for simplicity) that fully covers the LMC, extending 10 degrees in diameter, encompassing the bar (seen in W1 and W2) and the star formation regions (W3 and W4).    For the sky patch, we choose
a region that is equally between the LMC and the SMC so that we can use the same mean stellar ``sky" flux for both clouds.  The sky region is centered on (03:00:45, -71:42:10) and has a radius of 4.5 degrees. The resulting Galactic foreground mean surface brightness is 21.62, 21.38, 20.41 and 19.10 mag arcsec$^{-2}$ (Vega system) for the sky patch (W1, W2, W3 and W4, respectively), which is then subtracted from the surface brightness of each pixel in the LMC and the SMC.

One final note of interest with the big mosaic; there is a ``river" of emission extending from the north to south, west of the LMC, that appears to terminate in the Chameleon II SF complex of the Milky Way.  It is notably bright in W3 (hence, orange-reddish in the adopted Fig. 1 color scheme), which would suggest the emission arises from warm dust and polycyclic aromatic hydrocarbon (PAH) emission in the bridge region, or possibly a closer projection from the Milky Way itself.  Although you can clearly see ``ringing" artifact features throughout this large mosaic of the Magellanic clouds, the N-S filament is in fact real.
You can see the exact correspondence in deep optical imaging 
\citep[][using a wide-field luminance filter imaging]{Bes16},
and in {\it IRAS} maps, the so-called Galactic cirrus filaments -- perfectly consistent with \wise 12\m\ imaging -- 
and even in Planck  353 GHz polarization maps
The large-scale feature may be related to 
 the tidal interaction between the Clouds and the Milky Way, or simply a nearby 
 magnetic-field-collimated dust filament that is unrelated 
 to the Magellanic Clouds. In any case, it is a very interesting feature for further investigation.

\subsection{M\,31 and M\,33}

\begin{figure*}[ht!]
 \includegraphics[width=170mm]{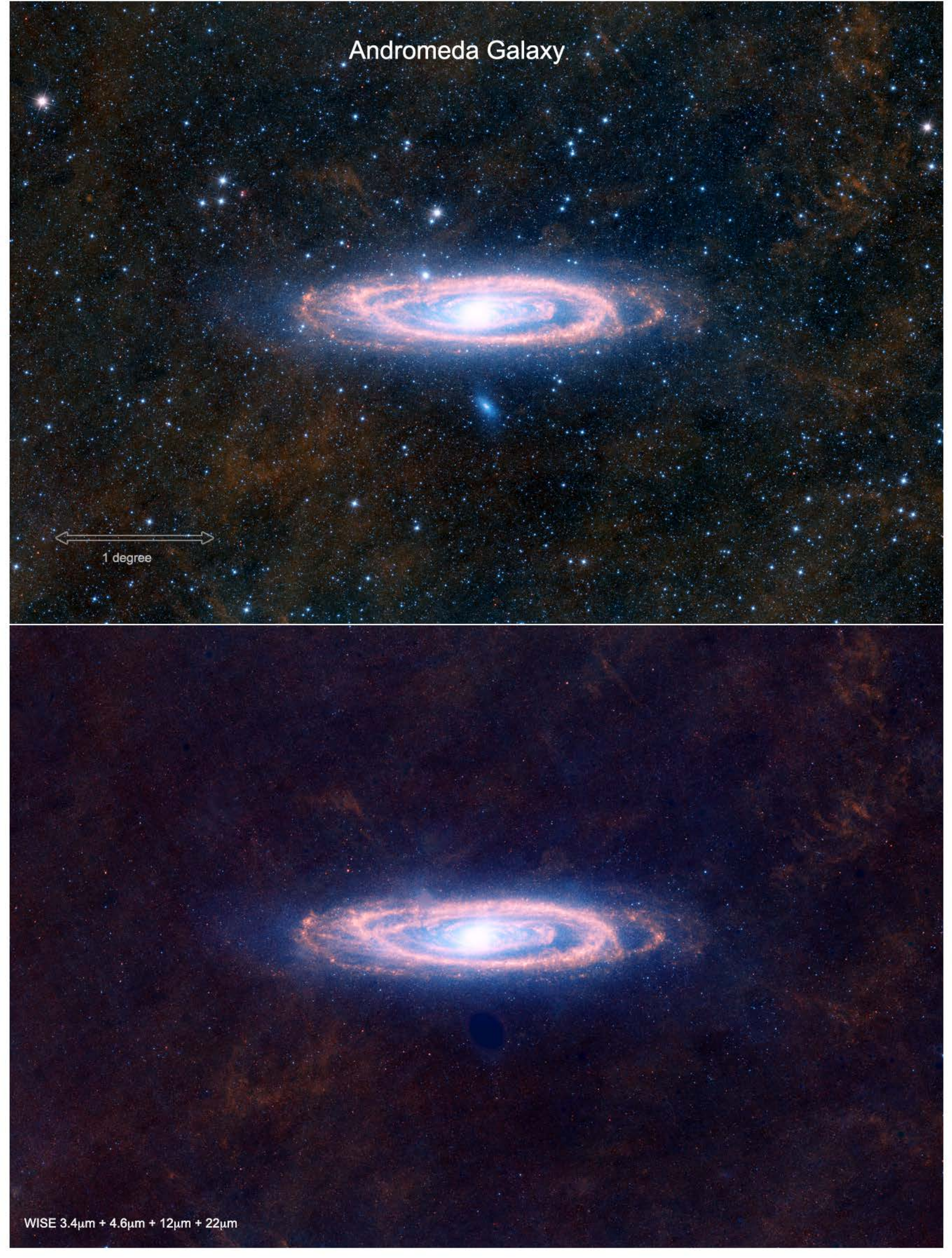}     
\caption{\wise four-color view of the Andromeda (M\,31) galaxy, projected in the ecliptic (spacecraft) orientation.  The upper panel shows the galaxy, its satellite galaxies (M\,32 and M\,110) and foreground Galactic stars, while the bottom panel shows only the galaxy with stars and satellites removed.  The outstanding feature with the infrared is that spiral arms embedded with molecular clouds and SF activity are in emission, as compared to absorption at optical wavelengths.
\label{fig:m31}}               
\end{figure*}

The extreme angular size of M\,31 and M\,33 required using a larger pixel scale (to conserve CPU memory, for example); here we construct with 1.5\arcs\ pixels.   The necessary angular extent of the images was first estimated by using the W1 stellar number density from the ALLWISE catalog, computed in shells that extended from the center of M\,31 out to large radii.  In close proximity to M\,31 the number density is high, while at large radii the value settles to the Milky Way mean value corresponding to about 3 degrees from the center of M\,31 \citep{Ch13}.  
In this way we determined that a 6 degree mosaic of M\,31 would be sufficient to measure the galaxy in the \wise bands; although M\,31 may yet extend beyond this range, we are not able to measure it, relative to the field, outside a radius of approximately 3 degrees.  Moreover, since M\,31 is fairly inclined and its orientation is diagonal in equatorial projection, we decided to construct the mosaic in \wise spacecraft mapping orientation (ecliptic). In this way, M\,31 disk appears horizontal, which is the optimal projection for minimizing the size of the required mosaic.  In the case of M\,33, since the galaxy is not nearly as large, we used the standard equatorial projection with a 2 degree extension along the north-south axis.

\begin{figure*}[ht!]
 \includegraphics[width=175mm]{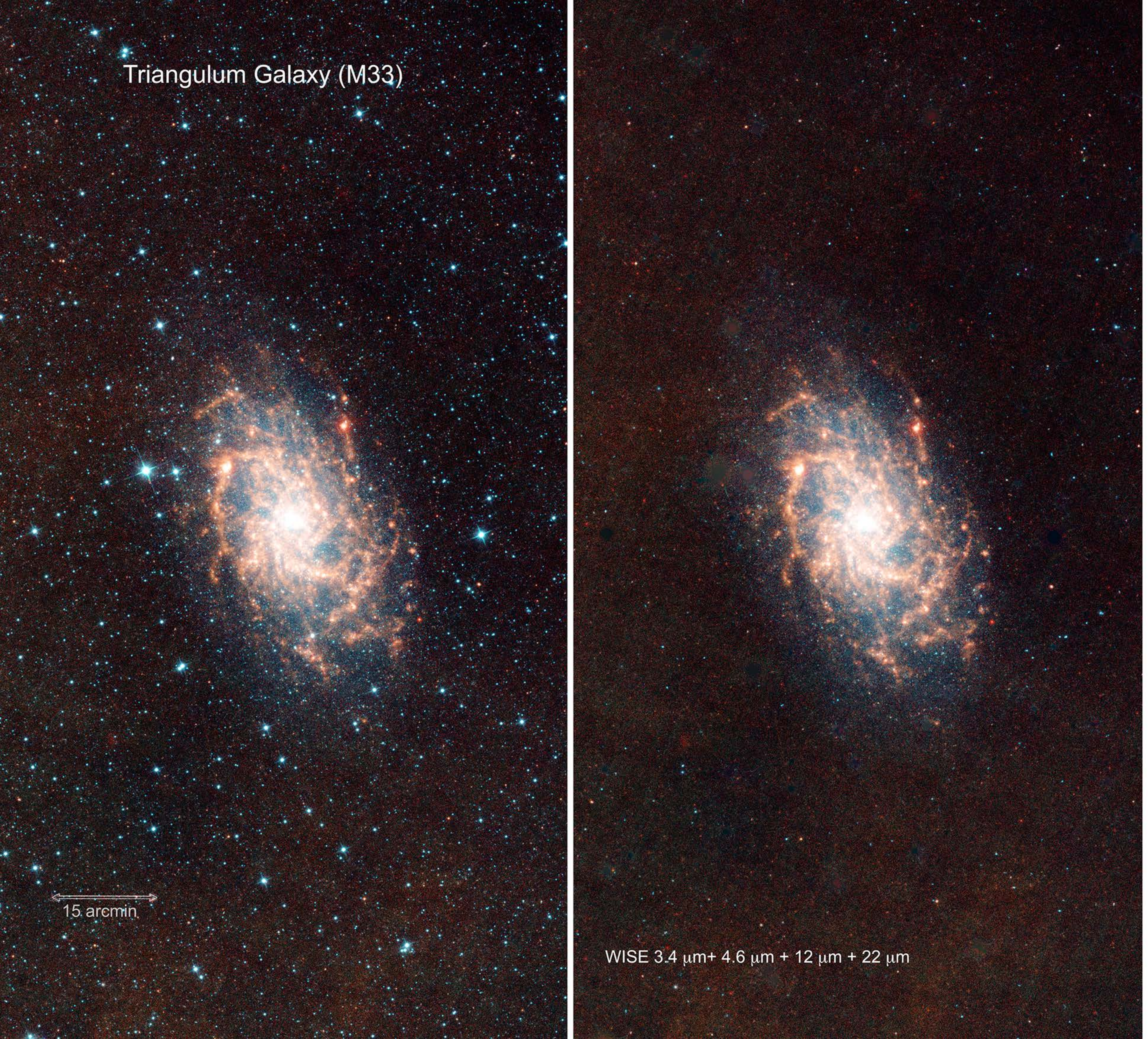}  
\caption{\wise four-color view of the Triangulum (M\,33) galaxy.  The left panel shows the galaxy and foreground Galactic stars, while the right panel shows only the galaxy.  M\,33 is one of the best targets in the sky to study because it is relatively face-on and presents parsec-scale SF regions for an entire range of spiral arm radii and environments.
\label{fig:m33}}               
\end{figure*}

It is not as straightforward to identify foreground Galactic stars from actual M\,31 sources due to the angular size of M\,31 and the large foreground Galactic population (centre of M\,31 is located at Galactic coordinates:  $l = 121.2\degr$, $b = -21.6\degr$).   The method we developed was part of a Masters dissertation \citep{Ch13}, and it uses in combination the stellar number density, proximity to the M\,31 disk, and photometric properties of the ALLWISE sources to determine the nature of the source.  It is a statistical method that determines the likelihood that a source belongs to M\,31 or to the Milky Way.  For example, well beyond the nucleus of M\,31, $>$2.7 deg, the field is completely dominated by Galactic stars, and the W1 star-counts and colors (W1$-$W2, W2$-$W3) have a signature of the Galactic population at the longitude and latitude of M\,31.  In closer proximity to the disk of M\,31, the source counts rise and the photometric properties begin to change; e.g., more faint stars are seen, from M\,31 itself, and colors redden because the K/M-giant populations are the only extragalactic stars detected because of the 
 distance of M\,31 relative to Milky Way stars; and hence, evolved and luminous stars appear much fainter (the distance modulus of M\,31 is about 24 magnitudes).  Exploiting these differences, we determine the most-likely Galactic foreground population, which are then removed from the M\,31 cleaned mosaic.   

The total number of foreground stars within the influence of M\,31 is impressive:  in a 11 deg$^2$ region centered on M\,31, over 100,000 foreground stars are identified.  We should note that near the nuclear bulge of M\,31, the density of sources is so high that confusion and crowding renders any source identification as impossible, and Galactic stars will remain in the bulge region.  Nevertheless, this is a relatively small area and the estimated flux of the total number of Galactic sources in the bulge is still less than 1\% of the total flux of M\,31 itself.  So with the sources that are identified as Galactic, they are removed from the image using the technique of PSF and area-masking, described earlier.       
The resulting mosaic of M\,31 is shown in Fig.~\ref{fig:m31} 

The same statistical method for star identification is used for M\,33, although not as challenging since M\,33 is considerably smaller ($\sim$1/3) relative to M\,31, and is located further from the Galactic Plane (coordinates:  $l = 133.6\degr$, $b = -31.3\degr$).   Crowding is also much less of a problem, as M\,33 has a weak stellar bulge.  The cleaned mosaic of M\,33 is shown in Fig.~\ref{fig:m33}.  

\section{The Largest Galaxies in the Sky}

We choose as the primary metric for angular size the \wise W1 (3.4\m) 
1\sig$_{\rm sky}$ isophotal radius (that is, the semi-major axis of the disk or tri-axial stellar distribution).  The W1 spectral window is sensitive to starlight from evolved stars, characterized by the stellar `bump' which peaks in the H-band (1.6\m) and extends into the mid-infrared through the Rayleigh-Jeans (R-J) tail.  Similar to using the K-band (2.2\m) as a tracer of the host galaxy stellar mass, the W1 band has been shown to be an excellent proxy for the stellar component of the baryonic mass \citep{Jar13, Mei14, Clu14, Pon17, Kett18, Hall18}.

In terms of surface brightness, \wise is relatively sensitive because it is a space-bound and it has large native pixels (2.5 arcsec).  Consequently, in the W1 band the typical RMS per pixel is about 23 mag  arcsec$^{-2}$ (Vega), and
 faint depths are reached with axi-symmetry averaging, typically 25 mag  arcsec$^{-2}$ (Vega), or $\sim$28 mag arcsec$^{-2}$ (AB), similar to those achieved with S$^4$G's {\it Spitzer}-IRAC \citep{She10}.  These depths are considerably fainter than the K$_s$-band in 2MASS which typically reaches 20 mag  arcsec$^{-2}$, and so W1 has much larger radii, ranging from 2 to 5 times larger than as measured by 2MASS K$_s$-band \citep{Clu14}.   Moreover, as will be shown below, the ``total" integrated flux in the W1 channel is 95\% contained within the isophotal aperture;  and hence, the W1 1-\sig$_{\rm sky}$ isophote is both robustly measured as well as being a good trace of the size and mass of the galaxy.

Based on the \wise W1 angular radius, we present the largest 100 galaxies in the sky in Table 1.  The table is sorted by the isophotal radius, and includes the Hubble Type, axis ratio (b/a), position angle orientation (East of North), and the integrated fluxes, quoted in flux density and in Vega magnitudes.  These parameters represent only a small subset of the full characterization that is completed for each galaxy (e.g., surface brightnesses, total fluxes, etc).  This additional data will be made available online with the full WXSC data release.

Not surprisingly, the largest galaxies in the sky are the Magellanic Clouds (Fig. 1), followed by the other Local Group galaxies M\,31 and M\,33 (Figs. 2, 3).  These galaxies have extents that can be measured in degrees, whereas the remaining galaxies in the universe have sizes in the arcmin and arcsec range.    We should note that it is challenging to measure the actual extent of these galaxies due to their sheer size relative to the imaging focal plane that they are measured with, thus requiring extremely large mosaic constructions.  Moreover, in the case of the Magellanic Clouds, they do not have well defined shapes; the LMC has an elliptical bar and a squarish-ISM (see Fig. 1), while the SMC is irregular in shape and extent.  Hence, the infrared size quoted here, 10 degrees diameter for the LMC and 5.6 degree for the SMC, has considerable uncertainty, $>$10\%.

The general galaxy properties and measured photometry are listed in Table 1. The first column (IS) indicates the radial size order, and the second column (IB) is the order based on the W1 integrated flux. The rank of the radius and integrated brightness correlates, as shown in Figure 5 and Figure 6.
 The LMC is the brightest galaxy in the sky, in terms of its integrated flux:  1828,       1064,       2377, and      5728 Jy, respectively for the four bands of {\it WISE}.  Next is Andromeda:  285,      152,       175, and  143 Jy, which just edges that of the SMC.  Both the Magellanic Clouds have larger W4 fluxes due to their current star formation  (see Section 5.4).  

Besides the famed LG galaxies, the largest galaxies includes many that are well-studied, including the local starburst NGC\,253 (\#\,5),  
Centaurus A (\#\,6), M\,101 (the Pinwheel; \#\,9), NGC\,1316 (Fornax A, \#\,12), the Sombrero (\#\,17), Virgo A (\#\,21),  LG dwarf NGC\,6822 (\#\,25), M\,83 (Southern Pinwheel, \#\,27),  M\,51 (Whirlpool, \#\,42), the Fireworks (\#\,49), knife-edge NGC\,5907 (\#\,63), the Seyfert galaxy Circinus (\#\,75)  located behind the Milky Way,  the starburst M\,82 (\#\,83), and spectacularly NGC\,1068 (\#\,85, M\,77), to name just a few.  Some of the largest-extent galaxies are not all that bright (e.g., the local dwarfs), correspondingly, some of the brightest galaxies are not in this list of large galaxies;  if we only consider the SF bands of W3 and W4, compact luminous IR galaxies would make the list of Top 100, such as  Arp\,220.

A large fraction of the LG galaxies appear in the list, however, a few notable ones are too small or too faint in surface brightness to detect in angular extent.  Hence, for completeness, we include the LG galaxies that \wise is able to detect and measure; note that many of the tiny and diffuse dwarf galaxies are not in play here.    Table 1, at the bottom, the list includes IC\,1613, M\,32,  UGC\,5373 and ESO\,245-007 (Phoenix\,Dwarf).


\begin{longrotatetable}
\begin{deluxetable}
{r | r | r | r | r | r | r | r | r | r | r | r | r | r | r | r | r | r | r | r | r | r | }

\def\arraystretch{0.95}%
\tabletypesize{\scriptsize}


\tablecaption{Photometry of the 100 Largest Galaxies Imaged by \wise}\label{table:big}

\tablenum{1}

\tablehead{
\colhead{iS} & \colhead{iB}  & \colhead{Galaxy} &  \colhead{R.A.} & \colhead{Dec} &
\colhead{Morph} & \colhead{ba  pa} &
\colhead{R$_{\rm W1}$} & 
\colhead{m1 $\pm \Delta$m1} &
\colhead{R$_{\rm W2}$} &
\colhead{m2 $\pm \Delta$m2} &
\colhead{R$_{\rm W3}$} &
\colhead{m3 $\pm \Delta$m3} &
\colhead{R$_{\rm W4}$} &
\colhead{m4 $\pm \Delta$m4} \\
\colhead{} & \colhead{} & \colhead{} & \colhead{deg} & \colhead{deg} & 
\colhead{} &  \colhead{} & 
\colhead{amin} & 
\colhead{mag} & 
\colhead{amin} & 
\colhead{mag} & 
\colhead{amin} & 
\colhead{mag} & 
\colhead{amin} & 
\colhead{mag} \\
\colhead{(1)} & \colhead{(2)} &\colhead{(3)} &\colhead{(4)} &\colhead{(5)} &\colhead{(6)} &\colhead{(7)} &\colhead{(8)} &\colhead{(9)} &
\colhead{(10)} & \colhead{(11)} & \colhead{(12)} &\colhead{(13)} &\colhead{14} &\colhead{(15)}  
}

\decimals

\startdata
  1 & 1 & LMC & 79.64101 & -68.81836  & SBm &  1.00 0.0    & 300.00 & -1.93 0.01 & 300.00 & -1.99 0.01 & 300.00 & -4.78 0.05 & 300.00 & -7.15 0.05 \\
  2 & 3 & SMC & 13.18660 & -72.82860  & SBm-pec & 1.00 0.0 & 166.67 &  0.56 0.01 & 166.67 & 0.62 0.01 & 166.67 & -0.93 0.01 & 166.67 & -3.48 0.01 \\
  3 & 2 & MESSIER31 & 10.68479 & 41.26907 & Sb & 0.35 37.3 & 111.44 &  0.08 0.01 & 111.44 & 0.13 0.01 & 111.44 & -1.95 0.01 & 111.44 & -3.09 0.01 \\
  4 & 5 & MESSIER33 & 23.46204 & 30.66022 & Sc & 0.59 18.9 &  31.64 &  3.09 0.01 & 31.64  & 3.02 0.01 & 31.64 & -0.17 0.01 & 31.64 & -1.94 0.01 \\
      5 &      6 &  NGC0253                    &   11.88844 &  -25.28827 &  SABc        &   0.28   53.0 &   21.13  &    3.46    0.01  &   21.13  &    3.27    0.01  &   21.13  &   -0.54    0.04  &   21.13  &   -2.94    0.04 \\
     6 &      4 &  NGC5128                    &  201.36481 &  -43.01954 &  S0          &   0.77   33.8 &   19.64  &    3.01    0.01  &   19.64  &    2.98    0.01  &   19.64  &    0.45    0.01  &   19.64  &   -1.25    0.01 \\
     7 &     36 &  NGC0055                    &    3.72192 &  -39.19775 &  SBm         &   0.19  108.9 &   19.49  &    5.45    0.01  &   19.49  &    5.36    0.01  &   19.49  &    2.80    0.01  &   19.49  &    0.29    0.01 \\
     8 &      7 &  MESSIER81                  &  148.88837 &   69.06528 &  Sab         &   0.56  156.7 &   14.67  &    3.60    0.01  &   14.67  &    3.61    0.01  &   14.67  &    1.84    0.03  &   14.67  &    0.45    0.03 \\
     9 &     27 &  MESSIER101                 &  210.80225 &   54.34893 &  SABc        &   0.92   33.3 &   14.50  &    5.22    0.01  &   14.50  &    5.10    0.01  &   14.50  &    1.62    0.01  &   14.50  &   -0.28    0.01 \\
    10 &     12 &  NGC4945                    &  196.36409 &  -49.46816 &  SBc         &   0.37   43.4 &   12.98  &    4.07    0.01  &   12.98  &    3.87    0.01  &   12.98  &    0.31    0.01  &   12.98  &   -1.56    0.01 \\
    11 &     10 &  IC0342                     &   56.70144 &   68.09635 &  SABcd       &   0.95   75.5 &   12.98  &    3.97    0.01  &   12.98  &    3.89    0.01  &   12.98  &    0.23    0.01  &   12.98  &   -1.81    0.01 \\
    12 &     23 &  NGC1316                    &   50.67380 &  -37.20796 &  S0          &   0.64   34.7 &   12.82  &    5.08    0.01  &   10.03  &    5.14    0.01  &    3.38  &    4.49    0.01  &    3.38  &    3.24    0.02 \\
    13 &     20 &  MESSIER49                  &  187.44489 &    8.00050 &  E           &   0.77  152.0 &   12.75  &    4.95    0.01  &    9.73  &    5.08    0.01  &    2.58  &    4.82    0.01  &    2.58  &    3.92    0.03 \\
    14 &     37 &  MESSIER86                  &  186.54897 &   12.94619 &  E           &   0.62  127.2 &   11.88  &    5.48    0.01  &    9.05  &    5.58    0.01  &    1.74  &    5.96    0.01  &    1.74  &    5.10    0.05 \\
    15 &      9 &  MAFFEI1                    &   39.14779 &   59.65487 &  E-S0        &   0.77   89.3 &   11.54  &    3.94    0.01  &   10.33  &    3.96    0.02  &    1.74  &    4.18    0.01  &    2.12  &    2.80    0.01 \\
    16 &     31 &  MESSIER110                 &   10.09197 &   41.68548 &  E           &   0.50  168.2 &   11.36  &    5.32    0.01  &   10.10  &    5.41    0.01  &   10.10  &    4.65    0.02  &   10.10  &    3.34    0.06 \\
    17 &     16 &  MESSIER104                 &  189.99765 &  -11.62305 &  Sa          &   0.46   90.7 &   11.04  &    4.62    0.01  &    9.28  &    4.66    0.01  &    3.67  &    3.74    0.01  &    3.67  &    2.57    0.02 \\
    18 &     43 &  NGC2403                    &  114.20885 &   65.60103 &  SABc        &   0.53  124.1 &   10.94  &    5.69    0.01  &   10.94  &    5.59    0.01  &   10.94  &    2.26    0.01  &   10.94  &    0.28    0.01 \\
    19 &     99 &  NGC0247                    &   11.78529 &  -20.76036 &  SABc        &   0.30  171.9 &   10.90  &    6.49    0.01  &   10.90  &    6.52    0.01  &   10.90  &    4.06    0.01  &   10.90  &    2.28    0.02 \\
    20 &     28 &  MESSIER106                 &  184.73955 &   47.30408 &  Sbc         &   0.46  152.0 &   10.67  &    5.23    0.01  &   10.67  &    5.19    0.01  &   10.67  &    2.75    0.01  &   10.67  &    1.09    0.01 \\
    21 &     30 &  MESSIER87                  &  187.70593 &   12.39113 &  E           &   0.79  152.9 &   10.37  &    5.32    0.01  &    6.88  &    5.46    0.01  &    1.86  &    5.14    0.01  &    1.86  &    3.84    0.02 \\
    22 &     48 &  NGC3628                    &  170.07085 &   13.58912 &  SBb         &   0.23  105.1 &   10.20  &    5.76    0.01  &    8.24  &    5.64    0.01  &    8.24  &    2.44    0.01  &    8.24  &    0.51    0.01 \\
    23 &     46 &  NGC0300                    &   13.72251 &  -37.68413 &  Scd         &   0.77  123.2 &   10.03  &    5.73    0.01  &   10.03  &    5.73    0.01  &   10.03  &    3.13    0.01  &   10.03  &    1.28    0.02 \\
    24 &     18 &  NGC4736                    &  192.72110 &   41.12020 &  SABa        &   0.89  105.6 &    9.64  &    4.77    0.01  &    7.19  &    4.78    0.01  &    7.19  &    2.03    0.01  &    7.19  &    0.32    0.01 \\
    25 &     51 &  NGC6822                    &  296.24115 &  -14.80224 &  IBm         &   0.82   20.5 &    9.53  &    5.80    0.01  &    9.53  &    5.80    0.01  &    9.53  &    3.50    0.01  &    9.53  &    1.20    0.02 \\
    26 &     90 &  NGC1532                    &   63.01807 &  -32.87421 &  SBb         &   0.24   32.2 &    9.44  &    6.43    0.01  &    7.40  &    6.41    0.01  &    7.40  &    3.61    0.01  &    7.40  &    2.00    0.01 \\
    27 &     13 &  NGC5236                    &  204.25296 &  -29.86598 &  Sc          &   0.95  161.4 &    9.33  &    4.24    0.01  &    9.33  &    4.12    0.01  &    9.33  &    0.32    0.01  &    9.33  &   -1.86    0.01 \\
    28 &    109 &  NGC0147                    &    8.30052 &   48.50901 &  E           &   0.57   28.7 &    9.28  &    6.57    0.01  &    7.52  &    6.69    0.01  &    7.52  &    4.84    0.01  &   null   &   null null     \\
    29 &     38 &  NGC6744                    &  287.44211 &  -63.85751 &  Sbc         &   0.70   11.9 &    9.14  &    5.49    0.01  &    9.14  &    5.47    0.01  &    9.14  &    2.27    0.01  &    9.14  &    0.68    0.01 \\
    30 &     29 &  MESSIER63                  &  198.95549 &   42.02931 &  Sbc         &   0.60  103.4 &    8.73  &    5.26    0.01  &    8.73  &    5.20    0.01  &    8.73  &    1.88    0.01  &    8.73  &    0.25    0.01 \\
    31 &     54 &  NGC1553                    &   64.04361 &  -55.78003 &  S0          &   0.79  153.1 &    8.70  &    5.89    0.01  &    6.21  &    6.00    0.01  &    2.25  &    5.38    0.01  &    2.25  &    4.27    0.02 \\
    32 &     52 &  NGC1399                    &   54.62093 &  -35.45040 &  E1pec       &   0.85   78.0 &    8.66  &    5.82    0.01  &    6.73  &    5.94    0.01  &    2.05  &    5.71    0.01  &    2.05  &    4.86    0.04 \\
    33 &    533 &  NGC4236                    &  184.17477 &   69.46757 &  SBdm        &   0.30  158.0 &    8.66  &    7.91    0.01  &    8.66  &    7.99    0.01  &    8.66  &    5.97    0.02  &    8.66  &    3.09    0.03 \\
    34 &     47 &  NGC4565                    &  189.08658 &   25.98755 &  Sb          &   0.19  136.3 &    8.54  &    5.76    0.01  &    8.54  &    5.71    0.01  &    8.54  &    3.21    0.01  &    8.54  &    1.66    0.01 \\
    35 &     17 &  Maffei2                    &   40.47926 &   59.60422 &  Sbc         &   0.44   27.7 &    8.50  &    4.76    0.01  &    8.50  &    4.58    0.01  &    8.50  &    1.07    0.01  &    8.50  &   -1.03    0.01 \\
    36 &     60 &  NGC4631                    &  190.53334 &   32.54174 &  SBcd        &   0.27   83.2 &    8.27  &    6.01    0.01  &    8.27  &    5.80    0.01  &    8.27  &    1.99    0.01  &    8.27  &   -0.06    0.01 \\
    37 &     33 &  MESSIER60                  &  190.91643 &   11.55272 &  E           &   0.82   99.3 &    8.23  &    5.38    0.01  &    6.28  &    5.47    0.01  &    2.17  &    5.04    0.01  &    2.17  &    3.98    0.04 \\
    38 &     65 &  NGC4636                    &  190.70770 &    2.68779 &  E           &   0.68  147.7 &    8.06  &    6.07    0.01  &    6.45  &    6.17    0.01  &    1.65  &    6.17    0.01  &    1.65  &    4.95    0.05 \\
    39 &    142 &  NGC2768                    &  137.90607 &   60.03756 &  E           &   0.32   91.7 &    7.98  &    6.76    0.01  &    5.89  &    6.83    0.01  &    1.80  &    6.67    0.01  &    1.80  &    5.62    0.06 \\
    40 &     77 &  NGC3585                    &  168.32115 &  -26.75477 &  E           &   0.57  109.0 &    7.94  &    6.28    0.01  &    5.76  &    6.40    0.01  &    1.34  &    6.30    0.01  &    1.34  &    5.35    0.06 \\
    41 &   1151 &  ESO270-G017                &  203.70584 &  -45.54919 &  SBm         &   0.18  107.5 &    7.94  &    8.55    0.01  &    6.05  &    8.72    0.01  &    6.05  &    6.27    0.02  &    6.05  &    3.98    0.03 \\
    42 &     24 &  MESSIER51a                 &  202.46959 &   47.19518 &  SABb        &   0.67   17.6 &    7.72  &    5.09    0.01  &    7.72  &    4.97    0.01  &    7.72  &    1.19    0.01  &    7.72  &   -0.61    0.01 \\
    43 &     41 &  NGC3115                    &  151.30800 &   -7.71870 &  E-S0        &   0.56   41.7 &    7.69  &    5.59    0.01  &    5.65  &    5.64    0.01  &    1.57  &    5.47    0.01  &    1.57  &    4.61    0.03 \\
    44 &     56 &  NGC3923                    &  177.75705 &  -28.80606 &  E           &   0.88   40.6 &    7.66  &    5.98    0.01  &    5.89  &    6.12    0.01  &    1.32  &    6.15    0.01  &    1.32  &    5.33    0.06 \\
    45 &     76 &  NGC4365                    &  186.11780 &    7.31775 &  E           &   0.68   35.6 &    7.53  &    6.27    0.01  &    5.48  &    6.44    0.01  &    0.85  &    6.86    0.01  &    0.85  &    5.82    0.05 \\
    46 &    136 &  NGC1313                    &   49.56689 &  -66.49796 &  SBcd        &   0.63  173.3 &    7.47  &    6.71    0.01  &    7.47  &    6.64    0.01  &    7.47  &    3.51    0.01  &    7.47  &    1.03    0.01 \\
    47 &     45 &  MESSIER84                  &  186.26555 &   12.88707 &  E           &   0.97   16.3 &    7.46  &    5.73    0.01  &    5.50  &    5.82    0.01  &    1.31  &    5.81    0.01  &    1.31  &    4.87    0.05 \\
    48 &     72 &  NGC0185                    &    9.74029 &   48.33795 &  E           &   0.74   49.0 &    7.45  &    6.23    0.01  &    5.81  &    6.34    0.01  &    2.06  &    5.52    0.01  &    2.06  &    4.65    0.04 \\
    49 &     21 &  NGC6946                    &  308.71796 &   60.15392 &  SABc        &   0.91   51.5 &    7.45  &    5.01    0.01  &    7.45  &    4.84    0.01  &    7.45  &    0.94    0.01  &    7.45  &   -1.04    0.01 \\
    50 &     93 &  NGC1395                    &   54.62383 &  -23.02732 &  E           &   0.81  146.5 &    7.42  &    6.45    0.01  &    5.63  &    6.57    0.01  &    1.32  &    6.41    0.01  &    1.32  &    5.47    0.07 \\
    51 &     25 &  IC0010                     &    5.07196 &   59.30388 &  IBm         &   0.87  125.8 &    7.33  &    5.11    0.01  &    7.00  &    5.04    0.01  &    3.87  &    2.35    0.01  &    3.87  &   -0.33    0.01 \\
    52 &    206 &  NGC4517                    &  188.18965 &    0.11525 &  Sc          &   0.18   81.9 &    7.22  &    7.08    0.01  &    7.22  &    6.99    0.01  &    7.22  &    3.91    0.01  &    7.22  &    2.11    0.01 \\
    53 &     32 &  NGC1291                    &   49.32742 &  -41.10804 &  S0-a        &   0.98   44.7 &    7.22  &    5.35    0.01  &    7.22  &    5.39    0.01  &    7.22  &    4.30    0.01  &    7.22  &    3.10    0.04 \\
    54 &     67 &  NGC2683                    &  133.17226 &   33.42194 &  Sb          &   0.19   42.0 &    7.16  &    6.14    0.01  &    6.26  &    6.10    0.01  &    6.26  &    3.79    0.01  &    6.26  &    2.65    0.02 \\
    55 &     59 &  NGC4697                    &  192.14961 &   -5.80060 &  E           &   0.81   49.9 &    7.13  &    6.00    0.01  &    5.27  &    6.10    0.01  &    1.32  &    6.02    0.01  &    1.32  &    4.81    0.04 \\
    56 &     35 &  NGC3521                    &  166.45242 &   -0.03587 &  SABb        &   0.62  165.4 &    7.13  &    5.44    0.01  &    4.93  &    5.38    0.01  &    4.93  &    1.98    0.01  &    4.93  &    0.33    0.01 \\
    57 &    598 &  NGC3109                    &  150.79233 &  -26.16096 &  SBm         &   0.31   91.7 &    7.09  &    8.00    0.01  &    7.09  &    8.01    0.01  &    7.09  &    6.66    0.04  &   null   &   null null     \\
    58 &     39 &  NGC0891                    &   35.63731 &   42.34778 &  Sb          &   0.25   22.9 &    7.09  &    5.51    0.01  &    7.09  &    5.30    0.01  &    7.09  &    1.97    0.01  &    7.09  &    0.23    0.01 \\
    59 &     50 &  MESSIER85                  &  186.35025 &   18.19108 &  S0-a        &   0.78   18.4 &    7.09  &    5.79    0.01  &    5.76  &    5.86    0.01  &    1.73  &    5.67    0.01  &    1.73  &    4.57    0.04 \\
    60 &    340 &  NGC4244                    &  184.37286 &   37.80739 &  Sc          &   0.19   46.5 &    7.08  &    7.51    0.01  &    7.08  &    7.46    0.01  &    7.08  &    5.34    0.01  &    7.08  &    3.27    0.03 \\
    61 &    221 &  NGC4762                    &  193.23326 &   11.23105 &  S0          &   0.29   27.8 &    7.08  &    7.11    0.01  &    5.16  &    7.19    0.01  &    1.61  &    6.96    0.02  &    1.61  &    6.18    0.10 \\
    62 &    155 &  NGC5084                    &  200.07019 &  -21.82725 &  S0          &   0.40   75.8 &    6.85  &    6.83    0.01  &    6.85  &    6.84    0.01  &    6.85  &    5.53    0.02  &    6.85  &    4.16    0.08 \\
    63 &     96 &  NGC5907                    &  228.97304 &   56.32876 &  SABc        &   0.19  154.3 &    6.76  &    6.47    0.01  &    6.76  &    6.35    0.01  &    6.76  &    3.24    0.01  &    6.76  &    1.57    0.01 \\
    64 &    461 &  NGC4395                    &  186.45364 &   33.54688 &  Sm          &   0.81  144.2 &    6.72  &    7.78    0.01  &    6.72  &    7.81    0.01  &    6.72  &    5.45    0.02  &    6.72  &    3.01    0.05 \\
    65 &     71 &  NGC1407                    &   55.04918 &  -18.58000 &  E           &   1.00  178.7 &    6.63  &    6.20    0.01  &    5.25  &    6.32    0.01  &    1.28  &    6.22    0.01  &    1.28  &    5.44    0.07 \\
    66 &     40 &  NGC3627                    &  170.06262 &   12.99155 &  Sb          &   0.53  177.3 &    6.59  &    5.56    0.01  &    5.16  &    5.47    0.01  &    5.16  &    2.03    0.01  &    5.16  &    0.02    0.01 \\
    67 &    148 &  NGC4438                    &  186.94028 &   13.00887 &  Sa          &   0.44   29.6 &    6.53  &    6.79    0.01  &    5.50  &    6.81    0.01  &    1.87  &    5.32    0.01  &    1.87  &    3.82    0.02 \\
    68 &     61 &  NGC1365                    &   53.40142 &  -36.14048 &  Sb          &   0.60   41.0 &    6.52  &    6.03    0.01  &    6.52  &    5.74    0.01  &    6.52  &    2.11    0.01  &    6.52  &   -0.46    0.01 \\
    69 &     42 &  NGC2903                    &  143.04214 &   21.50136 &  Sbc         &   0.49   21.5 &    6.48  &    5.67    0.01  &    6.48  &    5.58    0.01  &    6.48  &    2.07    0.01  &    6.48  &    0.09    0.01 \\
    70 &     87 &  NGC5846                    &  226.62210 &    1.60555 &  E           &   0.89   51.8 &    6.46  &    6.41    0.01  &    4.64  &    6.56    0.01  &    1.17  &    6.59    0.01  &    1.17  &    5.77    0.06 \\
    71 &     66 &  NGC4725                    &  192.61075 &   25.50079 &  SABa        &   0.63   40.0 &    6.44  &    6.09    0.01  &    5.12  &    6.15    0.01  &    5.12  &    3.99    0.01  &    5.12  &    2.62    0.02 \\
    72 &     83 &  NGC1549                    &   63.93811 &  -55.59235 &  E           &   0.80    5.3 &    6.43  &    6.35    0.01  &    5.68  &    6.46    0.01  &    1.90  &    6.09    0.01  &    1.90  &    5.21    0.03 \\
    73 &   1782 &  WLM                        &    0.49227 &  -15.46099 &  IB          &   0.33    2.9 &    6.43  &    8.93    0.01  &    6.43  &    9.00    0.01  &    6.43  &    8.36    0.20  &    6.43  &    5.47    0.26 \\
    74 &     53 &  NGC2841                    &  140.51106 &   50.97655 &  SBb         &   0.47  151.4 &    6.41  &    5.89    0.01  &    4.93  &    5.91    0.01  &    4.93  &    3.78    0.01  &    4.93  &    2.21    0.01 \\
    75 &     14 &  CircinusGalaxy             &  213.29146 &  -65.33923 &  Sb          &   0.67   20.9 &    6.37  &    4.43    0.01  &    6.37  &    3.76    0.01  &    9.19  &   -0.29    0.04  &    8.70  &   -2.58    0.04 \\
    76 &     82 &  NGC3621                    &  169.56879 &  -32.81405 &  SBcd        &   0.48  162.0 &    6.33  &    6.33    0.02  &    5.23  &    6.21    0.02  &    5.23  &    2.57    0.01  &    5.23  &    0.85    0.01 \\
    77 &    134 &  NGC5078                    &  199.95856 &  -27.41013 &  Sa          &   0.39  145.7 &    6.32  &    6.71    0.01  &    4.60  &    6.68    0.01  &    2.73  &    4.07    0.01  &    2.73  &    2.46    0.01 \\
    78 &     62 &  NGC1023                    &   40.10017 &   39.06337 &  E-S0        &   0.47   84.0 &    6.30  &    6.04    0.01  &    4.52  &    6.12    0.01  &    1.60  &    5.94    0.01  &    1.60  &    5.01    0.05 \\
    79 &     44 &  NGC7331                    &  339.26675 &   34.41580 &  Sbc         &   0.43  172.0 &    6.22  &    5.70    0.01  &    6.22  &    5.61    0.01  &    6.22  &    2.36    0.01  &    6.22  &    0.67    0.01 \\
    80 &     26 &  MESSIER64                  &  194.18185 &   21.68300 &  SABa        &   0.54  114.7 &    6.16  &    5.18    0.01  &    5.01  &    5.18    0.01  &    5.01  &    2.96    0.01  &    5.01  &    1.19    0.01 \\
    81 &     92 &  MESSIER59                  &  190.50938 &   11.64700 &  E           &   0.76  164.9 &    6.11  &    6.45    0.01  &    4.50  &    6.57    0.01  &    1.44  &    6.20    0.01  &    1.44  &    5.23    0.06 \\
    82 &    123 &  NGC4696                    &  192.20518 &  -41.31093 &  E           &   0.62   97.9 &    6.07  &    6.66    0.01  &    5.06  &    6.74    0.01  &    1.62  &    6.63    0.01  &    1.62  &    5.54    0.06 \\
    83 &     11 &  MESSIER82                  &  148.96646 &   69.67978 &  S?          &   0.79   54.1 &    6.06  &    4.07    0.01  &    6.06  &    3.61    0.01  &    8.01  &   -0.96    0.01  &    8.49  &   -4.14    0.01 \\
    84 &    538 &  ESO274-001                 &  228.55611 &  -46.80931 &  SAd         &   0.16   36.6 &    5.91  &    7.91    0.01  &    5.91  &    7.93    0.01  &    5.91  &    6.94    0.08  &    5.91  &    3.04    0.02 \\
    85 &     15 &  MESSIER77                  &   40.66975 &   -0.01340 &  Sb          &   0.87   80.2 &    5.87  &    4.59    0.01  &    4.73  &    2.89    0.01  &    5.91  &   -0.68    0.04  &    6.32  &   -2.45    0.04 \\
    86 &    149 &  NGC3077                    &  150.82985 &   68.73387 &  S?          &   0.73   38.5 &    5.84  &    6.80    0.01  &    4.51  &    6.76    0.01  &    2.64  &    4.12    0.01  &    2.64  &    1.64    0.01 \\
    87 &     55 &  MESSIER65                  &  169.73309 &   13.09245 &  Sa          &   0.34  171.3 &    5.84  &    5.90    0.01  &    4.91  &    5.92    0.01  &    4.91  &    4.16    0.01  &    4.91  &    2.78    0.02 \\
    88 &     98 &  NGC7213                    &  332.31741 &  -47.16676 &  Sa          &   0.97  174.6 &    5.83  &    6.48    0.01  &    3.99  &    6.47    0.01  &    2.21  &    4.36    0.01  &    2.21  &    2.48    0.01 \\
    89 &     49 &  IC0356                     &   61.94514 &   69.81254 &  Sb          &   0.72  103.0 &    5.80  &    5.76    0.01  &    4.98  &    5.78    0.01  &    4.98  &    3.78    0.01  &    4.98  &    1.95    0.02 \\
    90 &   1213 &  NGC1560                    &   68.20547 &   71.88394 &  Scd         &   0.22   24.4 &    5.79  &    8.59    0.01  &    4.51  &    8.75    0.01  &    1.56  &    7.71    0.04  &    1.56  &    5.55    0.04 \\
    91 &     74 &  NGC2663                    &  131.28387 &  -33.79462 &  E           &   0.65  112.2 &    5.78  &    6.26    0.01  &    5.03  &    6.36    0.01  &    1.22  &    6.45    0.01  &    1.22  &    4.95    0.03 \\
    92 &     85 &  NGC4216                    &  183.97667 &   13.14956 &  SABb        &   0.26   20.7 &    5.77  &    6.36    0.01  &    4.71  &    6.38    0.01  &    4.71  &    4.24    0.01  &    4.71  &    2.89    0.02 \\
    93 &    143 &  NGC1055                    &   40.43828 &    0.44376 &  SBb         &   0.30  103.8 &    5.69  &    6.76    0.01  &    5.69  &    6.64    0.01  &    5.69  &    3.08    0.01  &    5.69  &    1.32    0.01 \\
    94 &    303 &  NGC5170                    &  202.45326 &  -17.96646 &  Sc          &   0.17  127.3 &    5.67  &    7.43    0.01  &    4.74  &    7.41    0.01  &    4.74  &    5.24    0.01  &    4.74  &    3.79    0.02 \\
    95 &    135 &  MESSIER98                  &  183.45123 &   14.90056 &  SABb        &   0.28  155.7 &    5.64  &    6.71    0.01  &    5.64  &    6.67    0.01  &    5.64  &    3.83    0.01  &    5.64  &    2.21    0.02 \\
    96 &     68 &  NGC2997                    &  146.41144 &  -31.19094 &  SABc        &   0.78   99.2 &    5.62  &    6.15    0.01  &    5.62  &    6.03    0.01  &    5.62  &    2.35    0.01  &    5.62  &    0.54    0.01 \\
    97 &    101 &  NGC4125                    &  182.02449 &   65.17439 &  E           &   0.71   83.9 &    5.61  &    6.50    0.01  &    4.03  &    6.59    0.01  &    1.44  &    6.27    0.01  &    1.44  &    5.20    0.04 \\
    98 &    102 &  NGC7793                    &  359.45728 &  -32.59102 &  Scd         &   0.62   97.9 &    5.60  &    6.52    0.01  &    5.60  &    6.43    0.01  &    5.60  &    3.21    0.01  &    5.60  &    1.55    0.01 \\
    99 &    108 &  NGC5363                    &  209.03012 &    5.25490 &  S0-a        &   0.68  124.6 &    5.60  &    6.57    0.01  &    4.06  &    6.63    0.01  &    1.49  &    5.69    0.01  &    1.49  &    4.36    0.02 \\
   100 &    245 &  NGC4217                    &  183.96248 &   47.09152 &  ZZZ         &   0.17   49.3 &    5.59  &    7.23    0.01  &    4.20  &    7.09    0.01  &    4.20  &    3.74    0.01  &    4.20  &    2.07    0.01 \\
   -- &   1081 &  IC1613                     &   16.20089 &    2.11914 &  IBm         &   0.90  135.4 &    4.69  &    8.49    0.01  &    4.69  &    8.69    0.01  &   null   &   null null      &   null   &   null null     \\
   -- &     22 &  MESSIER32                  &   10.67439 &   40.86511 &  E           &   0.86  165.8 &    3.62  &    5.02    0.01  &    2.95  &    5.07    0.01  &    2.95  &    4.42    0.01  &    2.95  &    3.61    0.04 \\
   -- &   3902 &  UGC05373                   &  149.99974 &    5.33153 &  IB          &   0.75   97.0 &    2.89  &    9.76    0.01  &    1.76  &   10.04    0.02  &   null   &   null null      &    1.76  &    6.69    0.35 \\
   -- &   8996 &  ESO245-007                 &   27.77698 &  -44.44341 &  Sm          &   0.81   11.2 &    2.18  &   11.06    0.01  &    2.18  &   11.28    0.04  &   null   &   null null      &   null   &   null null     \\
\enddata

\tablecomments{columns:  (1) order of W1 3.4\m\ angular size; (2) order of W1 brightness; (3) Galaxy name;  (4, 5) J2000 coordinates;
(6) RC3 Hubble Type; (7) axis ratio and position angle (degrees, E of N);
(8) W1 angular semi-major axis (arcmin), see Fig.~\ref{fig:size}; 
 (9) W1 magnitude and associated error; 
 (10) W2 angular semi-major axis (arcmin); 
 (11) W2 magnitude and associated error; 
(12) W3 angular semi-major axis (arcmin); 
 (13) W3 magnitude and associated error; 
(14) W4 angular semi-major axis (arcmin); 
 (15) W4 magnitude and associated error.
The following galaxies had saturated cores, necessitating recovery of lost flux:   M\,82, NGC\,253, Circinus, NGC\,1068 (M\,77), and NGC\,2070 (LMC).   
 }




\end{deluxetable}
\end{longrotatetable}


In the following sections, we compare the properties of the largest galaxies,  including their derived physical properties (size, mass and star formation).  For some context, we also consider their mass and star formation properties relative to a large spectroscopic survey of local universe galaxies \citep{Jar17}.

\section{Photometric Properties of the Largest Galaxies}

In this section we present the source characterization for 100 largest galaxies, including 
isophotal and total flux measurements, colors, surface brightness and bulge-to-disk comparisons.

\subsection{Measurements of Integrated Flux and Radial Flux Distribution}

Isophotal fluxes are derived from the W1 (3.4\m) 1-\sig$_{\rm sky}$ isophote ($\sim$23 mag arcsec$^{-2}$)
fit with an elliptical aperture whose shape (axis ratio and position) was determined at higher S/N (3-\sig$_{\rm sky}$)
and, for simplicity, assumed constant orientation throughout all radii.  
Total fluxes are
estimated in two different ways:  (1) asymptotic apertures and (2) by fitting the axisymmetric radial flux distribution; see below.  

\begin{figure}[ht!]
 \begin{minipage}[b]{1.01\linewidth}
        \hspace*{-0.75cm}
    \includegraphics[width=1.12\linewidth]{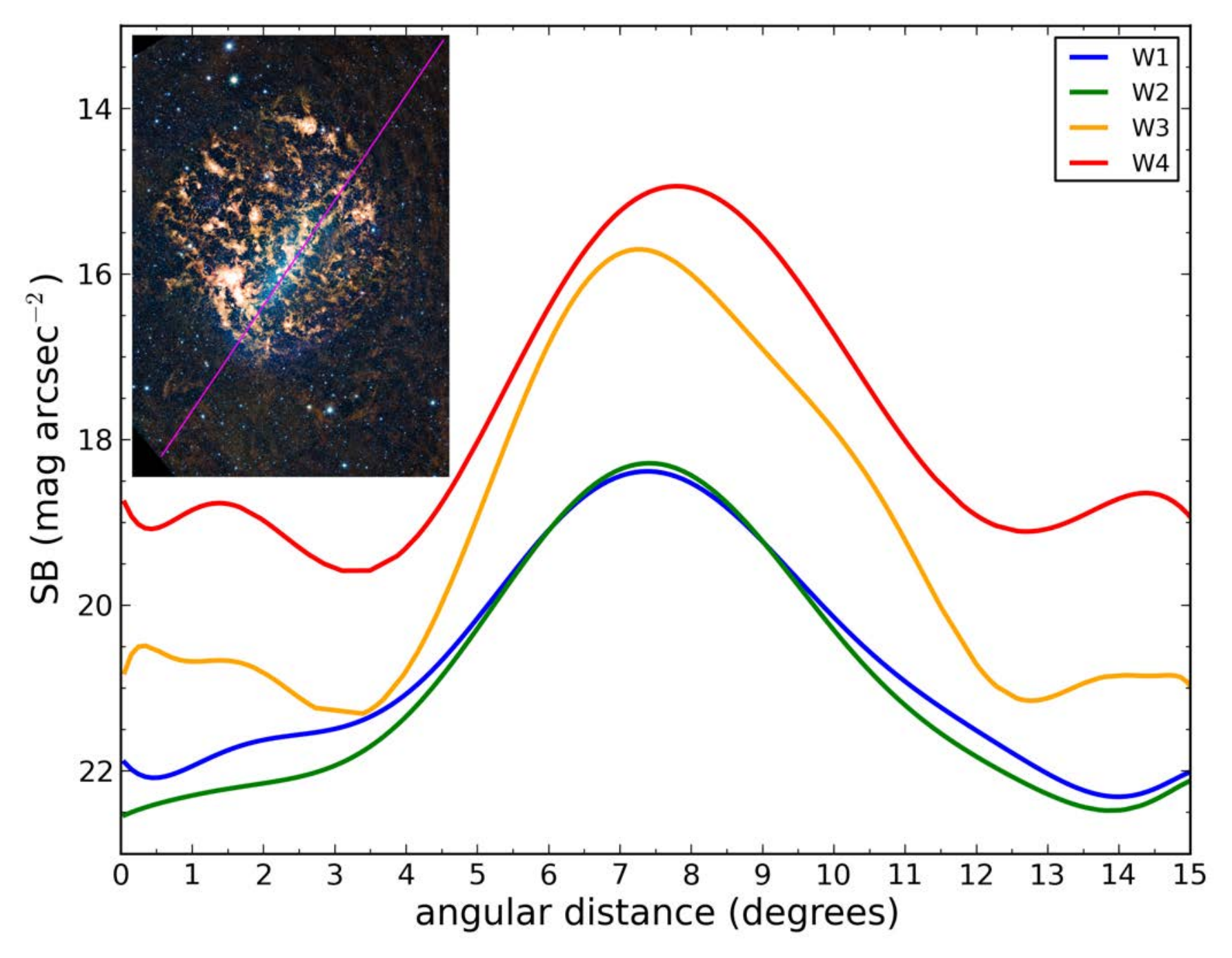} 
    \vspace*{-0.6cm}      
\caption{Mean surface brightness slice through the bar and central region of the LMC. The slice starts in the lower left of the LMC (see inset, magenta line), and extends 15 degrees ($\sim$15.7 kpc) diagonally along the stellar bar.  The mean surface brightness at 100$\times$100\,pc scale is first computed, followed by fitting an 11th order Legendre polynomial to produce the smooth (large-scale) result shown here.  The stellar bar surface brightness (W1) peaks at (05:25:36, -69:49:34) J2000 coordinates.
\label{fig:lmc-slice}}   
 \end{minipage}            
\end{figure}

\begin{figure}[!htbp]
  \begin{minipage}[b]{1.01\linewidth}
           \hspace*{-0.75cm}
    \includegraphics[width=1.12\linewidth]{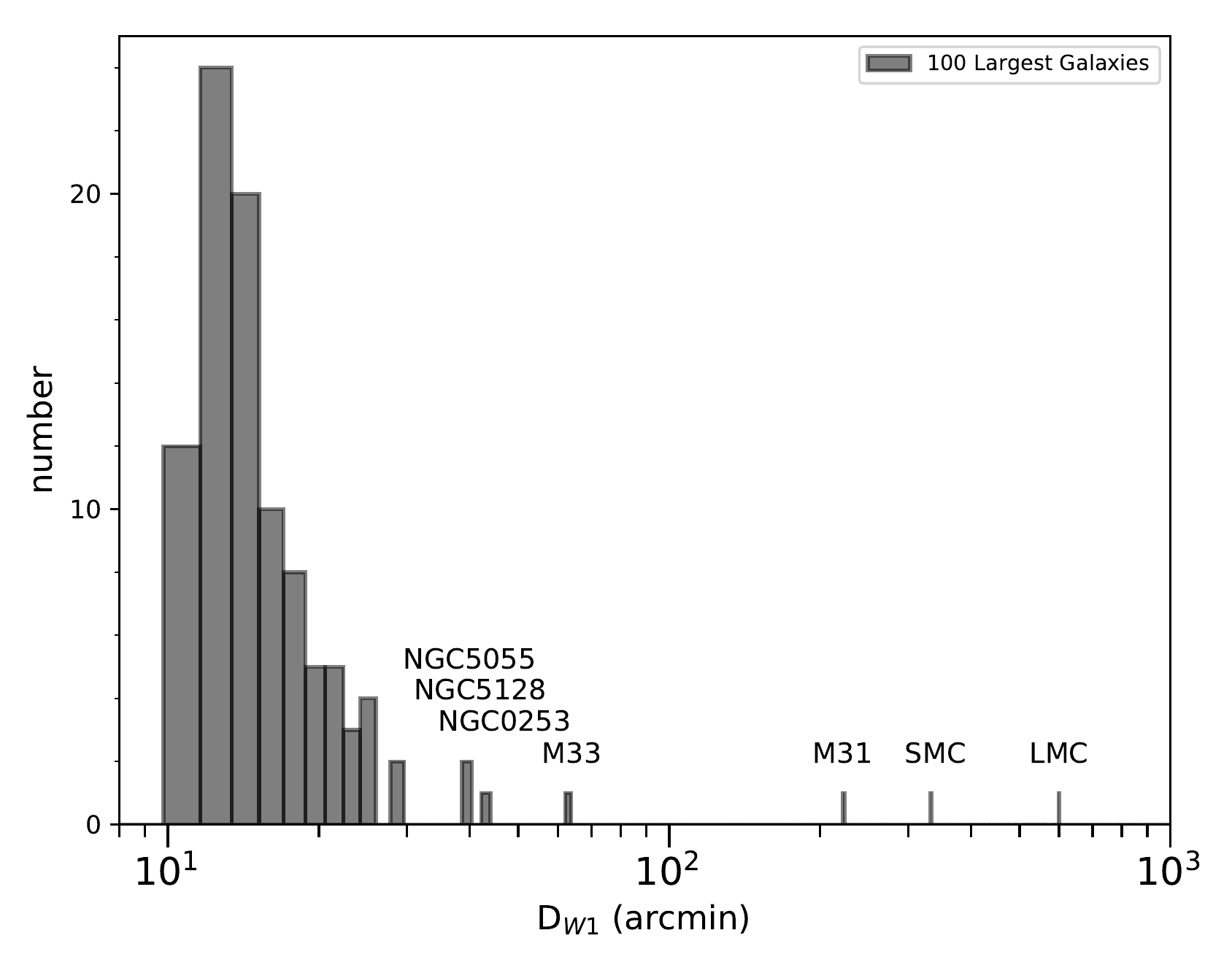} 
    \vspace*{-0.8cm}      
    \caption{Angular diameter size distribution of the 100 largest galaxies based on the \wise W1 imaging.  The size metric is from the 1-\sig$_{\rm sky}$ ($\sim$23 mag arcsec$^{-2}$) isophotal semi-major axis (R$_{W1}$).  The largest are the Local Group galaxies, followed by the nearest galaxy groups, (NGC\,253; Cen A), all within a few Mpc from the Milky Way.
    }
     \label{fig:size}
  \end{minipage}

  \begin{minipage}[b]{1.01\linewidth}
           \hspace*{-0.75cm}
    \includegraphics[width=1.12\linewidth]{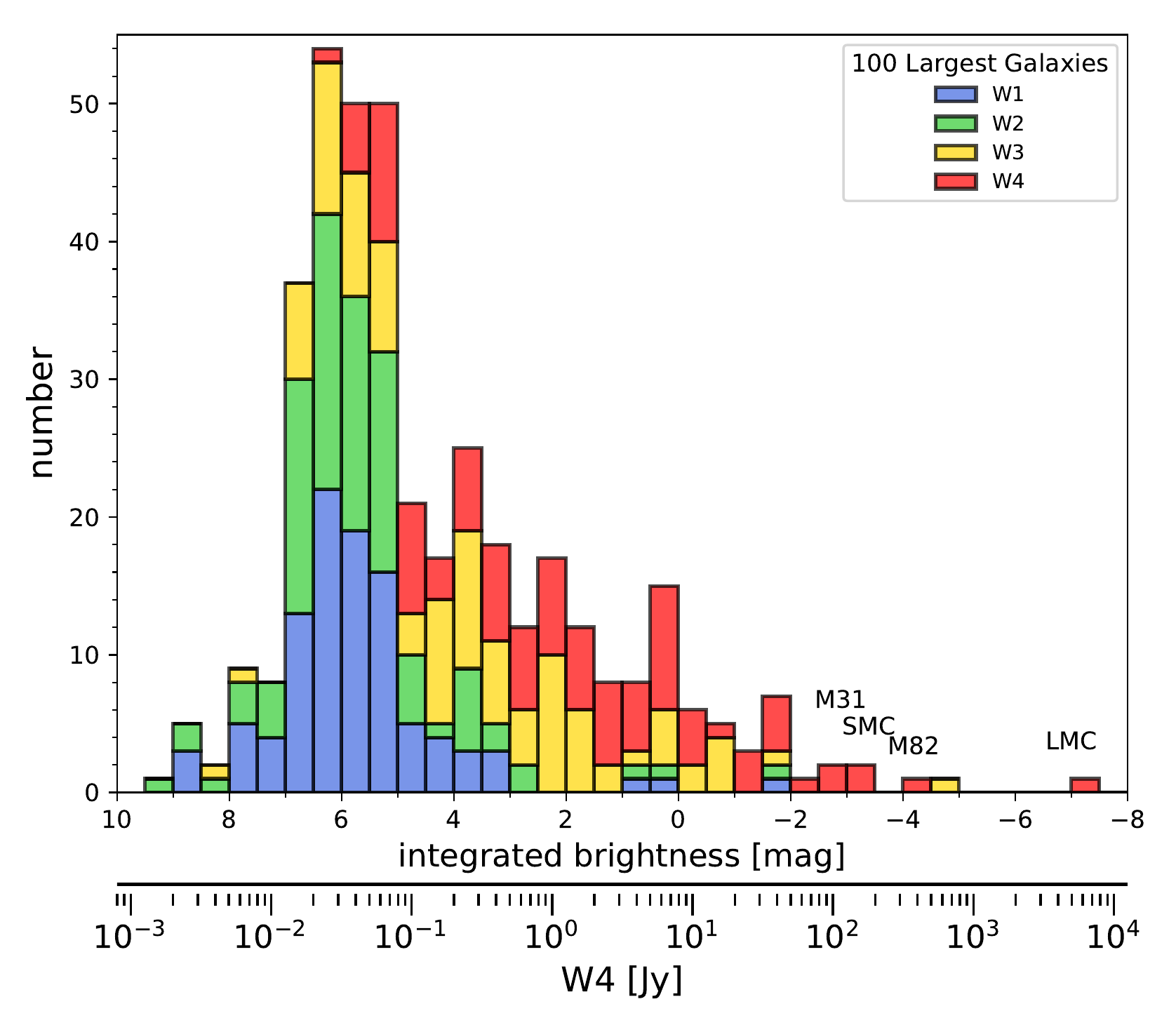} 
    \vspace*{-0.6cm}      
    \caption{Stacked histogram showing the integrated brightness distribution for the four \wise bands based on the isophotal measurements.  The brightest galaxies in W4 (22\m) are indicated, with the W4 flux density (in Jansky units) scale given for comparison to the magnitudes.   The LMC is the brightest owing to its proximity (50 kpc), but here the outstanding galaxy is M\,82, whose starburst is creating so much light in the infrared ($\sim$360 Jy) that it saturates in \wise (and {\it Spitzer}) imaging.
    }
    \label{fig:mag}
  \end{minipage}
\end{figure}

\begin{figure*}[htbp!]
\gridline{\leftfig{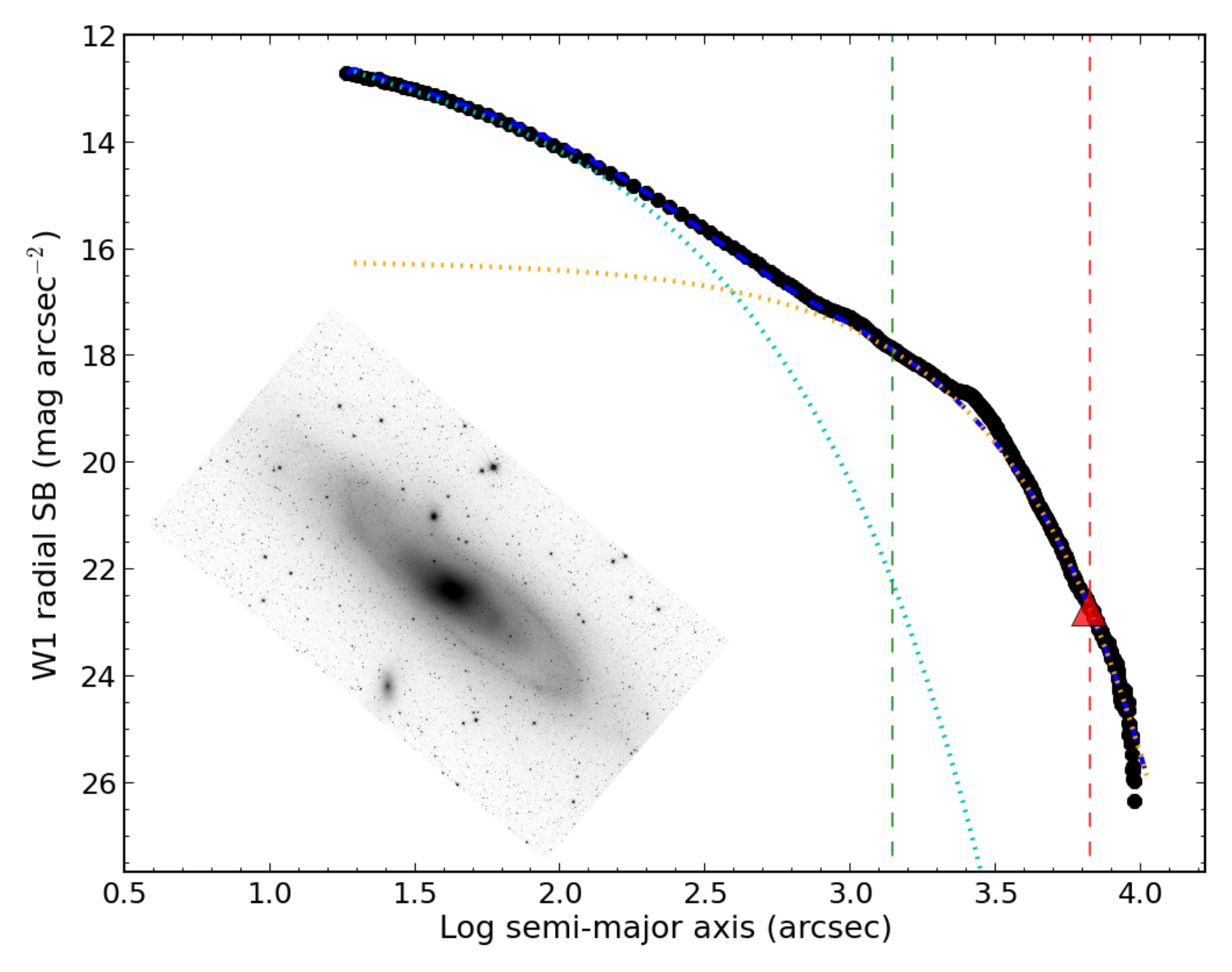}{0.5\textwidth}{(a)} 
              \rightfig{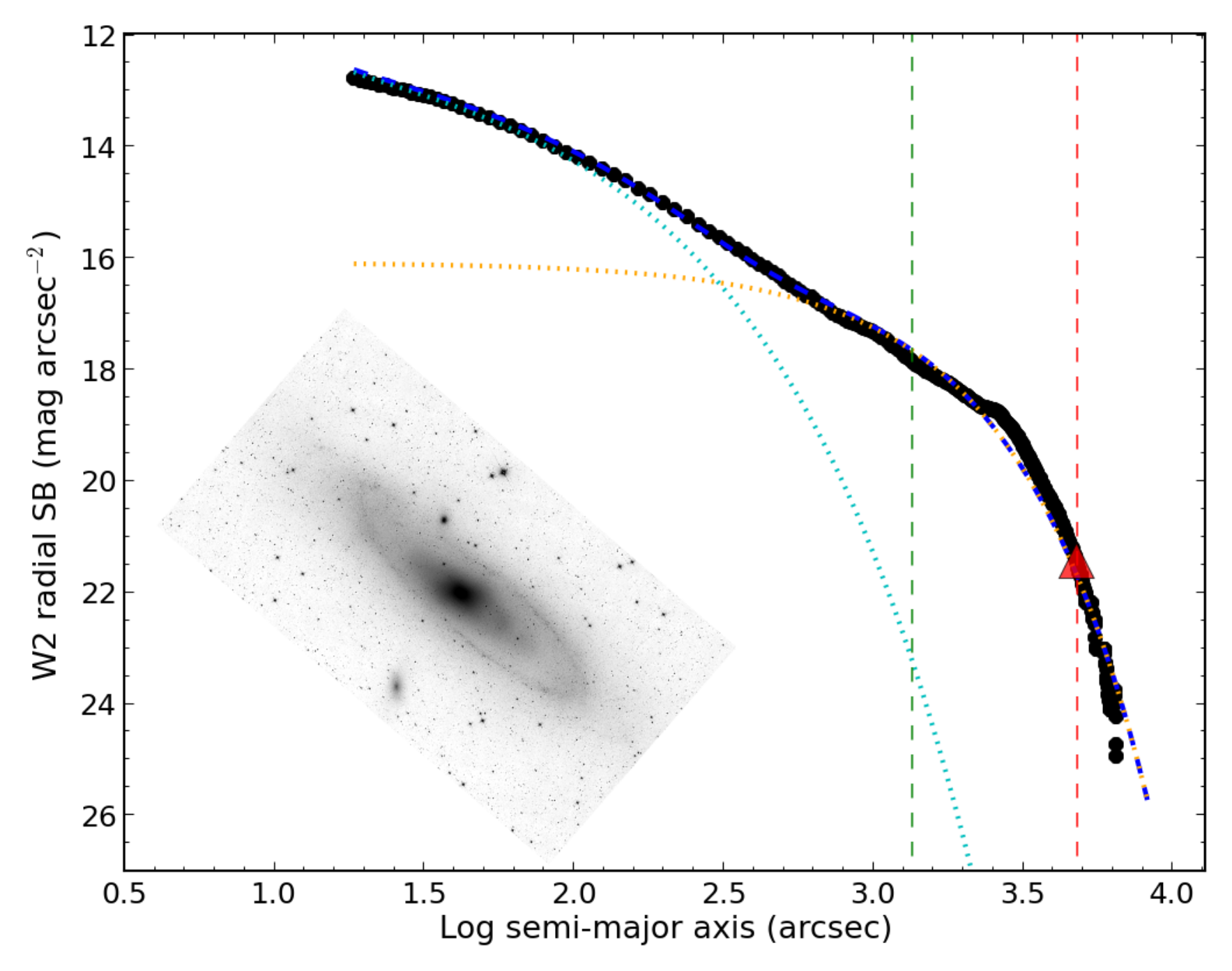}{0.5\textwidth}{(b)}}  
\gridline{\leftfig{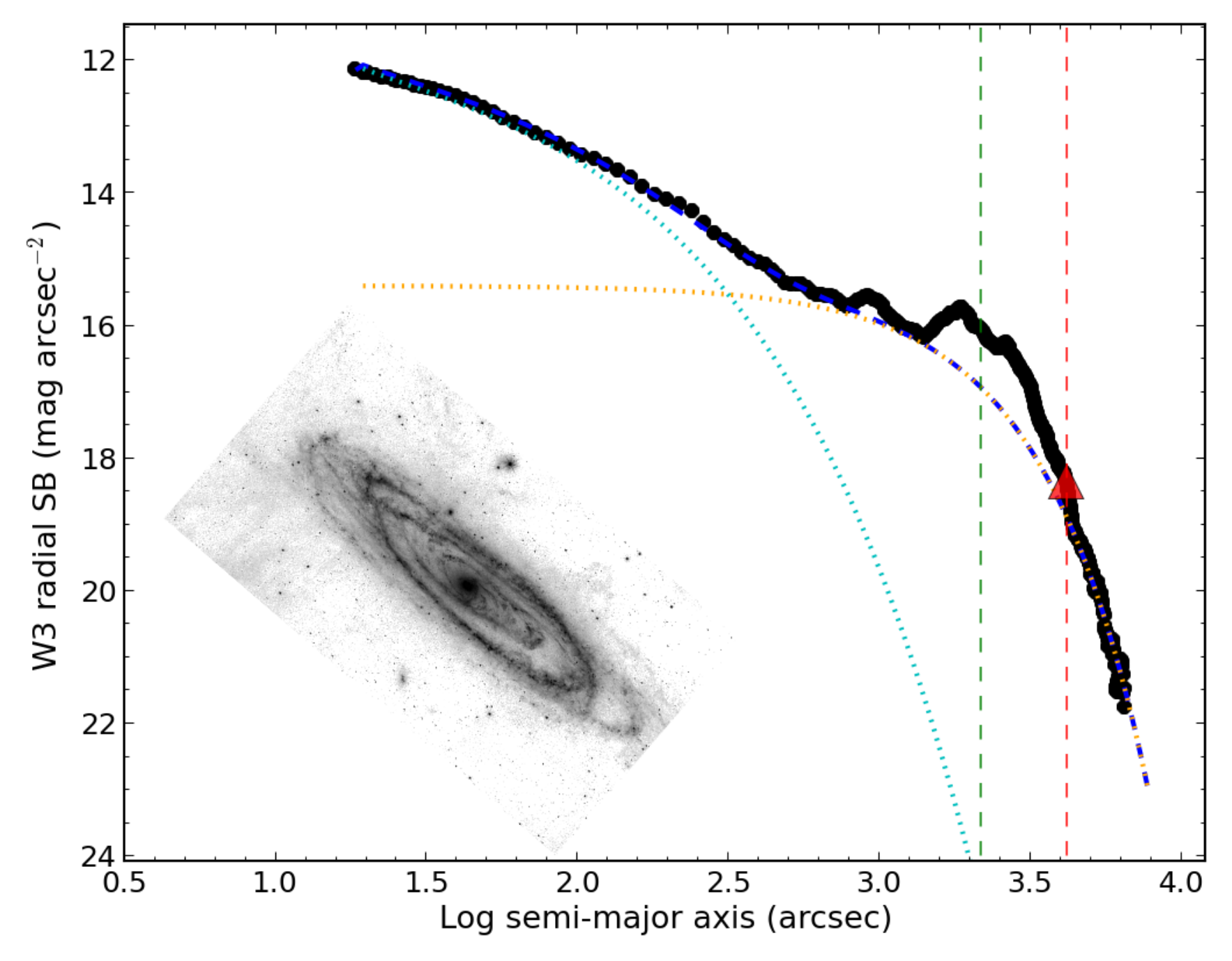}{0.5\textwidth}{(c)} 
              \rightfig{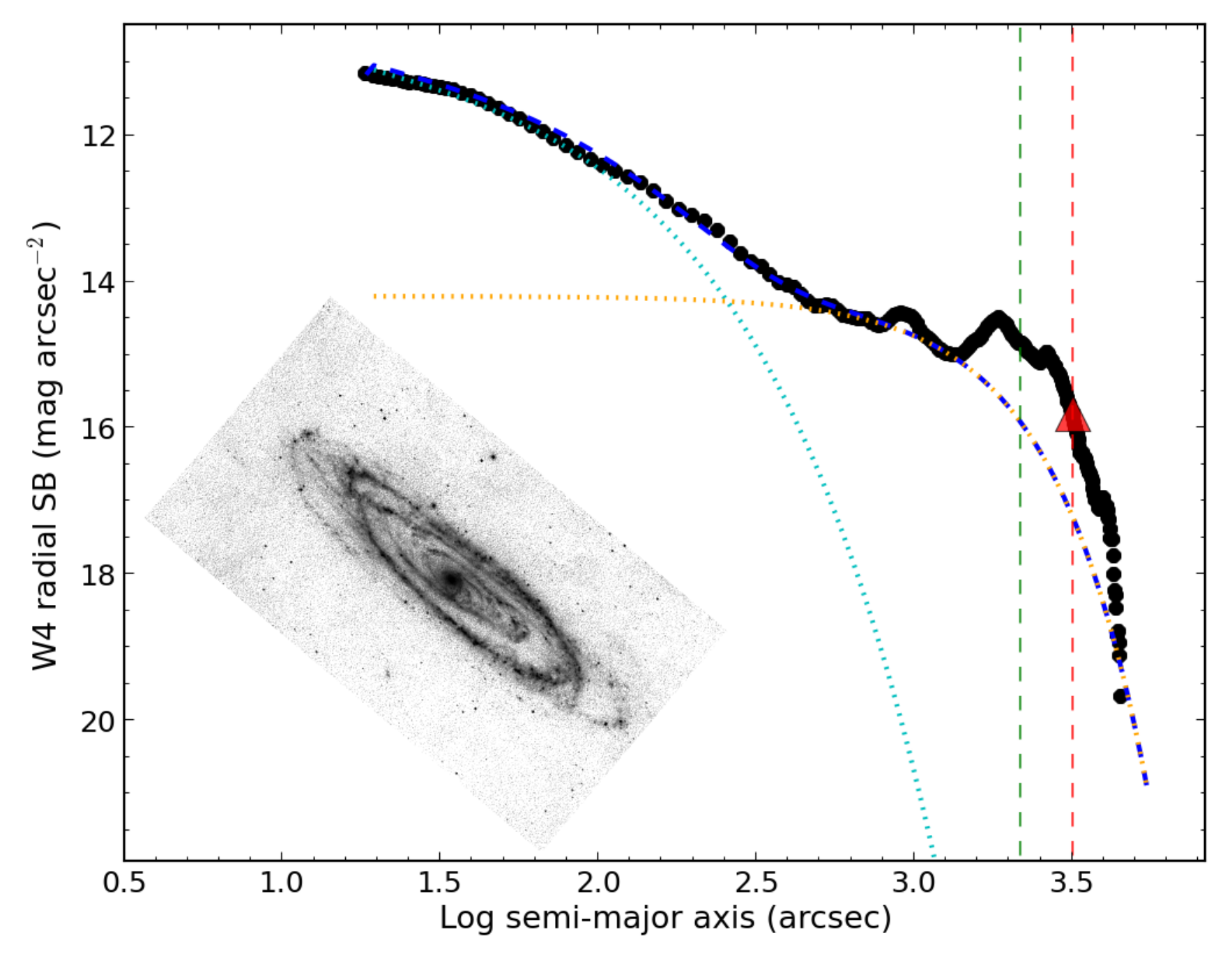}{0.5\textwidth}{(d)}}  
\caption{Axi-symmetric radial profiles of M\,31 as measured for the four bands of {\it WISE}.  The data measurements are indicated with black points, extending well beyond the 1-$\sigma_{\rm sky}$ isophotal radius (demarked by the red triangle and dashed red line) into the single pixel noise. The half-light radius is indicated by the green dashed line.  The data are fit with a double S\'{e}rsic  profile consisting of the inner bulge (blue dotted line) and the disk (orange dotted line).   The W1 B/T ratio is 0.67, indicating a prominent bulge.
The inset figure shows the galaxy in the individual monochromatic band.  Note that at the longer wavelengths, the assumption of axi-symmetry and smoothness is no longer valid due to the spiral arms and star formation regions.
 \label{fig:m31profile}}               
\end{figure*}

\vspace{-0.1in}
The angular size distribution ranges from 11 arcmin in diameter, to the extreme $\sim$10 degree size of the LMC.  
In fact, the size of the LMC is difficult to determine; it is neither symmetric in shape, nor does it have an easily-characterized radial distribution, as demonstrated in Figure~\ref{fig:lmc-slice}.  Here, slicing along the stellar bar, with lower-left endpoint (06:51:19, -71:32:57) to upper-right endpoint (04:14:11,  -65:19:19),  perpendicular width of 100\,pc (5.7 arcmin) and stretching 15 degrees in total angular extent, the mean surface brightness is computed in 100$\times$100\,pc blocks along the profile.  Fitting a high-order polynomial to the distribution produces a smoothed representation of the mean surface brightness (Fig.~\ref{fig:lmc-slice}).  The stellar bar is evident in all four bands, peaking at J2000 position (05:25:36, -69:49:34), and extending {\it at least} $\pm$5 degrees to the nominal background level. As noted in the previous section, there is significant structure in the local background to the Magellanic Clouds, including the N-S ``river" filament, which complicates the LMC characterization; consequently, the size and integrated flux measurements carry this additional uncertainty.

\begin{figure*}[htbp!]
\gridline{\leftfig{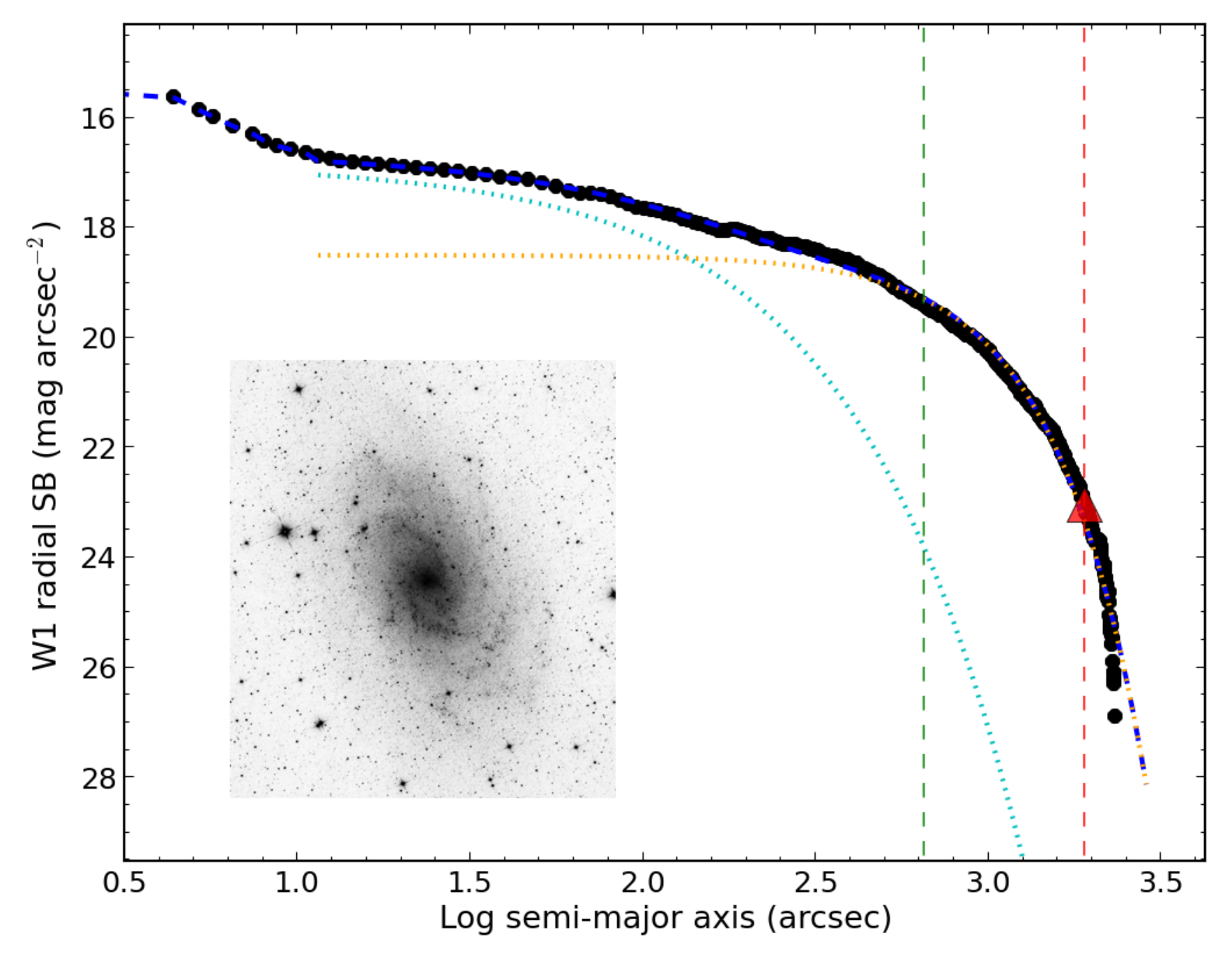}{0.5\textwidth}{(a)} 
              \rightfig{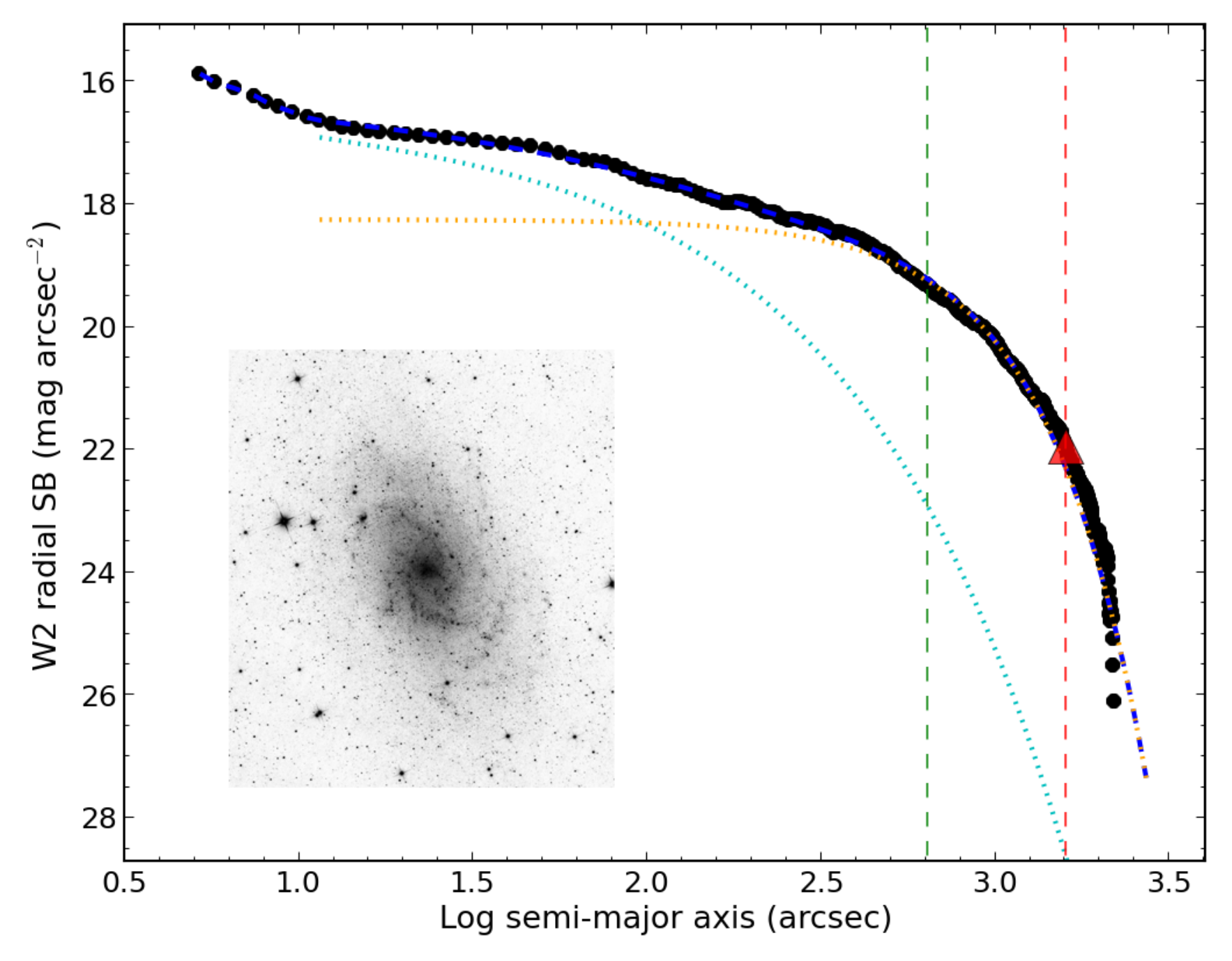}{0.5\textwidth}{(b)}}  
\gridline{\leftfig{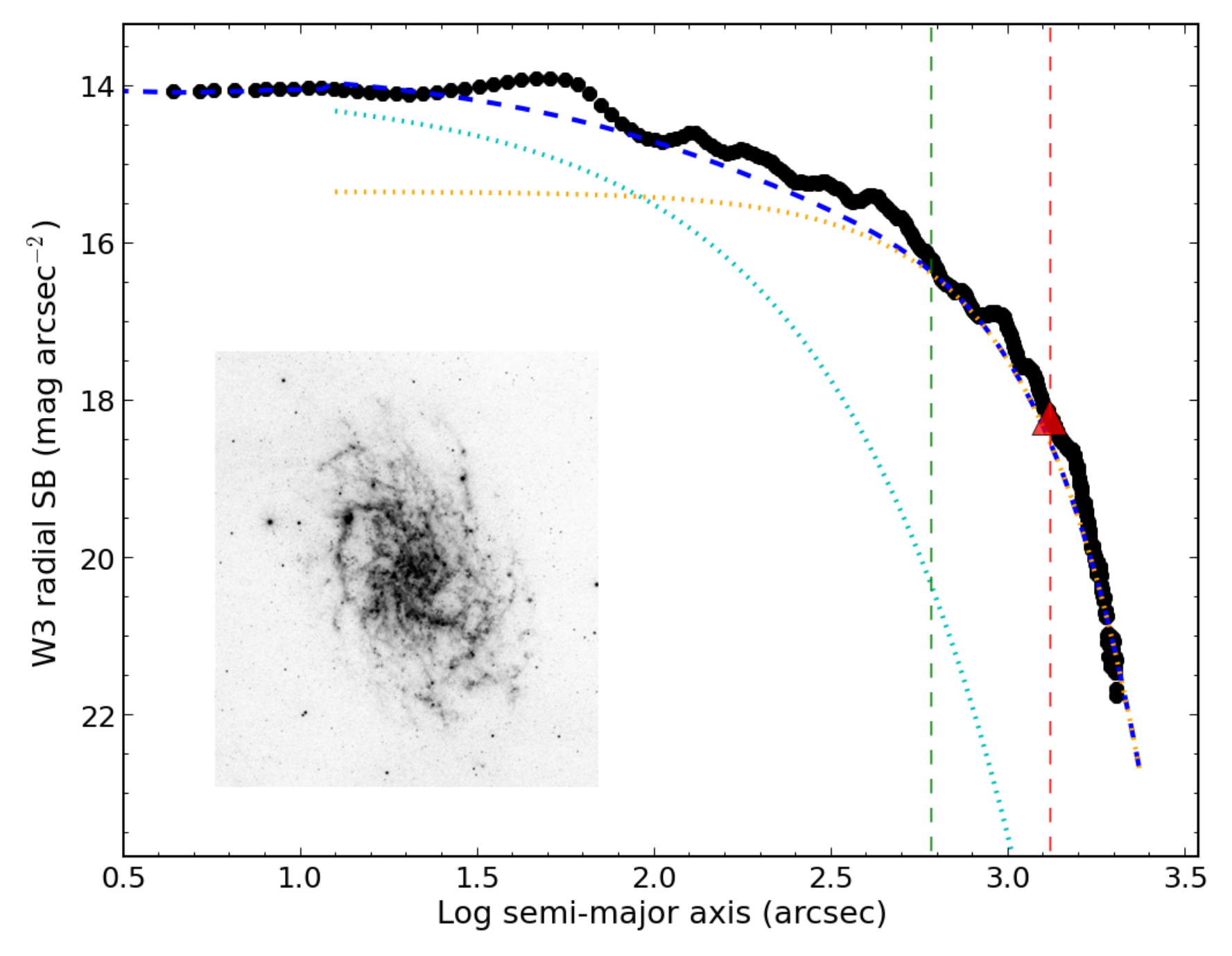}{0.5\textwidth}{(c)} 
              \rightfig{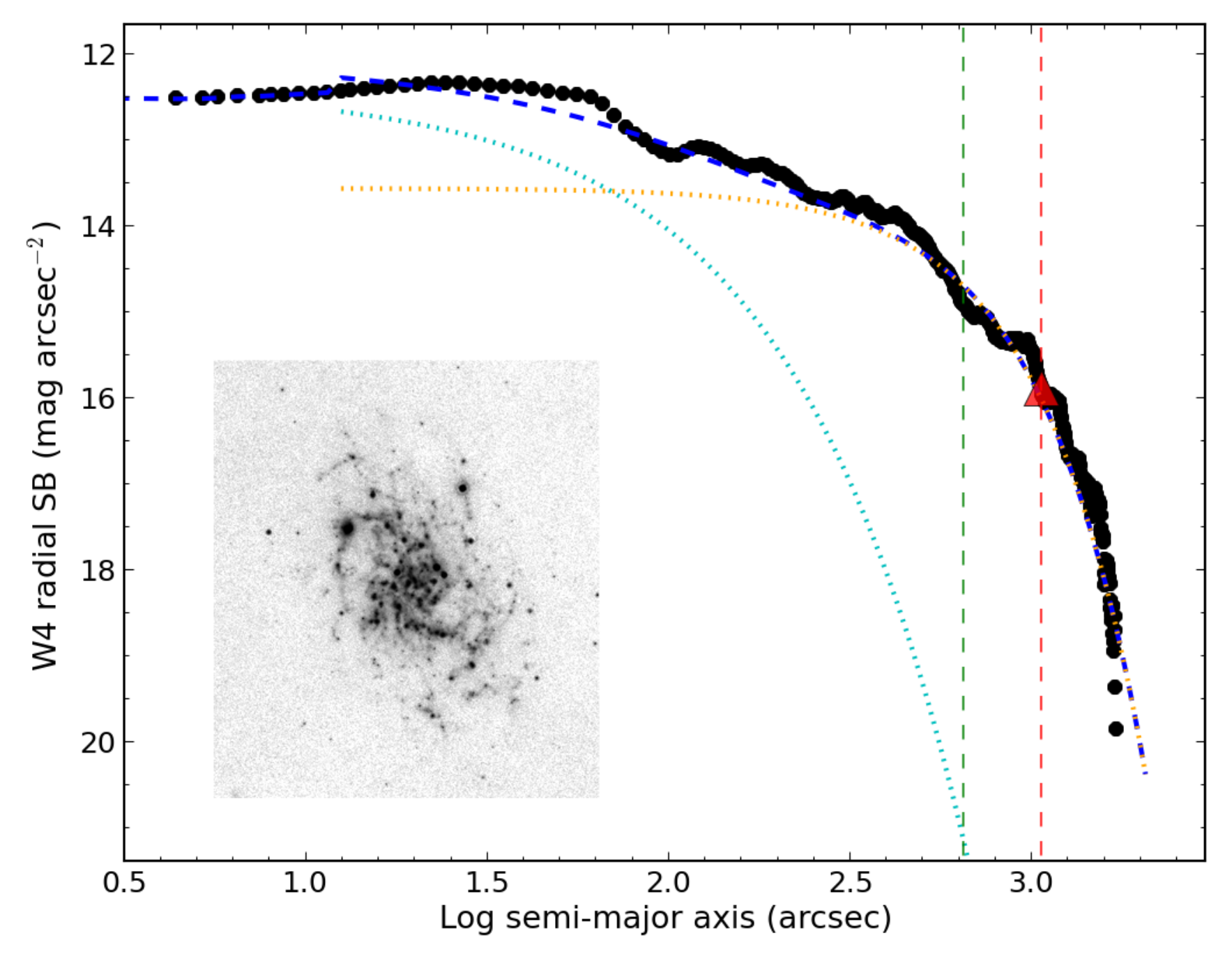}{0.5\textwidth}{(d)}}  
\caption{Axi-symmetric radial profiles of M\,33 as measured for the four bands of {\it WISE}.  See Fig.~\ref{fig:m31profile} for details.  The W1 B/T ratio is 0.33, indicating a prominent disk compared to
the bulge population (compare with the bulgy galaxy, M\,31).
 \label{fig:m33profile}}               
\end{figure*}

The 
mode of the size distribution, Figure~\ref{fig:size}, is about 13 arcmin in diameter.  
Other notably large galaxies include bright radio galaxy Cen A (NGC\,5128), nearby nuclear starburst NGC\,253, and the big-four LG galaxies.    The integrated flux ranges across $\sim$7 magnitudes for W1 and W2, but also an incredible 13-14 magnitudes for W3 and W4 due to the extremes in SF activity traced in these mid-IR bands (notably, M\,82 makes a strong appearance); see Figure~\ref{fig:mag}.

Total fluxes are estimated using both radial surface profile fitting and maximal curve-of-growth `asymptotic' apertures.    The latter is simply a large aperture that represents the maximum flux achieved through radial curve of growth within the error tolerance.  This aperture is less than the background annulus radius since it will converge before reaching the background level.  It has some advantages, including both simplicity and robustness for smaller angular-sized galaxies.  On the other hand, it is susceptible to background gradients (e.g., nearby bright stars) and unsubtracted foreground stars -- both of these effects can be systematic and are especially problematic for very large galaxies.  Asymptotic magnitudes are a common method for estimating the total flux, having been used in many studies; e.g., \citet{Neil14} used them for a large {\it WISE} sample Tully-Fisher relation study.
The other method to estimate the total flux comes from a technique developed for 2MASS \citep{Jar00}, by fitting a S\'{e}rsic profile to the axi-symmetric radial profile.  Since the mid-infrared (mid-IR) includes both a bulge and disk (SF) component, we employ a double S\'{e}rsic to better fit the distribution.   We demonstrate the performance of the fitting for M\,31 and M\,33 in Fig.~\ref{fig:m31profile} and Fig.~\ref{fig:m33profile} showing log-log plots of the radial surface brightness. The measured surface brightness as a function of radius (in symmetric shells) is denoted by the filled points, starting at a radius well beyond the PSF ($\sim$15\arcs) and extending several degrees (in the case of M\,31), to well beyond the 1-\sig$_{\rm sky}$ isophotal limit (denoted by the red triangle), to depths of around 26 mag arcsec$^{-2}$ (or about 28.7 mag arcsec$^{-2}$ in AB).  The S\'{e}rsic fit is shown with the blue dotted curve (bulge component) and the orange dotted curve (disk component).  

 \vspace*{-0.2cm}
\begin{table}[!ht]
\tablenum{2}
\caption{Comparison between Total and Asymptotic Magnitudes}\label{table:stats}
\begin{center}

\def\arraystretch{0.8}%
\footnotesize{
\begin{tabular}
{c| r | r | r | r|}
\hline
\hline
\multicolumn{4}{c|} {[Total - Asymptotic] Magnitudes}  \\
\wise Band & Ave  &  Median  &  St. Dev  \\
     --         &     mag      &          mag             &       mag  \\
\hline
W1 &  -0.01 &  -0.01   & 0.01 \\
W2 &  -0.01 &  -0.01   & 0.01 \\
W3 &  +0.05 &  -0.01   & 0.15 \\
W4 &  -0.06 &  -0.06   & 0.05 \\
 \hline
\end{tabular}
\tablecomments{The sample consists of the 100 largest galaxies (with the exception of the Magellanic Clouds, which are 
not relevant here);  the values correspond to the total  minus the asymptotic magnitudes, where the
total is estimated using a double-Sersic fitting to the radial surface brightness profile.}
}
\hfill{}
\end{center}
\end{table}

 \vspace*{-0.5cm}
 
Integrating out to three disk scale lengths beyond the 1-\sig$_{\rm sky}$ isophotal limit, a radius of about 2.8 degrees for M\,31 and 0.8 degrees for M\,33, the total integrated flux (isophotal flux $+$ extrapolation) for the four bands of \wise is 0.063,   0.104,
 -1.997, -3.132 mag, and
 3.078, 3.020 , -0.217, -2.015 mag for M\,31 and M\,33, respectively.  Comparing to the asymptotic aperture measurements, the M31 and M33 radii are 2.49 and 1.08 degrees respectively, with fluxes 0.078, 0.127, -1.950, -3.111 mag, and 3.032, 3.003, -0.174, -1.945 mag, respectively.  For these two LG galaxies, both total flux estimates appear to agree to better than 5\%.
 Likewise, comparing the total fluxes from the two methods for the 100 largest galaxies we find that there is very good correspondence, $\sim$1\%, for the sensitive \wise bands (W1 and W2), and 5 to 6\% for the W3 and W4 bands; see the tabulated statistical results in Table~\ref{table:stats}.  Interestingly there is a systematic in the median distribution, with profile fitting giving slightly (1\%) brighter results in comparison to asymptotic apertures, likely because the fitting goes beyond the image noise to capture more flux.  The exception is with W3, where the average (not the median) is skewed towards brighter fluxes.  Visual inspection of galaxies with large discrepancies between the two methods reveals that early-type galaxies with very little W3 emission (because only R-J stellar emission is seen for these types) have poor radial fits which do not extend much beyond the R$_{W1}$ radius.  It would appear that the asymptotic fluxes are more robust under these conditions, notably when the S/N is poor for radial fitting much beyond the  1-\sig$_{\rm sky}$ limit.   We list the total photometry results for the largest 100 galaxies, including the radial profile fitting and the asymptotic, in Appendix Table~\ref{table:total}.
 We will investigate the total flux differences in more detail and statistical clarity in Paper II when we include a much larger sample.

 \begin{figure}[!htbp]
  \begin{minipage}[b]{1.01\linewidth}
           \hspace*{-0.75cm}
    \includegraphics[width=1.12\linewidth]{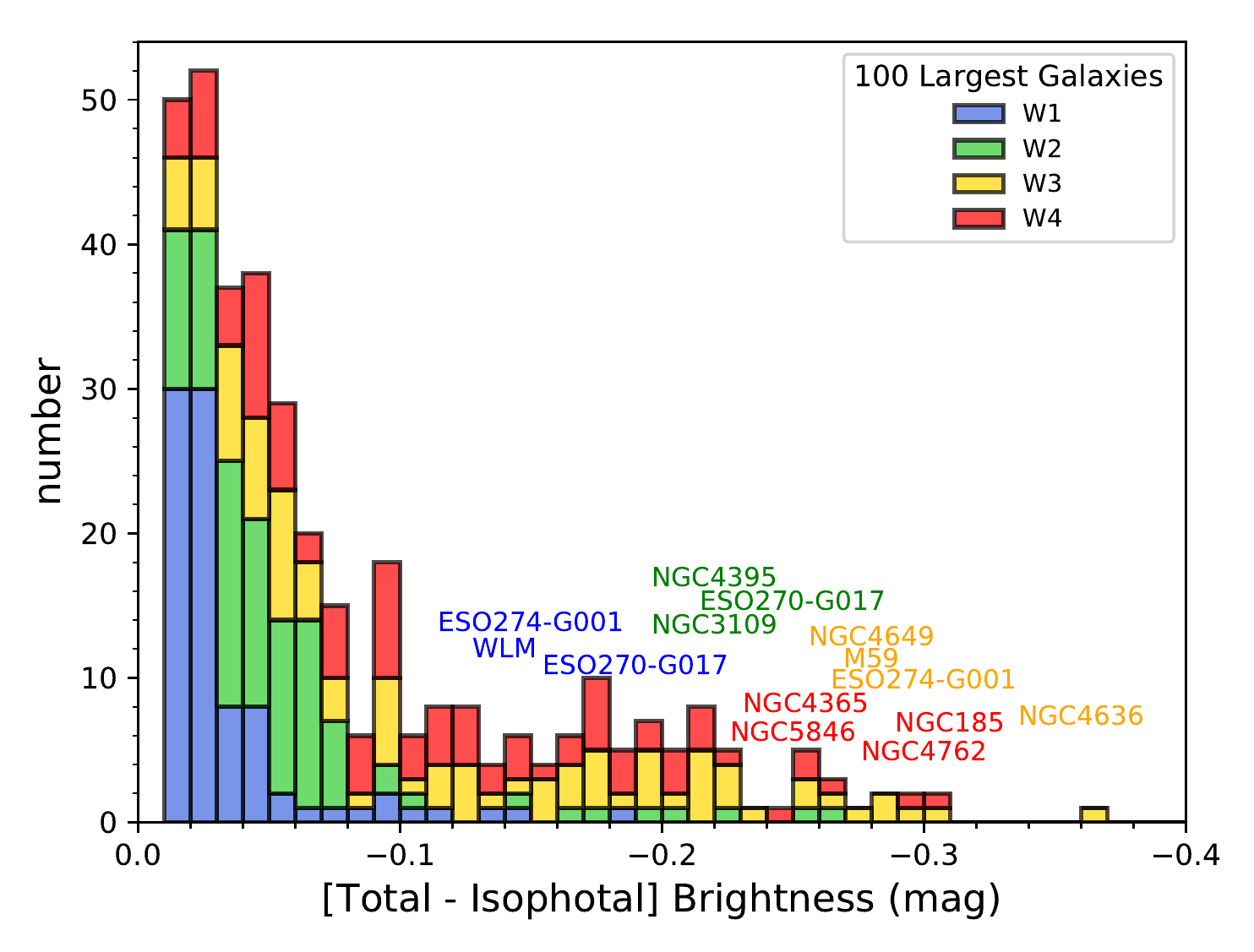} 
    \vspace*{-0.5cm}      
    \caption{Comparison of the total versus isophotal brightness for all four \wise bands.  It is notable that W1 isophotal and total fluxes are within 5\% of each other, indicating the depth of the 1-\sig\ photometry.
    Outliers are indicated:  low surface brightness galaxies tend to have large deviations for W1 and W2, early-types (spheroids) for W3 and W4.}
     \label{fig:total}
  \end{minipage}               
\end{figure}

  \vspace*{-0.5cm}
 Next, comparing the total fluxes from the radial fitting method to the 1-\sig$_{\rm sky}$ isophotal fluxes, these total values 
 are only 1.5, 2.3, 4.4 and 4.1\% brighter for bands W1, W2, W3 and W4, for M\,31,
 and 1.5,  0.5,    5.2,  7.6\% for M\,33.  There is relatively little flux beyond the W1 and W2 isophotes, as demonstrated in 
 Figure~\ref{fig:total}, which shows the difference (in magnitudes) between the total and isophotal mags for the largest galaxies.  For the most part, the total flux is within 4\% of the isophotal flux for W1 and 6\% for W2.  The exceptions ($>$20\%) tend to be very low surface brightness galaxies (e.g., WLM and ESO\,270-G017), in which there may be a significant component just below the 1-\sig$_{\rm sky}$ noise per pixel.

For the long wavelengths, W3 and W4, we observe a different behaviour.  The isophotal-to-total correction has a larger spread, easily ranging out to 25\% or more, with the largest deviations occurring for early-type galaxies (e.g., Maffei\,1; NGC\,4406; NGC\,4762), in which there is little emission in the W3 and W4 bands, only stellar (R-J) emission, and hence these bands miss a large fraction of the stellar and ISM light.  Consequently, total flux measurements  (either asymptotic or profile-fitting) are important for these bands when, e.g., measuring the global star formation rate; see next section.
 
\vspace*{0.1cm}

\subsection{Color Classification of the Largest Galaxies}

Similar to optical color analysis of galaxies (e.g., the red and blue sequence, green valley), 
mid-IR color diagnostics have proven to be useful in classifying the activity of galaxies, at least crudely separating quiescent, SF, and AGN populations \citep{Lac04,Jar11, Ste12, Yan13, Clu14, Clu17},
and as well for radio galaxy types  \citep{Chi17, Jar17}.  Here we consider the galaxy colors after correcting for spectral redshift using a diverse set of templates compiled by \citet{Br14a} fit to the 2MASS XSC \citep{Jar03} plus \wise photometric SED measurements from this work; the templates and fitting method are detailed in \citet{Jar17}, and some example SED templates with photometric fits given in Appendix B.  

\begin{figure*}[ht!]
\includegraphics[width=\linewidth]{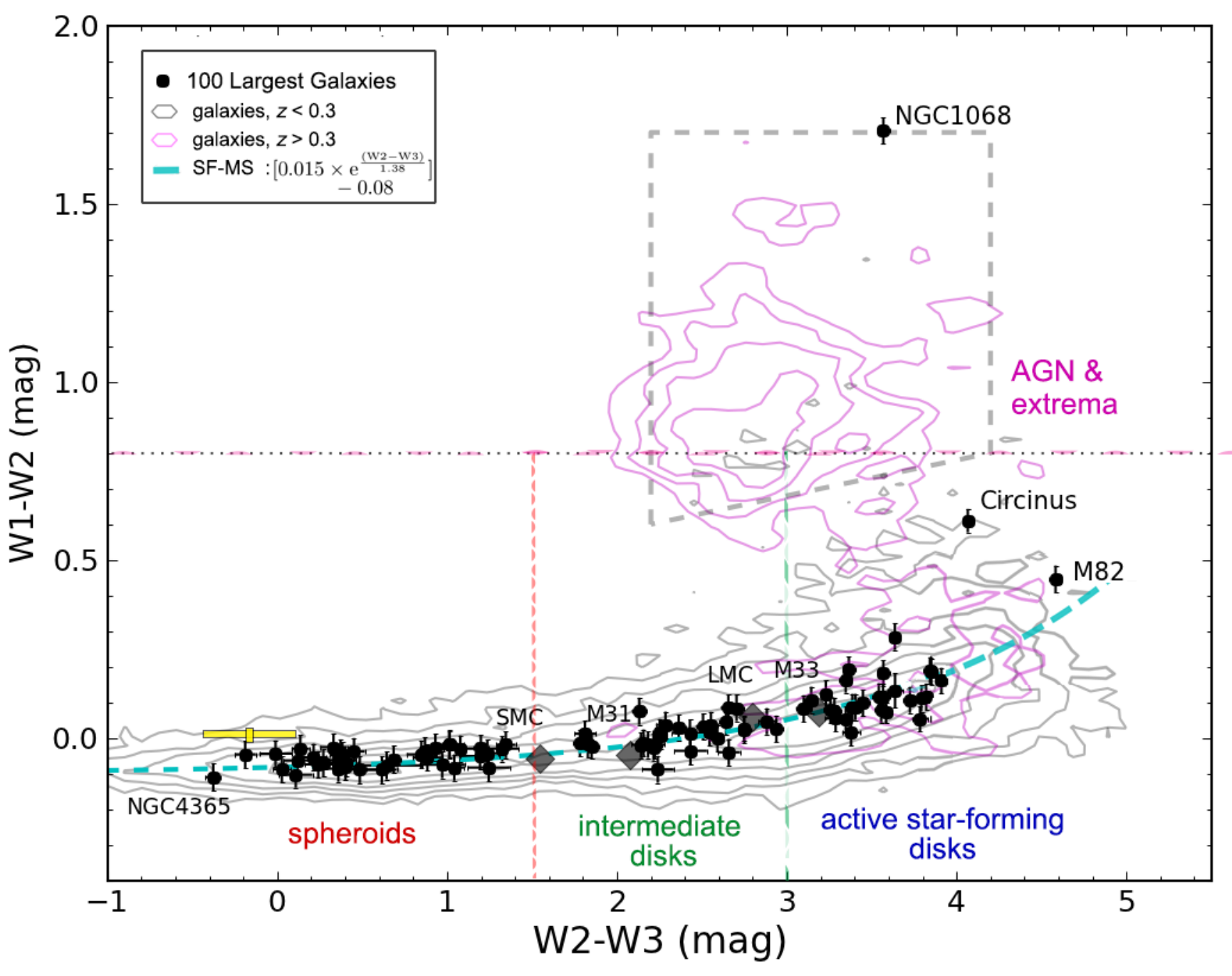}
\caption{\wise colors of the 100 largest galaxies.  The diagram attempts to delineate galaxies into four major groups:  spheroid-dominated, intermediate disk, star-formation dominated, and AGN-dominated populations.  The polygon dashed box is the AGN/QSO region defined in \citep{Jar11}.
 For comparison, the color distribution for $\sim$5000 WXSC (nearby, bright) galaxies is shown with the grey contours, and a set of WXSC galaxies that are at higher redshift (z $>$ 0.3), which accentuates the AGN and luminous (starbursting) types.  We have defined a "star formation sequence" (cyan dashed line) in which normal galaxies -- from quiescent to active SF -- follow a track in this infrared color-color space.
The yellow-cross, to the left, represents the mean colors and range of the brightest globular clusters, presented in Section 7.
\label{fig:colors}}               
\end{figure*}

The color-color location for the largest 100 galaxies is presented in Figure~\ref{fig:colors}.   For comparison and context, we show the \wise rest-frame-corrected colors for the $\sim$5000 brightest galaxies in the WXSC, to be presented in detail in Paper II. The largest galaxies, with their corresponding high signal to noise measurements, define a tight track within the color-color distribution of bright (low-redshift) WXSC galaxies (grey contours).   
The track ranges from
mid-IR `blue' stellar-dominated galaxies 
to mid-IR `red' or dusty, star-forming galaxies, 
with approximately four magnitudes of delineation in the W2$-$W3 color.  
Along the W1$-$W2 (Y-)axis, the trend is for the color to `warm' as the SF, or alternatively AGN, activity increases. In the case of the largest galaxies, this spans approximately 0.4 magnitudes.  

We use the rest-frame-corrected measurements from largest and highest S/N galaxies to fit the functional form of this \wise color-color ``sequence", given by:

\begin{equation}
{\rm (W1-W2)} = [0.015 \times {\rm e}^{ \frac{{\rm (W2-W3)}}{1.38} }]  -  0.08, \\ 
 \end{equation}

The track is roughly flat with W2$-$W3 color, curving upwards (in W1$-$W2 color) for the dusty, star-forming galaxies.
This curvature can be understood as warm dust from active star-formation elevating the W2 (4.6\m) band relative to the bluer W1 (3.4\m) band.  For example, the starburst galaxy M\,82 with a steeply rising mid-IR continuum exhibits a warmer W1$-$W2 color --  by 0.5 magnitudes -- compared to the early-type spiral galaxy M\,31.  More extreme galaxies, including luminous infrared galaxies and AGN lie well above this SF track, separating normal galaxies from active systems.

Considering just the 100 largest galaxies, there are several notable outliers:  starburst M\,82 has the reddest W2$-$W3 color of the sample, indicative of its active SF, while the two AGN of the sample, Circinus and NGC1068, have very warm W1-W2 colors, a signature of AGN torus-dust heating.  In fact, NGC\,1068 is so warm, W1$-$W2 $\sim$ 1.7 mag, it falls near the extreme limit of the AGN-QSO zone defined in \citet{Wri10} and \citet{Jar11}.  NGC\,1068 (M\,77), classified as a Seyfert-2 LIRG,  is not only exceptional given its close proximity to the Milky Way ($\sim$10 Mpc), but is clearly an outlier compared to other galaxies in the local universe as the grey contours ($z < 0.3$) are tracing, and the higher redshift galaxies ($> 0.3$) with luminous AGN appearing in significant numbers (the magenta contours).  

Following the prescription of \citet{Jar17}, we have divided the color-color plane into four zones that are inspired by infrared morphology considerations of bulge, disk and nuclear stellar populations \citep{Jo07,Wal10,Jar11,Jar13,Jar17}. 

The early-type (spheroids; E/S0) have `blue' colors, W2$-$W3 $<$ 1.5 mag. They are characterized by prominent bulges with stellar-dominated mid-IR emission arising from evolved (luminous) stars (see also next section, bulge-to-disk measurements).   They are some of the most massive and brightest galaxies in the mid-IR (see below), and tend to be located in dense environments:  brightest cluster galaxies,  such as Fornax A (NGC\,1316), Virgo A (M\,87) and many of the bright Virgo Cluster galaxies, live in this zone.  Since they tend to have little SF activity (unless undergoing a gas-rich `wet' merger or accretion event, consider the infrared excess in the core of Fornax A,  \citealt{Asa16}), they are often difficult to detect in W3 and W4 because of the rapidly diminishing R-J tail, and hence only nearby examples can be detected in all three \wise W1, W2, W3 bands, rendering a distance selection bias when studying these objects with \wise colors, also noted in \citet{Jar17}.   

In contrast, active SF galaxies, W2$-$W3 $>$ 3 mag,  have bright W3 emission and are easily detected in all four bands.  Here we find the late-type disks that have ongoing SF (e.g., M\,83), and starbursting galaxies (M\,82, NGC\,253, NGC\,1365).  Intermediate between the gas-depleted galaxies and the SF galaxies, 1.5 $<$ W2$-$W3 $<$ 3.0 mag, are the early-type spirals (e.g., M\,81, M\,31, Milky Way) characterized by prominent bulges, semi-quiescent SF-disks and pseudo-bulges.   Some of these galaxies may be undergoing quenching due to environmental effects, and maybe likened to `green valley' populations seen in the optical and infrared
\citep[e.g.,][]{Jo07,Wal10,Clu13}.

\begin{figure*}[ht!]
\gridline{\leftfig{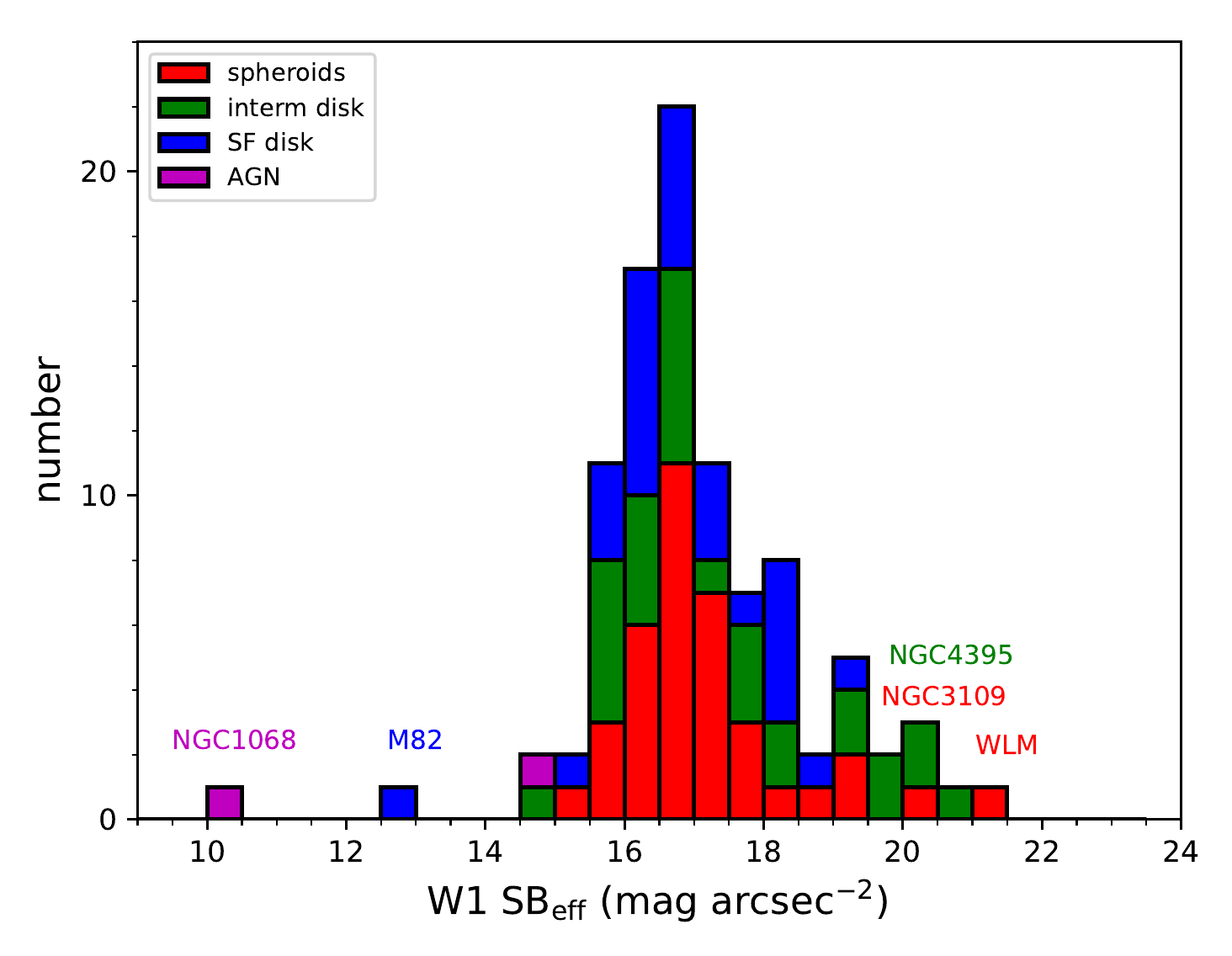}{0.5\textwidth}{(a)} 
              \rightfig{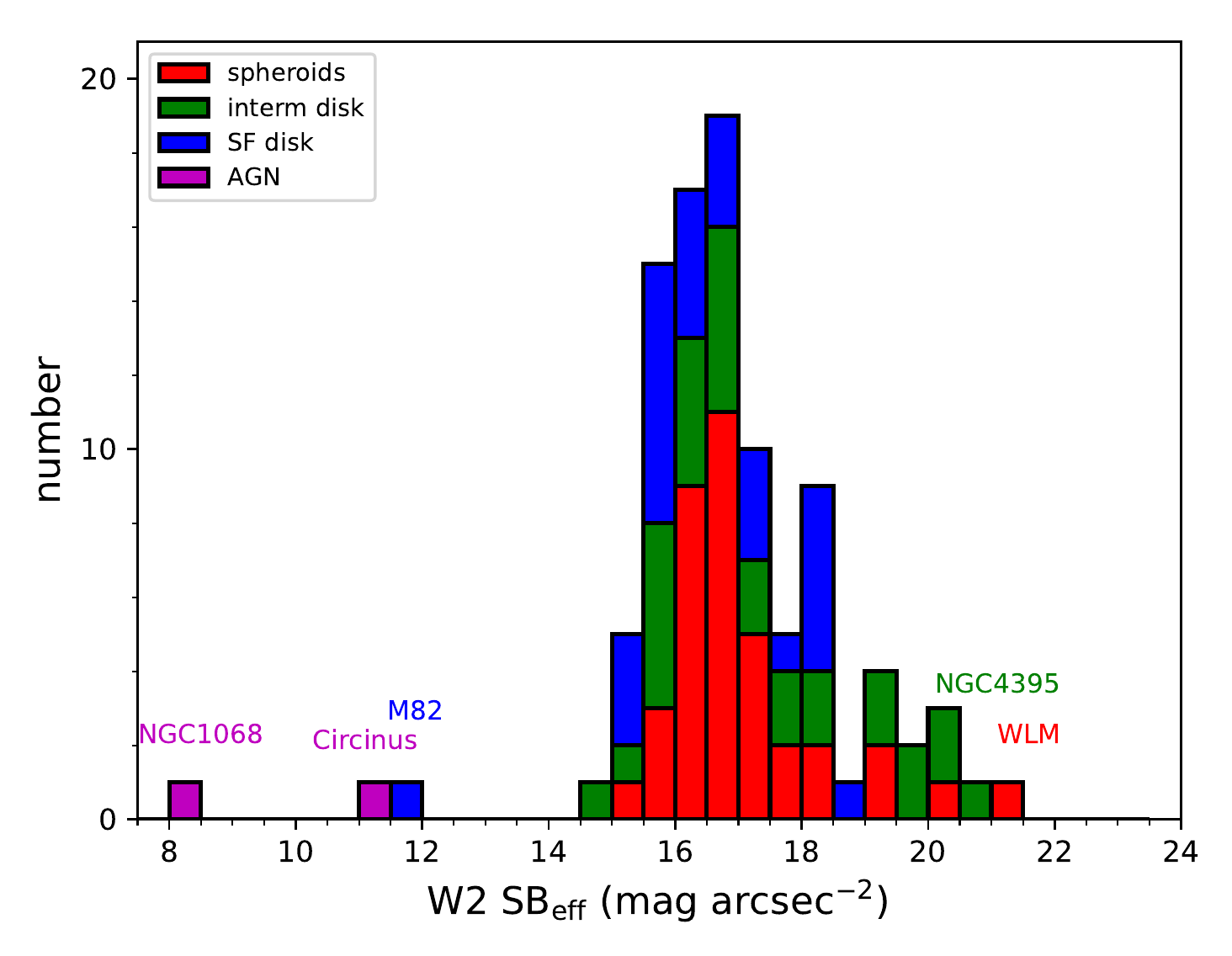}{0.5\textwidth}{(b)}}  
\vspace{-0.5cm}
\gridline{\leftfig{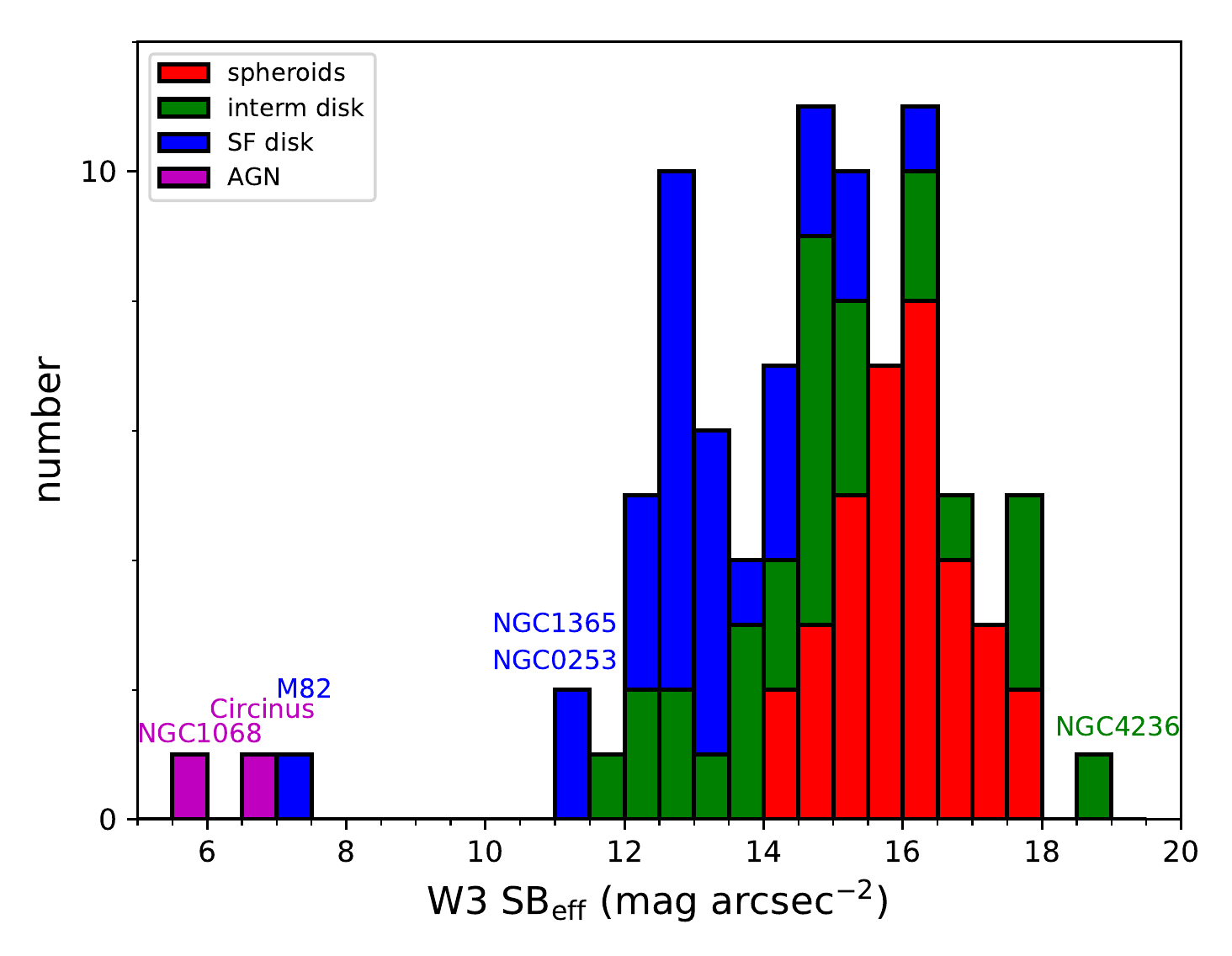}{0.5\textwidth}{(c)} 
              \rightfig{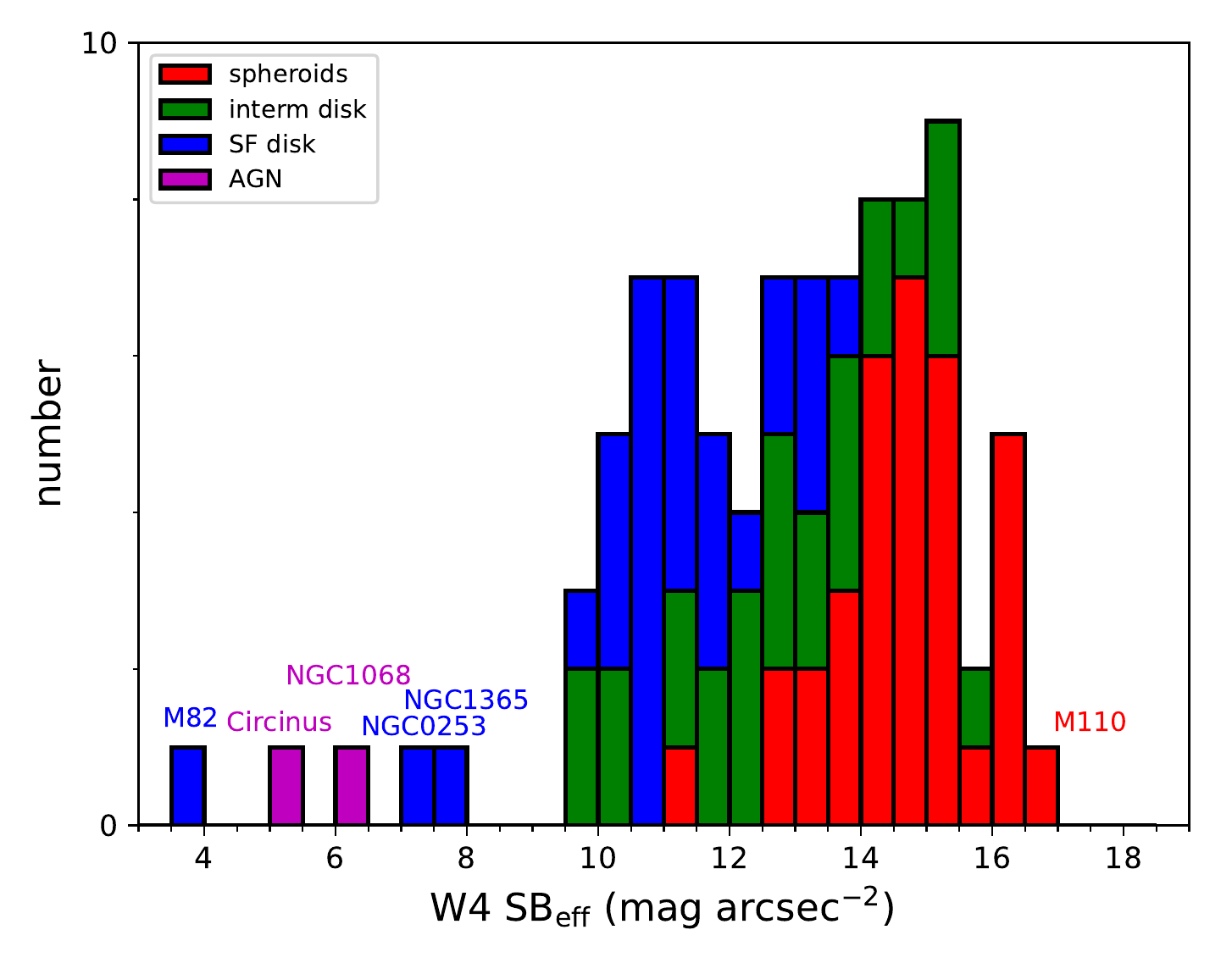}{0.5\textwidth}{(d)}}  
\caption{Half-light surface brightness (SB$_{eff}$) for the 100 largest galaxies based on the four bands of {\it WISE}. The galaxies are group-delineated by their \wise colors 
(see Fig.~\ref{fig:colors}), with notably bright and faint galaxies labeled, with four-band standouts 
NGC\,1068, Circinus and M\,82. 
 \label{fig:SB}}               
\end{figure*}

The final demarcation is for the extrema populations,  QSOs and various flavors of mid-infrared AGN activity, including high excitation, broad-line radio galaxies (HERGs and AeBs).  Those whose hosts are completely dominated by AGN emission tend to have W1$-$W2 colors greater than 0.8 \citep{Ste12}, while some Seyferts may lie below this line (e.g., Circinus) chiefly due to strong host SF emission, comparable to the AGN itself \citep{For12}. In order to track Circinus as an AGN (Seyfert 2), we relax the W1$-$W2 threshold to a value of 0.5 mag in order to include Circinus in the AGN group (along with NGC\,1068).  Current studies of AGN and X-ray/ infrared colors \citep{Ming16,Huang17} seem to indicate that Seyferts and low-power AGN can have W1$-$W2 colors well less than the 0.8 demarcation (e.g., 0.3 to 0.8), but still well above the SF track (Eq. 1).

Colors broadly reveal spectral features and continuum trends.  The actual mid-IR spectra for galaxies are far richer, encoding the early-to-late star-formation histories.  Examples of the broad \wise color classification are shown in Appendix B, where the SEDs reveal the rich emission features for each color class.

\subsection{Surface Brightness}

Employing the color classification, we investigate the half-light surface brightness, SB$_{\rm eff}$, of the sample.  The half-light, or effective radius is derived from the total flux measurement,  and corresponds to the radius at which half of the total light is enclosed within the ellipsoidal shape of the galaxy.   This radius can be compromised by the relatively large \wise PSF for small galaxies, but in the case of the largest galaxies in the sky, the radius is well-determined.  The effective surface brightness is the half-light flux normalized by the encompassed area.   This distant-independent value is sensitive to the Hubble Type, with bright surface brightnesses attributed to early-type bulge-dominated systems, and conversely, late-type spirals have relatively faint surface brightnesses.   

\begin{figure}[!htbp]
  \begin{minipage}[b]{1.01\linewidth}
           \hspace*{-0.75cm}
           \vspace*{-0.3cm}
    \includegraphics[width=1.12\linewidth]{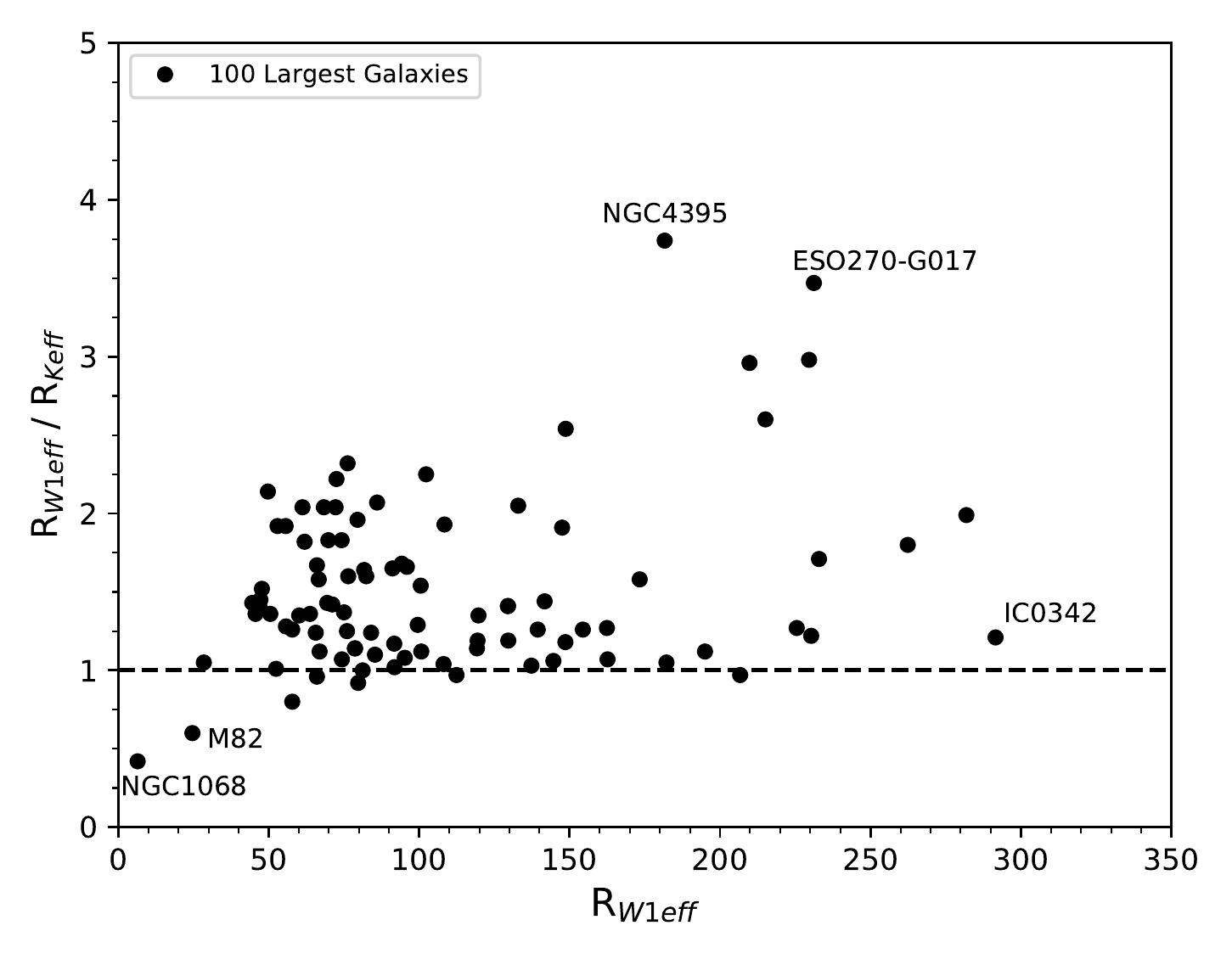} 
    \hspace*{-0.75cm}
    \vspace*{-0.3cm}
     \includegraphics[width=1.12\linewidth]{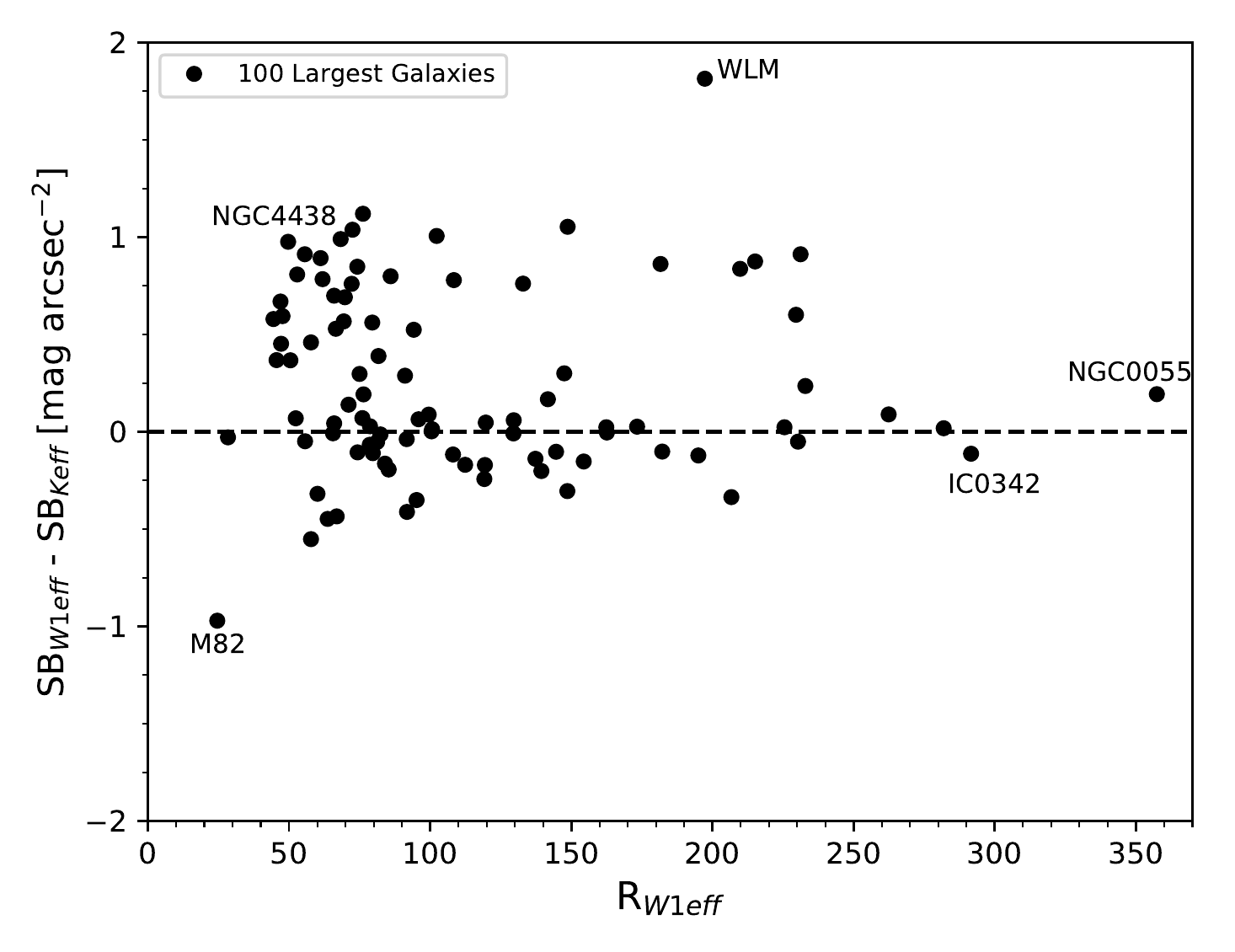} 
     \hspace*{-0.75cm}
     \vspace*{-0.3cm}
     \includegraphics[width=1.12\linewidth]{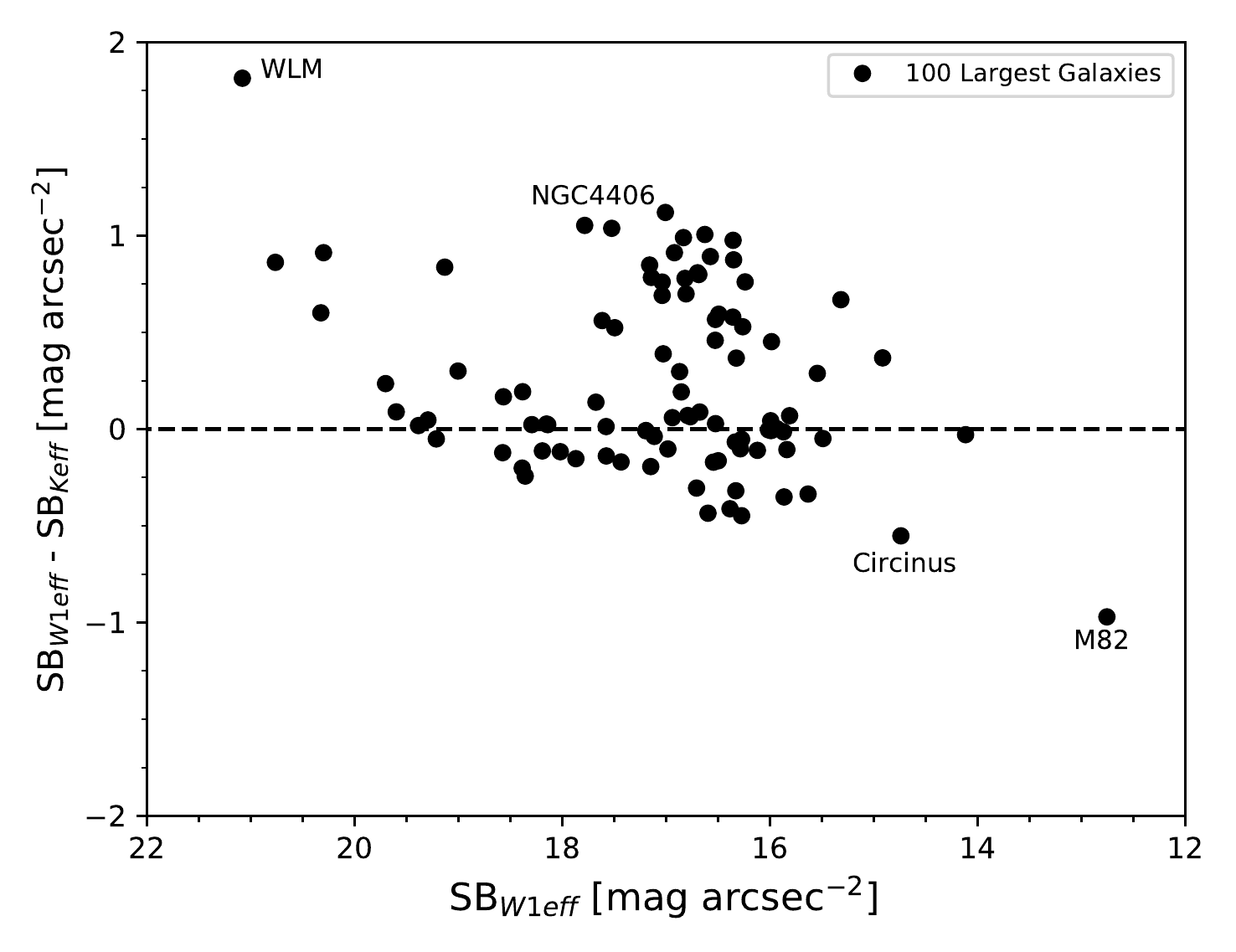} 
    \vspace*{-0.5cm}      
    \caption{Half light radius (top panel) and surface brightness (bottom two panels) comparison between \wise W1 and 2MASS K$_s$-band.
W1[3.4$\, \mu$m] is much more sensitive than 2MASS K$_s$[2.17$\, \mu$m]-band, and hence it has much larger radii (and fainter SBs) for dwarf (low-SB)  and
early-type  galaxies. 
    }
     \label{fig:SBcomp}
  \end{minipage}
  \end{figure}

The resulting half-light surface brightness for the 100 largest galaxies is presented in Figure~\ref{fig:SB}, which includes panels for each photometric band.   In the stellar bands, W1 and W2, there are a number of sources with brightnesses between 16 and 17 mag arcsec$^{-2}$, for all classification types, with more subtle delineations following the expected trends:  spheroids are typically brighter than SF disks, although there is a distinct population of spheroids that have low values, these are the dwarf spheroidals (see below).  Intermediate disk galaxies have the largest spread in surface brightness, 15-20 mag arcsec$^{-2}$, and finally, the brightest galaxies by several orders of magnitude are the AGN and nuclear starburst galaxies.  The faintest galaxies are the low surface brightness (LSB) dwarfs (hence, their moniker), including the ghostly ESO\,245-007.

More diversity is evident in the dust emission-sensitive bands of W3 and W4 
(Fig~\ref{fig:SB}c,d).  Spheroidals are now the faintest galaxies, and as noted previously, can be difficult to detect in the long wavelength channels of {\it WISE}.  Not surprisingly, the brightest galaxies are the active SF galaxies (e.g., M\,82, NGC\,0253, NGC\,1365) and the AGN `extrema' (NGC\,1068 and Circinus).  Intermediate disks again have a broad range in values,  indicating both semi-quiescent (low SB$_{\rm eff}$) and high SF (high SB$_{\rm eff}$) populations, but truly fill the `gap' between `dead' and SF galaxies, notably in the warm dust continuum band of W4 (Fig.~\ref{fig:SB}d).  

\begin{figure*}[ht!]
\gridline{\leftfig{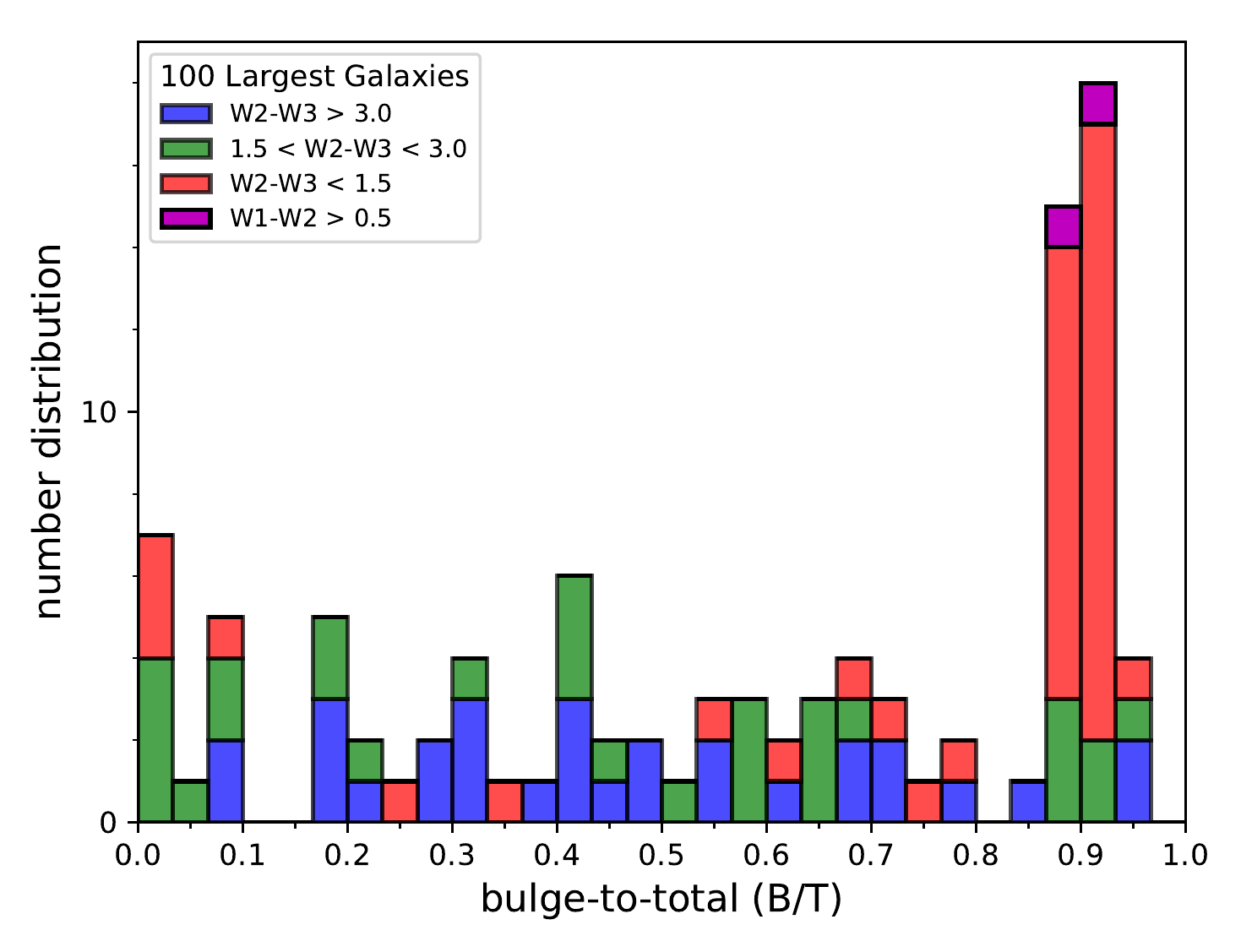}{0.5\textwidth}{(a)} 
              \rightfig{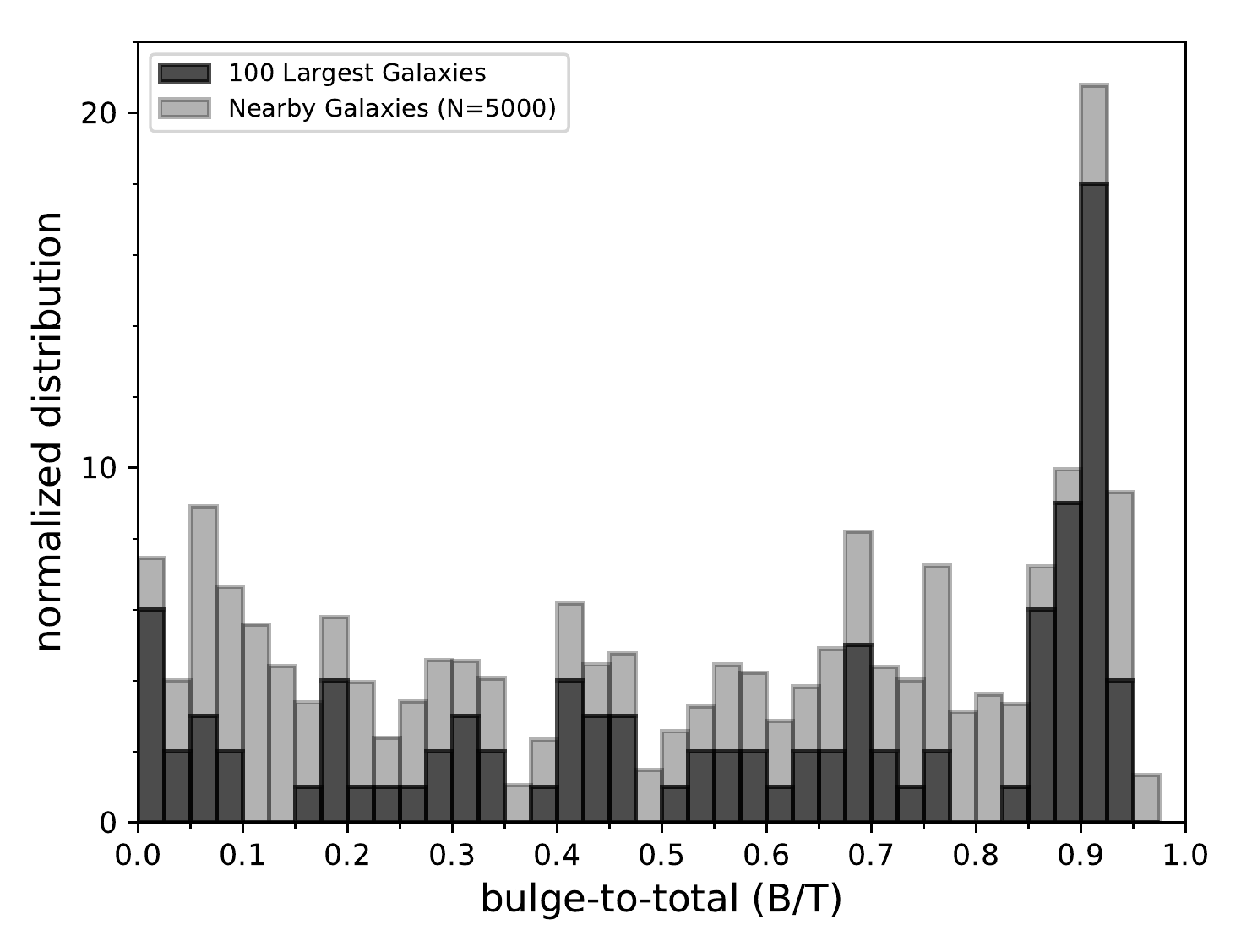}{0.5\textwidth}{(b)}}  
 \vspace{-0.5cm}
\caption{Bulge-to-Total light decomposition. (a) for the 100 Largest Galaxies in comparison to a large sample of nearby galaxies in the \wise Extended Source Catalogue.  The bulge and disk properties are derived from the axi-symmetric radial profile fit with a double S\'{e}rsic  Function.   Disk and spheroidal properties are correlated with W2-W3 color:  blue galaxies tend to have high B/T ratios, while red (SF) galaxies have considerably lower ratios.  It should be noted that the largest galaxies have a strong peak at B/T $\sim$ 0.9, which is not as apparent in a large sample of nearby galaxies (see panel b).
 \label{fig:BT}}               
\end{figure*}

It is instructive to compare the \wise half-light measurements with those from the 2MASS Large Galaxy Atlas (LGA) from \citet{Jar03}.  It has already been noted that \wise isophotal radii are 2 to 5 times larger than those from the LGA, and hence capturing more of the total light, notably for low surface brightness  (dwarf)  galaxies and for early types with R$^{1/4}$ profiles.   For these cases, we would expect half-light radii to be correspondingly larger, and surface brightnesses to be fainter.  Fig.~\ref{fig:SBcomp} presents the comparison of (effective) half-light  radii and surface brightness between the \wise and 2MASS measurements.   The top panel shows the results for the half-light radii; the dashed line shows the expected 1:1 correspondence if all things were equal.   The bottom panels show the effective surface brightness comparison with the dashed line representing equal half-light surface brightness.
What we see is that \wise 3.4$\, \mu$m half-light radii are, nearly all, much larger than those from the 2.2$\, \mu$m imaging, and in some extreme cases reach nearly 4 times larger, corresponding to the lowest SB galaxies (note bottom panels), thus demonstrating the relatively poor sensitivity of the LGA to dwarf galaxies (e.g., WLM and ESO270-G017).  Similarly, for the large and extended early-type galaxies, such as NGC4406, \wise detects far more total emission, giving rise to larger half-light radii and fainter surface brightnesses.
 Interestingly, for the nuclear active galaxies -- starburst M82 and AGN Circinus and NGC1068 -- the \wise half-light radii are smaller (more compact) and significantly higher SB (1 to 3 magnitudes, see bottom panels, Fig.~\ref{fig:SBcomp}) compared to 2MASS -- likely due to less extinction in the longer \wise bands and the SED differences between the 2.2$\, \mu$m and 3.4$\, \mu$m bands (see e.g., the SEDs in Fig 20, note the steep index power-law distribution for NGC1068).

\subsection{Bulge and Disk Populations}

On scales of kpc, well-resolved galaxies exhibit aggregate stellar populations that are typically decomposed into disk and bulge populations, which are distinct, not only in their spatial distribution, but also mean age, mass, and luminosity characteristics.  The near- and mid-infrared are sensitive to cooler populations of dwarf and giant stars, which dominate the total stellar mass of the host galaxy.  For the {\it WISE} colors, we have sub-divided the W2$-$W3 color to represent bulk populations, from spheroidal (early-type, typically the oldest populations), intermediate disks and young spiral galaxies, which in principle should be reflected in the light coming from the separate bulge and disk populations (see previous section on the infrared color-color diagram).

Here we employ a very simple decomposition based on axi-symmetry and a double S\'{e}rsic fit to the W1 (3.4\m) radial profile.  The inner profile represents the bulge population, constrained by a (relatively high) S\'{e}rsic index that may range between 2 and 4, and a smaller scale length relative to the disk. The disk component has an index between 0.8 and 2 (the ideal being a perfect exponential, index of 1).  An example of a radial decomposition can be seen in Figure 7, showing the bulge and disk populations of M\,31, which has a bulge-to-total ratio (B/T) of 0.67.

Figure~\ref{fig:BT}a presents the bulge-to-total ratios for the 100 largest galaxies, coded by their W2$-$W3 global colors and normalized to show relative differences.   The strongest trend is for the early-types, which have a strong peak at high ($\sim$0.9) B/T ratio, consistent with their dominant (global) spheroidal population. The two AGN in the sample also have a high ratio due to the unresolved nucleus emitting at mid-infrared wavelengths from hot dust accretion, thoroughly dominating the total light (see Fig. 11 showing the effective surface brightness).  Much less defined, the late-type spirals (W2$-$W3 $>$ 3 mag) range across the B/T scale, and similarly with the intermediate disks; only a slight trend towards lower B/T is seen for the more SF-active galaxies.   At face value, the B/T clearly delineates the early-type spheroids, but seems to have less power in decomposing the stellar populations in the more active (and presumably disky) galaxies, perhaps confusing the (low scale-height) pure disk, (larger scale-height) thick disk and pseudo-bulge populations that would comprise the SF-color galaxy set.      

We compare the B/T of the 100 largest galaxies with that of the nearby-galaxy WXSC, notably the 5000 brightest galaxies (all at low redshifts, $z <$ 0.1), in Fig~\ref{fig:BT}b, again normalized to show relative differences across the range.  The only difference between these two samples is that the 100 largest are the most well resolved, while the 5000 brightest may not always clearly resolve the different stellar populations.   We would expect little difference between the two sets; however, we see that the nearby galaxy sample does not have the prominent peak at B/T$=$0.9, while it does have a relative excess at low B/T, disky galaxies with 0.05 to 0.2 in ratio.  Either the larger nearby sample has a statistically significant set of disky galaxies (relative to the 100 largest), or conversely, the 100 largest sample is dominated by bulgy galaxies, or there is some biased difference, perhaps related to angular resolution.  The 100 largest is biased in the sense that we have chosen those with the largest angular extents, which could very well be the spheroidal galaxies whose $R^{\frac{1}{4}}$ light distribution is detectable at larger angular scales compared to disks with exponential light fall-off.  We will investigate the B/T properties of the nearby galaxies in Paper II, creating sub-samples that do not have such angular biases.

\section{Derived Physical Properties: Size, Mass and SFR}

Physical parameters presented in this section are derived from the isophotal measurements presented above, combined with the adopted distance to the object.  If available, we use redshift-independent distances 
(Cepheid, Tip of the Red Giant Branch, Tully-Fisher, etc) as tabulated in 
NED\footnote {\bf http://ned.ipac.caltech.edu/Library/Distances/}, 
adopting the median value as a robust estimate that minimizes the impact of  completely unphysical values, compared to the distribution, occasionally found in lists curated by NED. For those without redshift-independent distances, we instead compute the redshift luminosity distance in the CMB frame after correcting for LG motion.  In the case of physical diameters and beam sizes, we instead use the angular diameter distance.
For luminosities, we compute two different kinds: spectral $\nu$L$_\nu$, and in-band L$_\lambda$, the former is used for star formation rates and the latter for stellar mass, as is the tradition.  It should be emphasized that these two luminosities are distinctly different, the spectral luminosity is monochromatic and normalized by the bolometric luminosity of the Sun, while the in-band is normalized by the integrated Solar spectrum convolved with the \wise W1 band (as computed in Jarrett et al. 2013).
Key derived parameters are listed in 
Table~\ref{table:derived}.

\subsection{Physical Size Distribution}

\begin{figure}[!htbp]
  \begin{minipage}[b]{1.01\linewidth}
           \hspace*{-0.75cm}
    \includegraphics[width=1.12\linewidth]{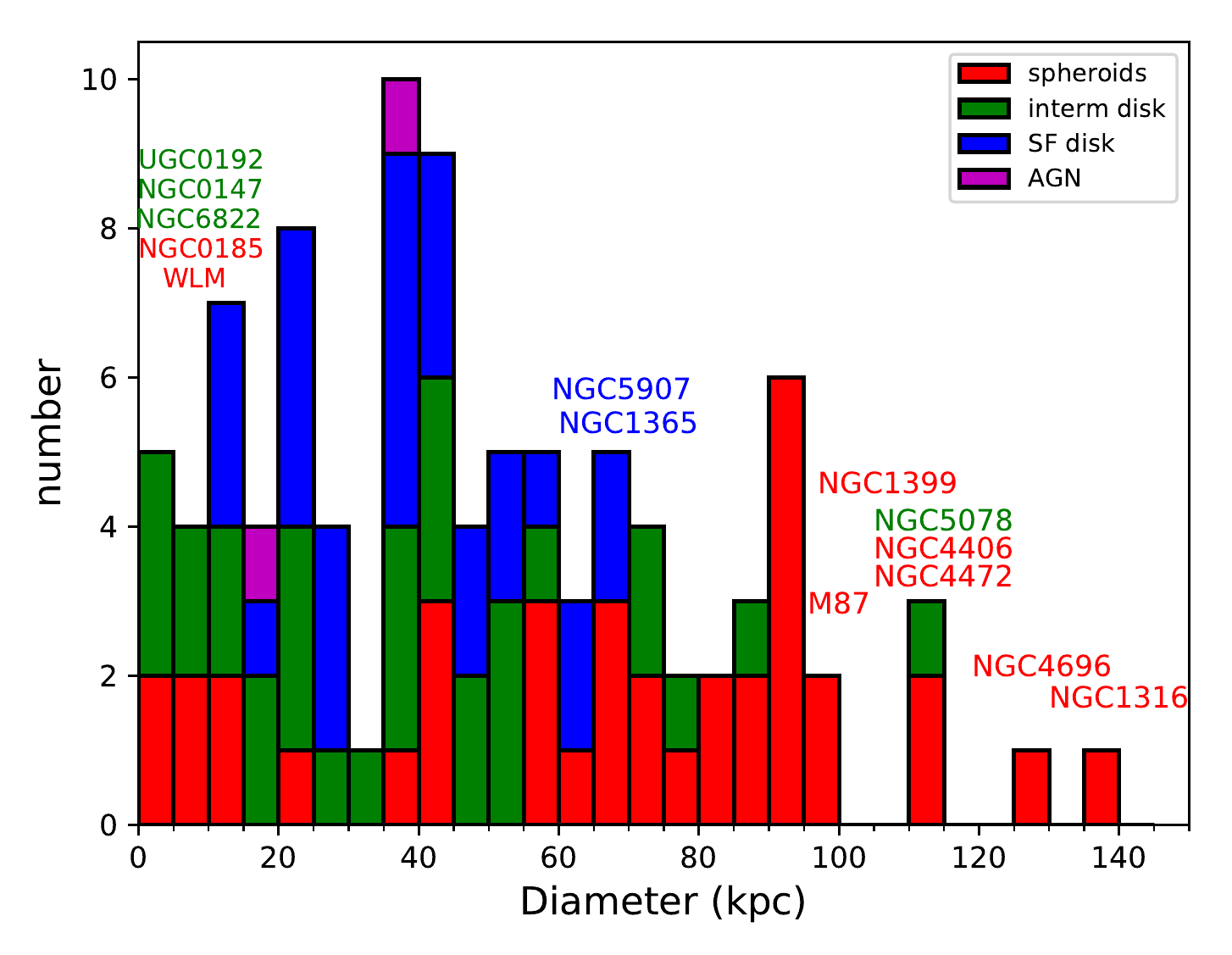} 
    \vspace*{-0.6cm}      
    \caption{Physical size (kpc) distribution for the 100 largest (angular) galaxies, based on the infrared W1 (3.4\m) size. The galaxies are group-delineated by their \wise colors 
(see Fig.~\ref{fig:colors}), with notable galaxies labeled.
\label{fig:Dkpc}}    
  \end{minipage}               
\end{figure}

We compute the diameters using the W1 (3.4\m) 1-\sig$_{\rm sky}$ isophotal radii and the redshift-independent distance or the angular diameter distance (as discussed above).  The distribution of the 100 largest (angular) galaxies is presented in Fig.~\ref{fig:Dkpc}, ranging from a few kpc to the well over 100 kpc.  Here we have once again divided the sample by the color classification (Fig.~\ref{fig:colors}) to discern any trends with the morphology proxy.

To begin with, there is a stark separation between the two flavours of early-type systems (bulge-dominated spheroids):   giant cluster (brightest or BCG) galaxies and the dwarf spheroidals.   Fornax A has the largest diameter, nearly 140 kpc in size, closely followed by the largest Virgo Cluster galaxies, which are all double or triple the size of large disk galaxies (e.g., Milky Way is $\sim$50 kpc in extent), and some local BCG galaxies (e.g., NGC\,4696 and NGC\,5078).   Intermediate disk galaxies (e.g., NGC\,5078) have the second largest distribution.
At the other end of the scale, the diminutive  satellite dwarf spheroidals (e.g., NGC\,0147) are the representative population, followed by dwarf intermediate-type galaxies (e.g., NGC\,6822).  Occupying the center stage, late-type disks uniformly range from small $\sim$10\,kpc to the broad 60\,kpc diameters.  
Setting aside the extreme size ($\sim$100\,kpc) of radio galaxy M\,87 (Virgo A),  
the two outstanding AGNs in the sample, NGC\,1068 (M\,77) and Circinus, have host galaxy sizes that are not particularly large, $<$25 kpc, indicating relatively modest or compact bulges and SF regions.  We will investigate this trend with the larger WXSC sample to see if Seyferts and AGN in general are more compact relative to SF disk galaxies.

\vspace{1cm}
\subsection{Spatial Resolution}

\wise does not have outstanding angular resolution, roughly 6\arcs\ in the shortest bands.  Nevertheless, the largest galaxies in the sky are also (typically) the nearest galaxies, and hence the physical spatial resolution is well-suited for the study of the internal components of galaxies.
Examining Column 4 in Table~\ref{table:derived}, the physical beam size (at 3.4$\, \mu$m) ranges from 1 to 2\,parsecs in the Magellanic Clouds, to $<$300 pc for Local Volume galaxies (e.g., M\,51), and 500\,pc for Virgo and Fornax Cluster galaxies, and up to 1\,kpc for the most distant galaxies in the 100 largest sample. The four independent bands of \wise\ at this physical resolution makes it a superlative imaging set to combine with complementary data sets (e.g., GALEX-UV; Herschel-PACS; radio continuum; neutral hydrogen) in order to study the internal star formation processes in GMCs and the general ISM; see for example the detailed study of M\,31 \citep{Tom19} ,  M\,33 \citep{Kam17, El19}, NGC\,253 \citep{Luc15}, and M\,83 \citep{Jar13, Hea16}.  Both the standard and mask/cleaned \wise\ mosaics (see Appendices C and D) of the largest galaxies will be publicly available to the astronomical community through NED and other astronomical archives.

\subsection{Global Stellar Mass}

As one of the most important physical parameters toward decoding the star formation history of a galaxy, estimating the aggregate stellar mass is one of the key priorities for near- and mid-infrared imaging extragalactic surveys because of their sensitivity to the evolved population R-J stellar bump between 1 and 4 microns; see \citep{Jar13} for discussion of the {\it Spitzer}-IRAC and {\it WISE} bands used to trace the stellar mass.  Accordingly, 
the galaxy host stellar mass is estimated using the W1 integrated (isophotal) flux density converted to in-band luminosity, $L_{\rm W1}$, and the M/L ratio based on the W1$-$W2 color relation \citep{Clu14}.
Here we assume the W1 (3.4\m) light arises from the evolved population, and that the short-lifetime dusty-phase populations (post-AGBs) are not significantly skewing the near-infrared brightness compared to the 
calibrated relation\footnote
{
The 3-5\m\ bands appear to remarkably trace the stellar mass compared to other independent metrics \citep{Clu14, Pon17, Kett18}, but see also
\citet{Q15} and \citet{Mei12} for altering views.}

\begin{figure}[!htbp]
  \begin{minipage}[b]{1.01\linewidth}
       \hspace*{-0.75cm}
    \includegraphics[width=1.12\linewidth]{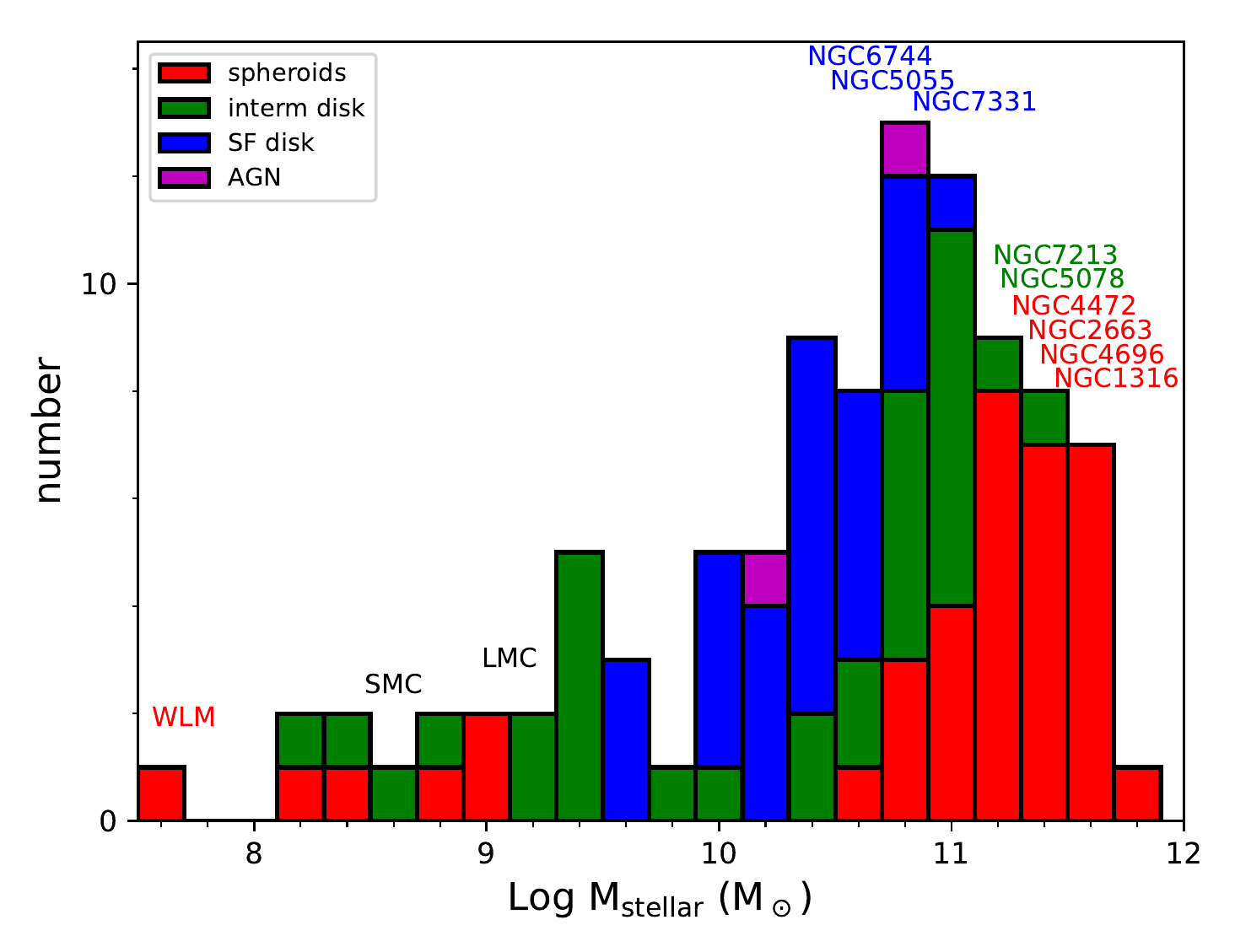} 
    \vspace*{-0.6cm}      
    \caption{Stellar mass distribution for the 100 largest galaxies.  
The galaxies are group-delineated by their \wise colors 
(see Fig.~\ref{fig:colors}), with a few of the exceptional labeled.   \label{fig:smass}}   
  \end{minipage}               
\end{figure}

We adopt the `nearby galaxy' M/L prescription from \citet{Clu14}, where the {\it WISE} W1 in-band luminosity was calibrated against GAMA stellar masses derived using stellar population synthesis models \citep{Tay11}. The relation is reproduced here:
\begin{equation}
{\rm log_{10}}\, M_{\star}/L_{\rm W1} = -2.54({\rm W1} - {\rm W2}) - 0.17, \\ 
 \end{equation}
 where $  L_{\rm W1}\ ( L_{\odot,W1}) = 10^{ -0.4 (M - M_{\rm \odot,W1})} $, M is the absolute magnitude of the source in W1, and M$_{\rm \odot,W1} = 3.24$ mag is the W1 in-band solar value; see 
\citet{Jar13}.  We place floor/ceiling limits on the W1$-$W2 color:  $-$0.05 to 0.2 mag -- corresponding M/L ranging from 0.21 to 0.91 -- to minimize the contaminating effects of AGN light, as well as unphysical blue colors due to low S/N in the W2 band (relative to the more sensitive W1 band).  For those dwarf galaxies with only W1 detections or colors with S/N $\le$ 3, we adopt a M/L value of 0.6 \citep{Kett18}.  In general, adopting a single M/L between 0.5 (for disky galaxies, see \citealt{Pon17}) and 0.7 (bulge-dominated types; see \citealt{Clu14}) is a good strategy for estimating the stellar mass using \wise W1 (3.4\m) light.

The distribution of stellar mass is presented in Figure~\ref{fig:smass}, with the sample delineated by color as before.  Unsurprisingly, the largest-diameter galaxies are also the most massive.  These bulge-dominated early-type systems typically have values $>$10$^{11} M_\odot$.  In terms of the stellar mass, the most massive galaxy is NGC\,1316 (Fornax A) weighing in at 5.2 $\times$10$^{11} M_\odot$, which is nearly an order of magnitude larger than the Milky Way at 6 $\times$10$^{10} M_\odot$  \citep{Lic15}.
At the other extreme, the dwarf spheroids have Log\,M$_{\star}$ values that range between 7.5 and 8.5. The least massive galaxies, 6 to 7, are the very low SB dwarfs (e.g., ESO\,245-007), which are only visible to \wise\ in the Local Volume, d $<$ 10 kpc.  The Magellanic Clouds are in the upper dwarf range (8.5 - 9.0).   Intermediate disks, such as M81 (Log\,M$_{\star}$ = 10.85)  and M\,31 (10.96), are the second most massive population (but also have a `dwarf' population near 9.4), followed by the late-type SF disks, which tend to have Log\,M$_{\star}$ $<$10.5, and are clearly still building their disks (see SFR and specific-SFR below and in Table 2).  The Milky Way falls within the intermediate range, with an estimated Log\,M$_{\star}$ = 10.78 \citep{Lic15}, somewhat smaller than its sister galaxy, M\,31.


\subsection{Star Formation Activity}

The ISM-sensitive bands of \wise, W3 (12\m) and W4 (23\m) are effective tracers of the dust-obscured SF activity \citep{Jar13, Clu14,Clu17}.  It turns out that W3 is particularly effective because, not only is it far more sensitive than the 23\m\ band, but it better traces the total infrared luminosity, and appears to be robust to metallicity variation. This is largely attributable to the breadth of the W3 band, resulting in it being dominated by the dust continuum and less susceptible to emission-line and PAH variations \citep{Clu17}, compared to, for example, the {\it Spitzer}-IRAC 8\m\ band.  We should stress that galaxies with little to no dust emission, or alternatively very low metallicities, will have correspondingly low SFRs based on the infrared emission. It is more appropriate to use H$\alpha$ and UV tracers of SF for such galaxies, which may be particularly important for dwarf galaxies.

\begin{figure*}[ht!]
\includegraphics[width=\linewidth]{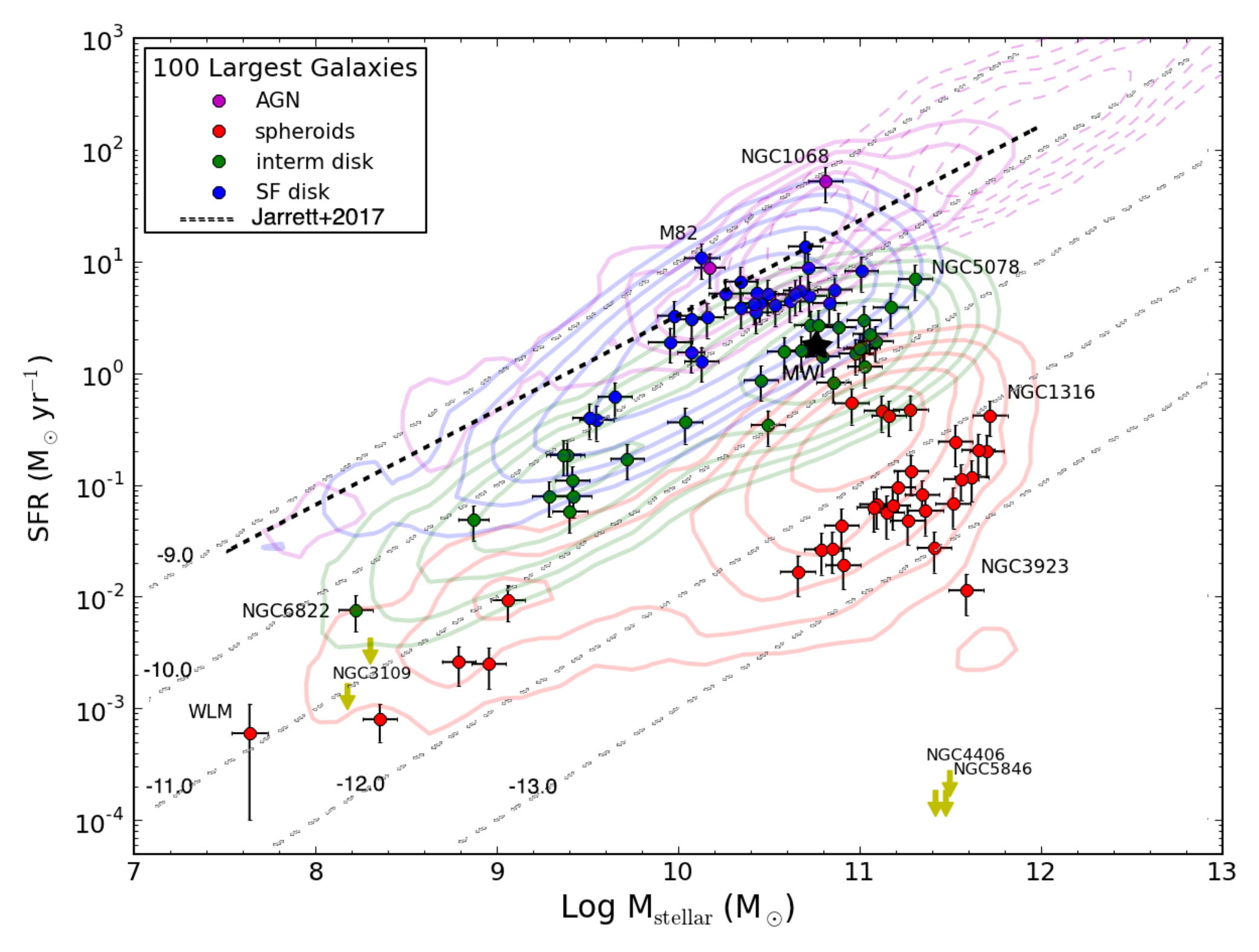}
\caption{Galaxy Main Sequence (GMS):  star-formation rate (based on W3, Eq. 3) versus stellar mass sequence for the 100 largest galaxies (points and error bars) and the bright galaxy sample from the WXSC (contours, where the higher redshifts are magenta dashed contours, tracking the AGN). 
The data are delineated by their \wise colors (Fig. 10), and generally fall below the sequence derived from a more distant (and luminous) sample as defined by the GAMA-G12 region \citet{Jar17},  dashed line.  
The Milky Way (MW, black star symbol) is located in the central spread of the "intermediate" disk galaxies (green points and contours).
The faint dashed lines represent lines of constant Log sSFR = -13, -12, -11, -10 and -9.0 yr$^{-1}$.
 \label{fig:gms}}   
\end{figure*}

In the low-$z$ universe, the W3 12\m\ band includes contributions from the warm T$\sim$150 K dust continuum, the 7.7, 8.5, and 11.3$\, \mu$m PAHs, and the 12.8$\, \mu$m [Ne\,II] and 15.7$\, \mu$m [Ne\,III] emission features. It is thus sensitive to SF activity (but also metallicity), as well as a small fraction of stellar light (R-J tail).  Although our sample does not contain any extreme IR-bright galaxies, the closest being NGC\,1068, we should note that the band also contains a 10\m\ silicate absorption, which is clearly present in the spectra of heavily dust-obscured galaxies, such as Arp\,220.  \citet{Clu17} does observe that LIRGS and ULIRGS have a slightly different SFR relation compared to normal galaxies, likely indicative of the extreme dust and ice(s) absorption. For these extreme infrared-bright galaxies, we recommend using the 23\m\ W4 band.

Relative to the near-IR bands of {\it WISE}, W1 and W2, the mid-IR W3 and W4 bands are not nearly as sensitive; this particularly affects stellar-continuum (bulge) dominated systems with predominately R-J emission at these wavelengths. Hence, SF activity can only be measured for (1) relatively nearby galaxies, or (2) relatively dusty galaxies.  Luminous infrared galaxies can be detected to high redshift (e.g., the Hyper-LIRGs discovered by WISE; see \citet{Tsai15}.   There are cases where the W3 flux is easily detected, but after subtraction of the stellar continuum (estimated from the W1 flux), there is nothing left representing the ISM emission (e.g., NGC\,2768). Nevertheless, for most of the largest galaxies in the sky, and for most WXSC galaxies within $z < 0.1$, SF activity can be estimated using {\it WISE}.

We employ the latest SFR calibration based on the total infrared luminosity -- and adopting a
Kroupa (2002) initial mass function (IMF) --
of a varied sample of nearby systems from the SINGS  \citep{Ken03}  and KINGFISH  \citep{Ken11} surveys, correlated to the corresponding mid-infrared luminosities \citep{Clu17}. The general prescription is that the 12 and 23$\, \mu$m SFRs are derived from the spectral luminosities: $\nu L_\nu$, where $\nu$ is the bandpass central frequency and is normalized by the bolometric luminosity of the Sun. \footnote{We note that the spectral luminosity is different from the in-band luminosity due to the bolometric versus W1 in-band normalization. See \citet{Jar13}.}

The total infrared SFR scaling relations of \citet{Clu17} are given by:

\begin{equation}
\begin{multlined}
{\rm Log_{10}\, SFR}_{\rm TIR} (M_\odot/{\rm yr}) = 0.873 (\pm 0.021)\, \times \\
{\rm Log_{10}}\, \nu L_{12\, \mu \rm m} (L_\odot)  -7.62 (\pm 0.18)
 \end{multlined}
 \end{equation}

 \begin{equation}
 \begin{multlined}
{\rm Log_{10}\, SFR}_{\rm TIR} (M_\odot/{\rm yr}) = 0.900 (\pm 0.027)\, \times \\
{\rm Log_{10}}\, \nu L_{22\, \mu \rm m} (L_\odot)  - 7.87 (\pm 0.24)
 \end{multlined}
\end{equation} 
where  $\nu L_{12\, \mu \rm m}$ and $\nu L_{22\, \mu \rm m}$ are the spectral luminosities, normalized by the solar luminosity ($L_\odot$),
and the stellar continuum removed using the W1 luminosity as a proxy for the stellar continuum, as follows:  15.8\%  in the W3 band, and 5.9\% in the W4 band (see \citet{Clu17} for details).  Note that the relation uncertainties, in particular the offsets, render SFR accuracy of at minimum 20 to 25\%, but can be several percent higher when 
propagating the formal photometric errors.  Moreover and as noted earlier, the mid-IR SFRs are based on the warmed ISM that arises from dust-obscured star formation, which does not account for unobscured UV emission from young stars that manages to escape the host galaxy.

The SFRs and stellar masses for each galaxy are presented in Table~\ref{table:derived}, which includes the redshift-independent or luminosity distance (column 3), the physical resolution (4), 
the colors (5, 6), the spectral luminosities, continuum-subtracted (7, 8), the W1 in-band luminosity (9), the estimated stellar mass (10), and the TIR SFRs from the W3 and W4 (11,12) spectral luminosities.  Here the spectral luminosities, with the stellar subtraction, represent the 12\m\ aggregate and 23\m\ dust emission after stellar continuum subtraction.  Finally, the specific star formation rate (sSFR, column 13) is the SFR normalized by the stellar mass.


To graphically show the SFR results, we relate
the SFRs (based on the W3 relation) with the corresponding host stellar masses,  creating the so-called galaxy star formation  ``main sequence" (hereafter referred to as the GMS), Fig.~\ref{fig:gms}.
This GMS sequence gives a crude representation of the past-to-present star formation history in which ever larger SFRs track with ever larger stellar masses, which appears to hold even at higher redshifts 
 (see for example Elbaz et al. 2007; Noeske et al. 2007; Bouch\'{e} et al. 2010).   
The slope of the sequence therefore represents a secular evolution path, with deviations due to events from 
interactions, bar formation, 
AGN and superwind feedback, starburst and quenching events.  
For reference, we show lines of constant specific star formation ratio(sSFR = SFR/mass, column 13 of Table 3) , ranging from 10$^{-13}$ to 10$^{-9}$ yr$^{-1}$, covering the range that is seen for nearby galaxies, from active building  (10$^{-9}$  to 10$^{-10}$), to semi-quiescent (10$^{-10}$  to 10$^{-11}$) and quenched or depleted ($<10^{-11}$).

To put the largest 100 galaxies into context, 
we show the results for the larger WXSC sample, some 5000 galaxies, contours coded by color classification as before,
and the overall trend in the GMS derived from a higher redshift  ($z$ complete up to 0.3) sample from the GAMA G12 region (Jarrett et al. 2017).   

A number of interesting features are seen in Figure~\ref{fig:gms}. The dusty galaxies with their strong W3 emission (blue points in the plot) are generally massive in stars,  Log\,M$_{\star}$ $>$10 and relatively high ($>$2) SFRs, defining the upper end of the sequence.  A maximum SFR is reached where the disk galaxies appear to turn over at stellar mass of Log\,M$_{\star}$ between 10.4 to 10.7, perhaps signifying a gas-depletion phase.  
Sitting well above the sequence is the starburst galaxy M\,82, clearly demonstrating how young (Myr) starburst events move galaxies upward, outpacing the stellar mass Gyr build-up time scale. A cautionary note:  the highest SFR galaxy is NGC\,1068 (see also the Circinus Galaxy), which represents an upper limit since the AGN accretion emission from hot dust is heavily boosting the \wise fluxes -- while even more luminous galaxies with AGN (e.g., Mrk\,231, see Cluver et al. 2017) move long this upper AGN-sequence as represented by the dashed magenta contours, and should be viewed with discretion: neither the SFR nor the stellar mass of mid-infrared AGN is properly measured or modeled with simple \wise fluxes and colors.

As galaxies build their mass and consume their gas, they move to the right and down from the SF sequence, as demonstrated by the intermediate disk galaxies (green points and contours).   Andromeda (M\,31), M\,81 and the Milky Way, values from \citet{Lic15}, are example galaxies that have built the bulk of their stellar backbone, while continuing to form stars at a relatively semi-quiescent rate, $\sim$1 to 2 M$_\odot$ yr$^{-1}$.  The final evolutionary sequence is dominated by the spheroidals (red points), with examples of dwarf ($<$ 10$^9$ M$_\odot$) and massive ($>$10$^{11}$ M$_\odot$) varieties in the 100 largest galaxies, whose SF rates are several orders of magnitude smaller than those of disk galaxies, and correspondingly low sSFRs ($<$10$^{-12}$ yr$^{-1}$).
Presumably the gas has been depleted -- or accretion shutdown, so-called strangulation (Peng et al. 2015) -- halting the SF and creating a downward descent from the GMS, reaching a peak in stellar mass around Log\,M$_{\star}$(M$_\odot$) $\sim$11.5.   The SF activity is so low for many of these early types that no \wise W3 emission is detected, and hence only upper limits are indicated in the GMS plot, for example the Heart of Virgo, NGC\,4406 (M86). As previously noted, galaxies with low dust content may poorly trace the ongoing SF that is better detected and traced using UV {\it GALEX} imaging or H$\alpha$ spectro-imaging.
  
\pagebreak
In Figure~\ref{fig:gms}, we can also compare the GMS of the bright and nearby galaxy with that of a deep and complete redshift survey (dashed line), 
GAMA-G12 $z < 0.3$ (Jarrett et al. 2017).   Dwarf spheroidals,  such as NGC\,0185,  have the lowest SFRs, 1 or 2 orders lower than the low-mass dwarfs in G12.  It should be noted that G12 is optically selected (r-band) and preferentially finds redshifts of emission line systems at ever larger distances (hence relatively-unobscured SF disk galaxies, which would have higher SFRs than dwarf spheroidals). The corresponding specfic-SFRs  
of the dwarf spheroids is quite long, $>$10$^{-11}$ yr$^{-1}$, confirming their quiescent state.
The late-type SF galaxies, such as M83, have unsurprisingly, the highest SFR rates and lie within the G12 sequence (i.e., consistent with the GAMA selection), with $>$10$^{-9}$ yr$^{-1}$ rapid-building rates.   In the 100-largest sample, the most rapidly-building galaxy is M\,82, its current starburst phase is perhaps just one of many that will occur in its lifetime.
Meanwhile, the massive, high-surface bright spheroids, such as NGC\,4696, have rapidly declining specific-SFRs, 10$^{-12}$ yr$^{-1}$,  indicating they are `dead', having consumed their gas and fallen off the sequence long ago.


\vspace*{2.5cm}

\begin{table}
\tablenum{3}
\caption{Derived Global Stellar Mass and SFRs}\label{table:derived}
\vspace*{0.5cm}


\def\arraystretch{0.75}%
\hspace*{-3.5cm}
\footnotesize{
\begin{tabular}
{r | r | r | r | r | r | r | r | r | r | r | r |r|}
\hline
\hline
iS & Galaxy & d & beam &  1$-$2 &  2$-$3 & Log L$_{\rm W1}$ & Log M$_\star$ &Log $\nu$L$_{3}$ & Log $\nu$L$_{4}$ &  SFR$_{W3}$ & SFR$_{W4}$  &Log\,sSFR\\
\, &  \,         &   Mpc  & pc &    mag    &     mag   &  L$_{\odot,W1}$       &   M$_\odot$  &    L$_\odot$             &     L$_\odot$        & 
 M$_\odot$ yr$^{-1}$         & M$_\odot$ yr$^{-1}$ & yr$^{-1}$  \\
 (1) &  (2) &  (3) &  (4) &  (5) &  (6) &  (7) &  (8) &  (9) & (10) & (11) & (12) & (13) \\
\hline
 1 & LMC & 0.05 & 1 & 0.06 & 2.79 &   9.47 0.01 &   9.15 0.11 &   7.62 0.01 &   7.76 0.01 &   0.12 0.03 &   0.03 0.14 & -10.07  \\
               2 & SMC & 0.06 & 2 & -0.05 & 1.55 &   8.63 0.01 &   8.59 0.09 &   6.06 0.01 &   6.43 0.01 &   0.01 0.01 &   0.01 0.01 & -10.59  \\
               3 & MESSIER31 & 0.77 & 22 & -0.03 & 2.09 &  11.04 0.02 &  10.95 0.14 &   8.82 0.02 &   8.48 0.03 &   1.20 0.11 &   0.58 0.07 & -10.88  \\
               4 & MESSIER33 & 0.86 & 25 & 0.07 & 3.18 &   9.97 0.02 &   9.61 0.14 &   8.33 0.02 &   8.22 0.03 &   0.45 0.04 &   0.33 0.04 & -9.96  \\
               5 & NGC0253 & 3.01 & 88 & 0.19 & 3.83 &  10.88 0.01 &  10.24 0.09 &   9.54 0.01 &   9.64 0.01 &   5.36 1.86 &   6.52 2.56 & -9.51  \\
               6 & NGC5128 & 3.65 & 106 & 0.01 & 2.55 &  11.23 0.01 &  11.02 0.09 &   9.25 0.02 &   9.12 0.01 &   2.98 1.03 &   2.16 0.85 & -10.55  \\
               7 & NGC0055 & 2.10 & 61 & 0.09 & 2.66 &   9.77 0.01 &   9.39 0.09 &   7.88 0.02 &   8.06 0.01 &   0.19 0.06 &   0.23 0.09 & -10.11  \\
               8 & MESSIER81 & 3.65 & 106 & -0.01 & 1.81 &  10.99 0.01 &  10.85 0.09 &   8.62 0.04 &   8.44 0.02 &   0.82 0.29 &   0.52 0.20 & -10.94  \\
               9 & MESSIER101 & 7.22 & 210 & 0.12 & 3.54 &  10.91 0.01 &  10.45 0.10 &   9.43 0.01 &   9.36 0.01 &   4.22 1.46 &   3.61 1.42 & -9.82  \\
               10 & NGC4945 & 3.22 & 94 & 0.18 & 3.57 &  10.71 0.01 &  10.07 0.09 &   9.26 0.01 &   9.15 0.01 &   3.03 1.05 &   2.31 0.91 & -9.59  \\
               11 & IC0342 & 3.14 & 91 & 0.05 & 3.80 &  10.74 0.01 &  10.43 0.09 &   9.33 0.02 &   9.23 0.01 &   3.49 1.21 &   2.75 1.08 & -9.88  \\
               12 & NGC1316 & 18.92 & 550 & -0.03 & 0.74 &  11.82 0.01 &  11.72 0.10 &   8.28 0.61 &   8.60 0.06 &   0.42 0.14 &   0.73 0.29 & -12.09  \\
               13 & MESSIER49 & 15.28 & 444 & -0.09 & 0.47 &  11.66 0.01 &  11.62 0.10 & $<$4.28 &   7.73 0.28 & $<$0.01 &   0.12 0.05 & $<$-12.17  \\
               14 & MESSIER86 & 16.32 & 475 & -0.05 & -0.18 &  11.53 0.01 &  11.48 0.10 & $<$4.33 & $<$4.04 & $<$0.01 & $<$0.01 & $<$-10.84  \\
               15 & MAFFEI1 & 3.39 & 99 & -0.04 & -0.03 &  10.85 0.01 &  10.79 0.10 & $<$2.97 &   7.02 0.23 & $<$0.01 &   0.03 0.01 & $<$-11.43  \\
               16 & MESSIER110 & 0.82 & 24 & -0.05 & 0.89 &   9.00 0.01 &   8.96 0.09 &   5.75 0.31 &   5.85 0.06 &   0.01 0.00 &   0.00 0.00 & -10.96  \\
               17 & MESSIER104 & 9.38 & 273 & -0.03 & 1.01 &  11.38 0.01 &  11.28 0.09 &   8.34 0.19 &   8.30 0.04 &   0.47 0.16 &   0.39 0.15 & -11.61  \\
               18 & NGC2403 & 3.18 & 93 & 0.09 & 3.38 &  10.04 0.01 &   9.65 0.09 &   8.47 0.01 &   8.43 0.01 &   0.61 0.21 &   0.51 0.20 & -9.86  \\
               19 & NGC0247 & 3.44 & 100 & -0.04 & 2.67 &   9.79 0.01 &   9.72 0.09 &   7.84 0.03 &   7.73 0.04 &   0.17 0.06 &   0.12 0.05 & -10.49  \\
               20 & MESSIER106 & 7.31 & 213 & 0.03 & 2.51 &  10.93 0.01 &  10.68 0.09 &   8.94 0.02 &   8.80 0.02 &   1.60 0.55 &   1.12 0.44 & -10.47  \\
               21 & MESSIER87 & 15.80 & 460 & -0.07 & 0.42 &  11.57 0.01 &  11.53 0.10 & $<$4.30 &   8.08 0.11 & $<$0.01 &   0.24 0.10 & $<$-12.49  \\
               22 & NGC3628 & 10.45 & 304 & 0.12 & 3.24 &  11.02 0.01 &  10.53 0.09 &   9.40 0.01 &   9.33 0.01 &   4.05 1.40 &   3.40 1.33 & -9.93  \\
               23 & NGC0300 & 1.93 & 56 & -0.00 & 2.59 &   9.58 0.01 &   9.41 0.10 &   7.62 0.02 &   7.55 0.01 &   0.11 0.04 &   0.08 0.03 & -10.37  \\
               24 & NGC4736 & 5.20 & 151 & 0.02 & 2.75 &  10.81 0.01 &  10.58 0.10 &   8.93 0.02 &   8.80 0.01 &   1.56 0.54 &   1.11 0.44 & -10.39  \\
               25 & NGC6822 & 0.47 & 14 & -0.02 & 2.45 &   8.34 0.01 &   8.22 0.09 &   6.30 0.04 &   6.39 0.05 &   0.01 0.00 &   0.01 0.00 & -10.22  \\
               26 & NGC1532 & 14.25 & 415 & 0.05 & 2.86 &  11.02 0.01 &  10.73 0.09 &   9.20 0.02 &   9.02 0.02 &   2.69 0.93 &   1.76 0.69 & -10.30  \\
               27 & NGC5236 & 4.51 & 131 & 0.12 & 3.82 &  10.90 0.01 &  10.43 0.10 &   9.53 0.01 &   9.56 0.01 &   5.17 1.79 &   5.42 2.12 & -9.72  \\
               28 & NGC0147 & 0.73 & 21 & -0.02 & 1.85 &   8.41 0.01 &   8.30 0.09 &   6.06 0.04 & $<$1.34 & $<$0.01 & $<$0.01 & $<$-9.69  \\
               29 & NGC6744 & 9.51 & 277 & 0.01 & 3.35 &  11.04 0.01 &  10.83 0.09 &   9.43 0.02 &   9.20 0.01 &   4.24 1.47 &   2.57 1.01 & -10.20  \\
               30 & MESSIER63 & 9.81 & 285 & 0.05 & 3.35 &  11.16 0.01 &  10.86 0.09 &   9.56 0.01 &   9.38 0.01 &   5.60 1.94 &   3.75 1.47 & -10.11  \\
               31 & NGC1553 & 17.18 & 500 & -0.06 & 0.69 &  11.39 0.01 &  11.34 0.10 &   7.47 1.42 &   8.04 0.08 &   0.08 0.03 &   0.22 0.09 & -12.44  \\
               32 & NGC1399 & 19.61 & 570 & -0.08 & 0.39 &  11.56 0.01 &  11.51 0.10 & $<$4.49 &   7.47 0.41 & $<$0.01 &   0.07 0.03 & $<$-11.90  \\
               33 & NGC4236 & 4.47 & 130 & -0.09 & 2.32 &   9.44 0.01 &   9.40 0.10 &   7.31 0.04 &   7.56 0.02 &   0.06 0.02 &   0.08 0.03 & -10.62  \\
               34 & NGC4565 & 12.05 & 351 & 0.04 & 2.56 &  11.14 0.01 &  10.88 0.09 &   9.18 0.02 &   8.99 0.01 &   2.58 0.89 &   1.67 0.66 & -10.47  \\
               35 & Maffei2 & 2.79 & 81 & 0.07 & 3.57 &  10.42 0.01 &  10.07 0.08 &   8.93 0.01 &   8.82 0.01 &   1.53 0.53 &   1.17 0.46 & -9.89  \\
               36 & NGC4631 & 7.35 & 214 & 0.19 & 3.84 &  10.63 0.01 &   9.98 0.09 &   9.30 0.01 &   9.27 0.01 &   3.26 1.13 &   2.95 1.16 & -9.46  \\
               37 & MESSIER60 & 17.36 & 505 & -0.05 & 0.68 &  11.60 0.01 &  11.56 0.10 &   7.64 1.61 &   8.14 0.11 &   0.11 0.04 &   0.27 0.11 & -12.52  \\
               38 & NGC4636 & 14.85 & 432 & -0.07 & 0.33 &  11.19 0.01 &  11.15 0.10 & $<$4.25 &   7.38 0.22 & $<$0.01 &   0.06 0.02 & $<$-11.80  \\
               39 & NGC2768 & 20.60 & 599 & -0.03 & 0.34 &  11.20 0.01 &  11.09 0.10 & $<$4.54 &   7.46 0.18 & $<$0.01 &   0.07 0.03 & $<$-11.83  \\
               40 & NGC3585 & 18.14 & 528 & -0.07 & 0.25 &  11.31 0.01 &  11.26 0.10 & $<$4.43 &   7.30 0.34 & $<$0.01 &   0.05 0.02 & $<$-11.74  \\
               41 & ESO270-G017 & 15.37 & 447 & 0.01 & 2.43 &  10.24 0.01 &  10.04 0.10 &   8.21 0.04 &   8.31 0.03 &   0.36 0.13 &   0.40 0.16 & -10.48  \\
               42 & MESSIER51a & 9.12 & 265 & 0.11 & 3.81 &  11.16 0.01 &  10.71 0.09 &   9.79 0.01 &   9.66 0.01 &   8.77 3.04 &   6.77 2.65 & -9.77  \\
               43 & NGC3115 & 9.64 & 280 & -0.02 & 0.33 &  11.03 0.01 &  10.91 0.10 & $<$3.88 &   6.87 0.48 & $<$0.01 &   0.02 0.01 & $<$-11.22  \\
               44 & NGC3923 & 20.08 & 584 & -0.10 & 0.15 &  11.52 0.01 &  11.47 0.10 & $<$4.51 & $<$4.22 & $<$0.01 & $<$0.01 & $<$-10.82  \\
               45 & NGC4365 & 21.68 & 631 & -0.11 & -0.37 &  11.46 0.01 &  11.42 0.10 & $<$4.58 & $<$4.28 & $<$0.01 & $<$0.01 & $<$-10.76  \\
               46 & NGC1313 & 4.12 & 120 & 0.05 & 3.30 &   9.86 0.01 &   9.55 0.09 &   8.24 0.02 &   8.37 0.02 &   0.38 0.13 &   0.45 0.18 & -9.97  \\
               47 & MESSIER84 & 17.70 & 515 & -0.03 & 0.12 &  11.50 0.01 &  11.41 0.10 & $<$4.40 &   7.04 0.97 & $<$0.01 &   0.03 0.01 & $<$-11.42  \\
               48 & NGC0185 & 0.64 & 19 & -0.04 & 0.89 &   8.43 0.01 &   8.35 0.09 &   5.22 0.29 &   5.13 0.07 & $<$0.01 & $<$0.01 & $<$-9.48  \\
               49 & NGC6946 & 5.89 & 171 & 0.16 & 3.91 &  10.84 0.01 &  10.26 0.09 &   9.52 0.01 &   9.46 0.01 &   5.14 1.78 &   4.43 1.74 & -9.55  \\
               50 & NGC1395 & 22.03 & 641 & -0.07 & 0.28 &  11.40 0.01 &  11.36 0.10 & $<$4.59 &   7.40 0.34 & $<$0.01 &   0.06 0.02 & $<$-11.83  \\
\hline
\end{tabular}

\tablecomments{columns:  (1) order of W1 (3.4\m) angular size;  (2) Galaxy name;  (3) distance; (4) physical resolution, based on 6\arcs beam; 
(5) W1-W2 color; (6) W2-W3 color;
(7) W1 in-band luminosity ;  (10)  stellar mass derived from Log L$_{\rm W1}$ and the color-dependent mass-to-light;
(9, 10) W3 and W4 spectral luminosities, Log $\nu$L$_\nu$, after subtraction of the stellar continuum;    (11, 12) TIR star formation rates based on W3 and W4. }

}

\hfill{}
\end{table}

\clearpage{}

\begin{table}[!ht]
\tablenum{3}
\caption{... continued:  Derived Global Stellar Mass and SFRs}\label{table:derived2}
\vspace*{0.5cm}


\def\arraystretch{0.70}%
\hspace*{-3.8cm}
\footnotesize{
\begin{tabular}
{r | r | r | r | r | r | r | r | r | r | r | r | r |}
\hline
iS & Galaxy & d & beam &  1$-$2 &  2$-$3 & Log L$_{\rm W1}$ & Log M$_\star$ &Log $\nu$L$_{3}$ & Log $\nu$L$_{4}$ &  SFR$_{W3}$ & SFR$_{W4}$  &Log\,sSFR\\
\, &  \,         &   Mpc  & pc &    mag    &     mag   &  L$_{\odot,W1}$       &   M$_\odot$  &    L$_\odot$             &     L$_\odot$        & 
 M$_\odot$ yr$^{-1}$         & M$_\odot$ yr$^{-1}$ & yr$^{-1}$  \\
 (1) &  (2) &  (3) &  (4) &  (5) &  (6) &  (7) &  (8) &  (9) & (10) & (11) & (12) & (13) \\
\hline
               51 & IC0010 & 0.77 & 22 & 0.03 & 2.73 &   9.10 0.01 &   8.87 0.08 &   7.22 0.02 &   7.42 0.01 &   0.05 0.02 &   0.06 0.02 & -10.17  \\
               52 & NGC4517 & 10.63 & 309 & 0.08 & 3.15 &  10.51 0.01 &  10.13 0.09 &   8.83 0.01 &   8.72 0.01 &   1.27 0.44 &   0.94 0.37 & -10.02  \\
               53 & NGC1291 & 10.54 & 307 & -0.05 & 1.20 &  11.18 0.01 &  11.12 0.10 &   8.33 0.12 &   8.19 0.05 &   0.46 0.16 &   0.31 0.12 & -11.46  \\
               54 & NGC2683 & 8.47 & 246 & 0.03 & 2.37 &  10.70 0.01 &  10.45 0.09 &   8.64 0.02 &   8.29 0.02 &   0.87 0.30 &   0.38 0.15 & -10.51  \\
               55 & NGC4697 & 10.50 & 305 & -0.05 & 0.22 &  10.94 0.01 &  10.90 0.10 & $<$3.95 &   7.26 0.17 & $<$0.01 &   0.04 0.02 & $<$-11.67  \\
               56 & NGC3521 & 8.60 & 250 & 0.09 & 3.40 &  11.00 0.01 &  10.61 0.09 &   9.44 0.01 &   9.26 0.01 &   4.37 1.51 &   2.92 1.14 & -9.97  \\
               57 & NGC3109 & 1.25 & 36 & -0.02 & 1.35 &   8.30 0.01 &   8.18 0.10 &   5.60 0.09 & $<$1.80 & $<$0.01 & $<$0.00 & $<$-9.23  \\
               58 & NGC0891 & 7.09 & 206 & 0.19 & 3.37 &  10.81 0.01 &  10.16 0.09 &   9.28 0.01 &   9.13 0.01 &   3.16 1.09 &   2.20 0.86 & -9.66  \\
               59 & MESSIER85 & 14.38 & 418 & -0.04 & 0.38 &  11.27 0.01 &  11.21 0.10 & $<$4.22 &   7.63 0.15 & $<$0.01 &   0.10 0.04 & $<$-12.03  \\
               60 & NGC4244 & 4.21 & 122 & 0.04 & 2.29 &   9.55 0.01 &   9.29 0.09 &   7.46 0.03 &   7.49 0.03 &   0.08 0.03 &   0.07 0.03 & -10.38  \\
               61 & NGC4762 & 13.86 & 403 & -0.05 & 0.42 &  10.71 0.01 &  10.66 0.10 & $<$4.19 &   6.80 0.27 & $<$0.01 &   0.02 0.01 & $<$-11.22  \\
               62 & NGC5084 & 17.88 & 520 & -0.03 & 1.41 &  11.06 0.01 &  10.95 0.09 &   8.41 0.09 &   8.27 0.04 &   0.54 0.19 &   0.36 0.14 & -11.22  \\
               63 & NGC5907 & 17.06 & 496 & 0.11 & 3.15 &  11.16 0.01 &  10.72 0.09 &   9.50 0.01 &   9.33 0.01 &   4.92 1.70 &   3.39 1.33 & -10.03  \\
               64 & NGC4395 & 4.51 & 131 & -0.04 & 2.47 &   9.50 0.01 &   9.42 0.10 &   7.46 0.04 &   7.62 0.04 &   0.08 0.03 &   0.09 0.04 & -10.52  \\
               65 & NGC1407 & 25.41 & 739 & -0.09 & 0.34 &  11.63 0.01 &  11.59 0.10 & $<$4.72 &   6.62 3.39 & $<$0.01 & $<$0.01 & $<$-11.06  \\
               66 & NGC3627 & 10.02 & 291 & 0.10 & 3.45 &  11.09 0.01 &  10.67 0.09 &   9.56 0.01 &   9.51 0.01 &   5.49 1.90 &   4.97 1.95 & -9.93  \\
               67 & NGC4438 & 11.32 & 329 & 0.01 & 1.57 &  10.69 0.01 &  10.49 0.09 &   8.19 0.06 &   8.06 0.03 &   0.34 0.12 &   0.23 0.09 & -10.96  \\
               68 & NGC1365 & 17.92 & 521 & 0.28 & 3.65 &  11.38 0.01 &  10.70 0.09 &  10.01 0.01 &  10.19 0.01 &  13.62 4.72 &  20.39 8.00 & -9.56  \\
               69 & NGC2903 & 9.98 & 290 & 0.08 & 3.55 &  11.02 0.01 &  10.64 0.09 &   9.52 0.01 &   9.46 0.01 &   5.07 1.76 &   4.42 1.74 & -9.94  \\
               70 & NGC5846 & 25.23 & 734 & -0.06 & 0.10 &  11.54 0.01 &  11.50 0.10 & $<$4.71 & $<$4.42 & $<$0.01 & $<$0.00 & $<$-10.82  \\
               71 & NGC4725 & 12.74 & 371 & -0.03 & 2.20 &  11.08 0.01 &  10.98 0.09 &   8.91 0.03 &   8.68 0.02 &   1.50 0.52 &   0.86 0.34 & -10.80  \\
               72 & NGC1549 & 17.66 & 514 & -0.09 & 0.54 &  11.22 0.01 &  11.18 0.10 & $<$4.40 &   7.45 0.20 & $<$0.01 &   0.07 0.03 & $<$-11.88  \\
               73 & WLM & 0.94 & 27 & -0.08 & 0.64 &   7.68 0.01 &   7.64 0.10 & $<$1.90 &   5.23 0.16 & $<$0.01 &   0.00 0.00 & $<$-8.44  \\
               74 & NGC2841 & 14.08 & 410 & -0.01 & 2.16 &  11.22 0.01 &  11.09 0.09 &   9.04 0.03 &   8.91 0.01 &   1.93 0.67 &   1.40 0.55 & -10.80  \\
               75 & Circinus & 4.21 & 122 & 0.61 & 4.07 &  10.85 0.01 &  10.17 0.08 &   9.79 0.01 &   9.80 0.01 &   8.79 3.04 &   8.98 3.52 & -9.23  \\
               76 & NGC3621 & 7.01 & 204 & 0.13 & 3.66 &  10.46 0.01 &   9.95 0.12 &   9.03 0.01 &   8.87 0.01 &   1.89 0.65 &   1.29 0.51 & -9.68  \\
               77 & NGC5078 & 31.13 & 906 & 0.05 & 2.64 &  11.59 0.01 &  11.30 0.09 &   9.67 0.02 &   9.50 0.01 &   6.94 2.40 &   4.80 1.88 & -10.46  \\
               78 & NGC1023 & 10.41 & 303 & -0.05 & 0.37 &  10.89 0.01 &  10.85 0.10 & $<$3.94 &   7.03 0.24 & $<$0.01 &   0.03 0.01 & $<$-11.46  \\
               79 & NGC7331 & 14.72 & 428 & 0.08 & 3.28 &  11.37 0.01 &  11.01 0.09 &   9.75 0.01 &   9.59 0.01 &   8.19 2.84 &   5.79 2.27 & -10.09  \\
               80 & MESSIER64 & 10.02 & 291 & 0.00 & 2.26 &  11.24 0.01 &  11.06 0.09 &   9.11 0.03 &   9.02 0.01 &   2.25 0.78 &   1.78 0.70 & -10.70  \\
               81 & MESSIER59 & 16.32 & 475 & -0.08 & 0.61 &  11.12 0.01 &  11.08 0.10 & $<$4.33 &   7.43 0.16 & $<$0.01 &   0.06 0.03 & $<$-11.86  \\
               82 & NGC4696 & 36.83 & 1071 & -0.06 & 0.36 &  11.74 0.01 &  11.70 0.09 & $<$5.04 &   7.98 0.20 & $<$0.01 &   0.20 0.08 & $<$-12.40  \\
               83 & MESSIER82 & 3.70 & 108 & 0.45 & 4.58 &  10.81 0.01 &  10.13 0.10 &   9.89 0.01 &  10.29 0.01 &  10.68 3.70 &  25.27 9.91 & -9.10  \\
               84 & ESO274-001 & 3.18 & 93 & -0.03 & 1.28 &   9.15 0.01 &   9.06 0.09 &   6.40 0.11 &   7.31 0.01 & $<$0.01 &   0.05 0.02 & -11.06  \\
               85 & MESSIER77 & 10.28 & 299 & 1.71 & 3.57 &  11.49 0.01 &  10.81 0.10 &  10.67 0.01 &  10.51 0.01 &  52.00 18.00 &  40.60 15.90 & -9.09  \\
               86 & NGC3077 & 3.82 & 111 & 0.08 & 2.71 &   9.75 0.01 &   9.37 0.09 &   7.88 0.02 &   8.03 0.01 &   0.19 0.06 &   0.22 0.09 & -10.09  \\
               87 & MESSIER65 & 12.65 & 368 & -0.02 & 1.82 &  11.15 0.01 &  11.02 0.09 &   8.78 0.04 &   8.60 0.02 &   1.14 0.40 &   0.73 0.29 & -10.97  \\
               88 & NGC7213 & 26.66 & 776 & 0.08 & 2.14 &  11.53 0.01 &  11.17 0.10 &   9.38 0.03 &   9.37 0.03 &   3.87 1.34 &   3.64 1.44 & -10.58  \\
               89 & IC0356 & 11.75 & 342 & -0.02 & 2.20 &  11.14 0.01 &  11.02 0.09 &   8.97 0.03 &   8.87 0.02 &   1.69 0.59 &   1.30 0.51 & -10.79  \\
               90 & NGC1560 & 3.01 & 88 & -0.08 & 1.04 &   8.83 0.01 &   8.79 0.09 &   5.77 0.20 &   6.22 0.03 & $<$0.01 & $<$0.01 & $<$-10.79  \\
               91 & NGC2663 & 28.49 & 829 & -0.09 & 0.03 &  11.69 0.01 &  11.65 0.09 & $<$4.82 &   8.00 0.17 & $<$0.01 &   0.21 0.08 & $<$-12.43  \\
               92 & NGC4216 & 15.50 & 451 & -0.02 & 2.22 &  11.12 0.01 &  11.00 0.09 &   8.96 0.03 &   8.71 0.02 &   1.65 0.57 &   0.93 0.36 & -10.78  \\
               93 & NGC1055 & 14.16 & 412 & 0.12 & 3.60 &  10.88 0.01 &  10.42 0.09 &   9.42 0.01 &   9.27 0.01 &   4.18 1.45 &   3.01 1.18 & -9.80  \\
               94 & NGC5170 & 22.07 & 642 & 0.02 & 2.26 &  11.00 0.01 &  10.79 0.09 &   8.89 0.03 &   8.70 0.02 &   1.42 0.49 &   0.90 0.35 & -10.64  \\
               95 & MESSIER98 & 15.98 & 465 & 0.03 & 2.91 &  11.01 0.01 &  10.77 0.09 &   9.20 0.02 &   9.03 0.01 &   2.70 0.94 &   1.82 0.71 & -10.34  \\
               96 & NGC2997 & 11.32 & 329 & 0.11 & 3.73 &  10.93 0.01 &  10.49 0.09 &   9.52 0.01 &   9.40 0.01 &   5.08 1.76 &   3.88 1.52 & -9.79  \\
               97 & NGC4125 & 21.99 & 640 & -0.04 & 0.44 &  11.35 0.01 &  11.28 0.10 & $<$4.59 &   7.79 0.12 & $<$0.01 &   0.13 0.05 & $<$-12.18  \\
               98 & NGC7793 & 3.95 & 115 & 0.08 & 3.27 &   9.88 0.01 &   9.51 0.09 &   8.26 0.01 &   8.09 0.01 &   0.40 0.14 &   0.25 0.10 & -9.91  \\
               99 & NGC5363 & 20.30 & 591 & -0.02 & 1.02 &  11.29 0.01 &  11.16 0.10 &   8.28 0.18 &   8.29 0.04 &   0.42 0.14 &   0.38 0.15 & -11.54  \\
               100 & NGC4217 & 18.57 & 540 & 0.16 & 3.34 &  10.93 0.01 &  10.35 0.09 &   9.38 0.01 &   9.20 0.01 &   3.85 1.33 &   2.59 1.02 & -9.76  \\
               - & IC1613 & 0.73 & 21 & -0.19 & $<$3.30 &   7.63 0.01 &   7.58 0.10 & $<$1.63 & $<$1.34 & $<$0.01 & $<$0.01 & $<$-8.58  \\
               - & MESSIER32 & 0.82 & 24 & -0.05 & 0.69 &   9.12 0.01 &   9.07 0.09 &   5.27 1.24 &   5.55 0.12 & $<$0.01 & $<$0.01 & -11.07  \\
               - & UGC05373 & 1.42 & 41 & 0.11 & $<$2.79 &   7.70 0.01 &   7.25 0.10 & $<$2.21 &   5.07 0.14 & $<$0.01 & $<$0.01 & $<$-6.91  \\
               - & ESO245-007 & 0.43 & 13 & -0.22 & $<$1.39 &   6.14 0.01 &   6.10 0.14 & $<$1.17 & $<$2.58 & $<$0.01 & $<$0.01 & $<$-6.95  \\
\hline
\end{tabular}

\tablecomments{columns:  (1) order of angular size;  (2) Galaxy name;  (3) distance; (4) physical resolution, based on 6\arcs beam; 
(5) W1-W2 color; (6) W2-W3 color;
(7) W1 in-band luminosity ;  (10)  stellar mass derived from Log L$_{\rm W1}$ and the W1-W2 color-dependent mass-to-light;
(9, 10) W3 and W4 spectral luminosities, Log $\nu$L$_\nu$, after subtraction of the stellar continuum;    (11, 12) W3 and W4 star formation rates; (13) specific SFR(W3). }

}
\hfill{}
\end{table}

\clearpage{}


\section{\wise Extended Source Catalogue}

\subsection{The Largest Angular-size Galaxies}

We have presented the largest galaxies in the sky, based on the angular apparent size, as measured in the mid-infrared.  Compared to large ($>5000$) samples of bright nearby galaxies, the 100 largest have physical properties that are generally unremarkable.
Nevertheless, because these galaxies are either LG members or  tend to be very close to the Milky Way, $<$ 50 Mpc, we have access to the lowest mass satellites and dwarf galaxies (e.g., NGC\,6822, NGC\,3109), while also graced by two galaxy clusters that have some singularly-large galaxies (e.g., Fornax A and Virgo A, and many other BCGs such as M\,86). 

We have shown that the apparent mid-IR colors have distinct trends that can be exploited to discern galaxies in different states of their star formation history, including semi-quiescent/passive types (intermediate or green-valley disks), actively building SF types (late-type disks), `dead' and gas-poor types (spheroidals), and even incredibly slow yet building types (dwarf spheroidals).   There are two stand-out galaxies with AGN: NGC\,1068 and Circinus, both of which have extraordinarily high surface brightnesses, several magnitudes brighter than typical galaxies, indicative of their intense nuclear activity.    In the following we discuss how the 100 largest and the greater WXSC can be used to further our understanding of galaxy evolution.

\subsection{Legacy of Large Galaxy Archives}  

The 100 largest are interesting in the sense that they are nearby and may be be studied in precise -- resolution and multi-wavelength -- detail; they can have very small physical sizes, or be giants that live within dense environments.  This special sample is but a small set of a much larger one that is being constructed, the whole-sky \wise Extended Source Catalogue (WXSC), which will have tens of thousands of galaxies drawn from the local universe, many of which are legacy galaxies from early surveys such as the Messier and NGC (RC3) catalogs, but also fainter galaxies that enter the realm (targets) of large-area spectroscopic surveys such as SDSS-V (in the north; Kollmeier et al. 2017) and Taipan (in the south; da Cunha et al. 2016). 


The legacy extragalactic data from the WXSC is timely particularly in the southern hemisphere, given the historic paucity of studies below the equator, and the rapid rise of large telescopes in the south.  The Square Kilometre Array (SKA), for example, is driving radio surveys of the southern hemisphere first through the pathfinders (e.g., ASKAP, MeerKAT, both now underway) and later next decade, by the SKA itself. These future surveys notably target the early universe, but will sweep across the sky and image the local universe with clarity (resolution),   quality (S/N), and will range across the vast mass spectrum of galaxies that is so pivotal in connecting how baryons are cycled through the cosmic web, to halos and into galaxies.

The goal of the WXSC archive is to make available data products that are directly useful to these surveys, including source catalogs of global properties, large-area imaging showing the environments of galaxies,  internal measurements (bulge-disk decompositions), and diagrams that summarize physical characteristics.  Ultimately our aiml is to measure all resolved galaxies as discerned in the W1 (3.4\m) band, estimated to be 1 to 2 million galaxies, which may be combined with the ALLWISE point source catalog \citep{Cut12} and the deeper CATWISE catalog (Eisenhardt et al. 2019), to form a complete whole-sky census of galaxies down to 15 micro-Jy, and sensitive to early galaxy formation at redshifts beyond one.

The 100 Largest and WXSC are comprised of several data products -- images, profiles, SEDs and tables -- that fully characterize the measurements from WISE.   As part of a value-add effort to compare and contrast across galaxy properties, we 
create graphic diagrams that depict the key global measurements and derived values -- colors, surface brightness, size, star formation rate --
with the goal to be able to categorize and discern groups of galaxies that have similar properties and SF histories.  We refer to these as ``pinwheel" diagrams, described below.

\subsection{Galaxy Physical Properties:  the PINWHEEL DIAGRAM}  

\begin{figure*}[ht!]   
  \includegraphics[width=190mm]{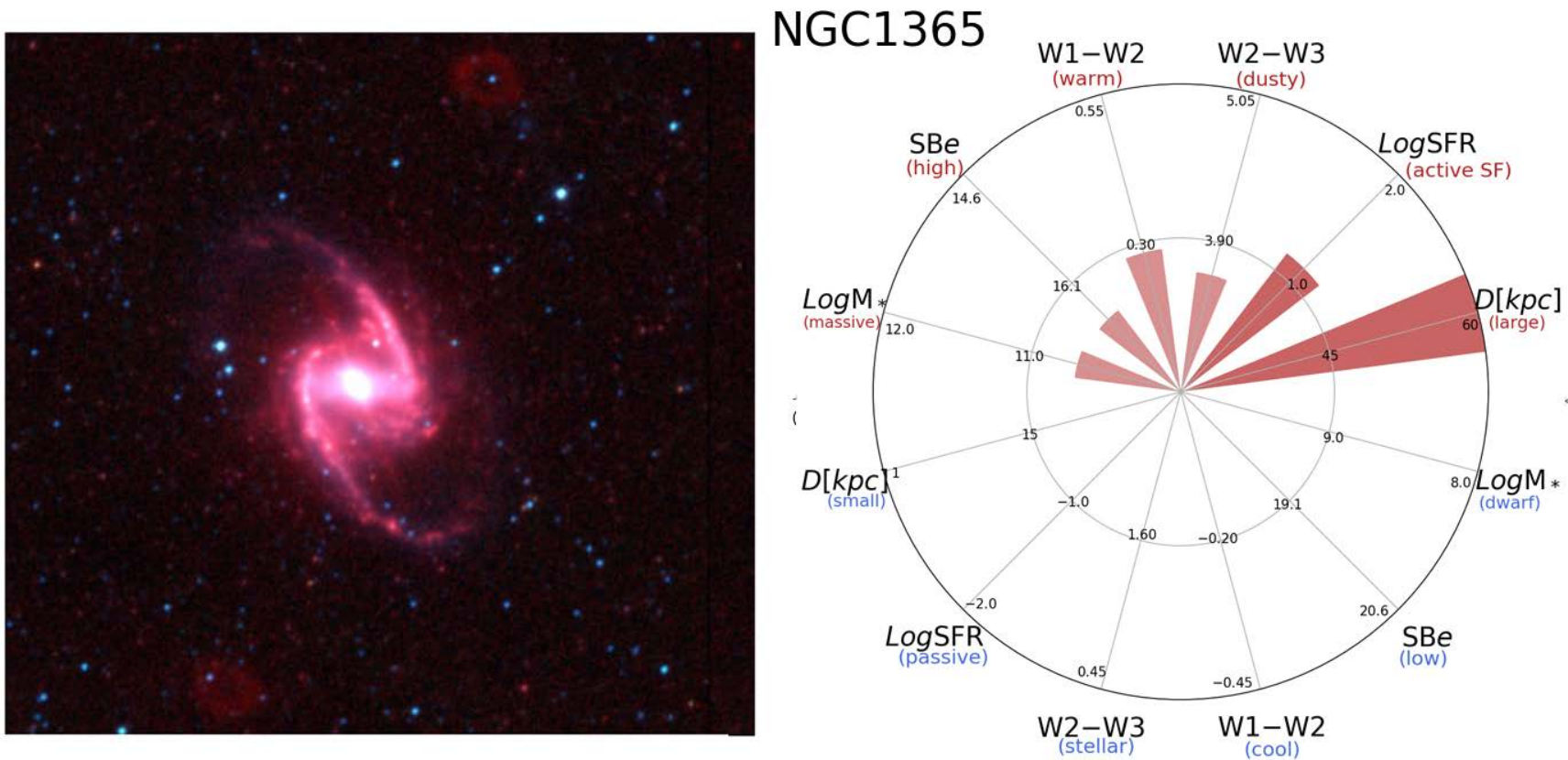}   
 \includegraphics[width=190mm]{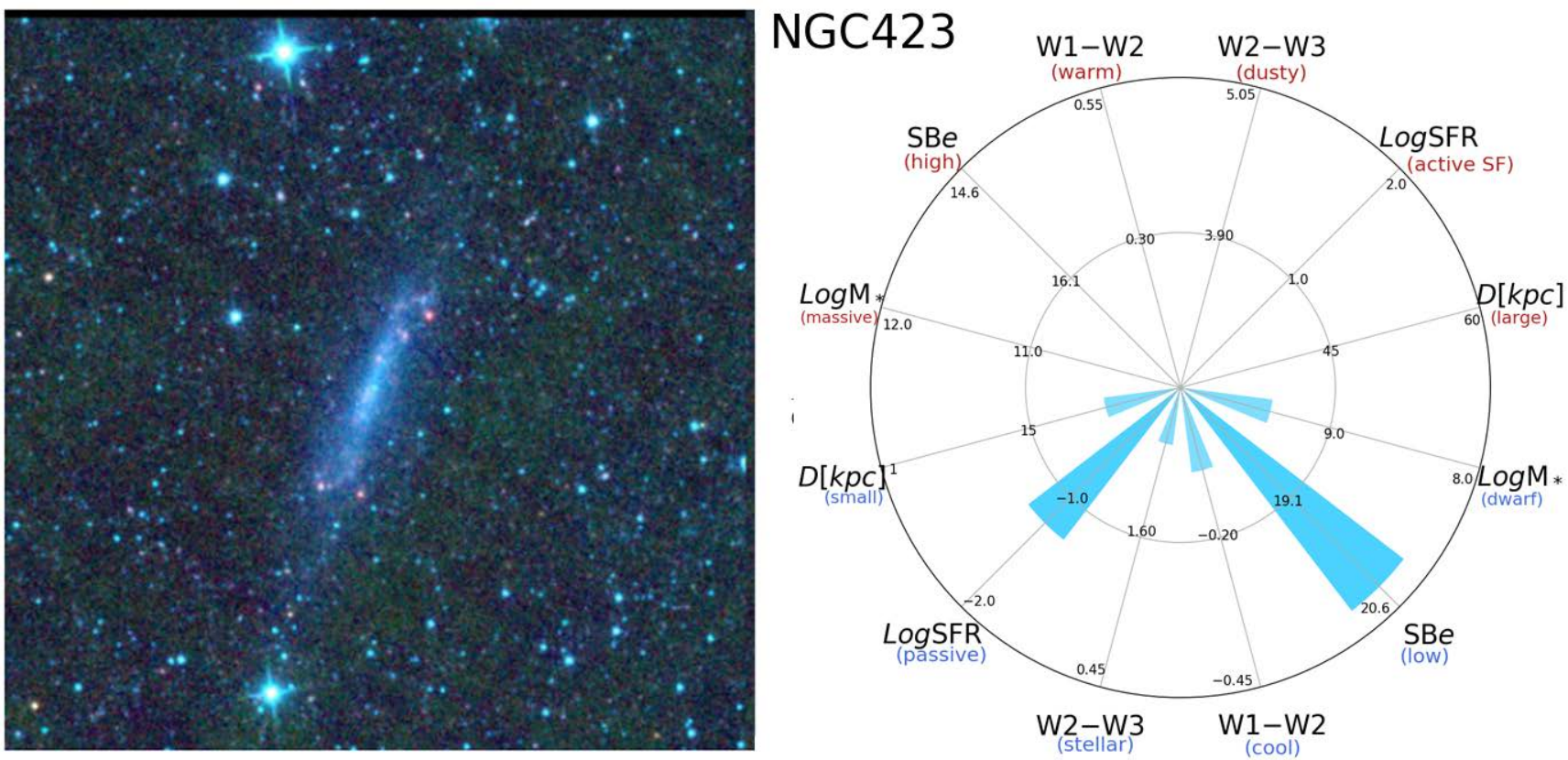}     
  \caption{Physical properties of barred spiral NGC\,1365 (top panel), and 
low surface brightness, highly-inclined galaxy NGC\,4236 (bottom panel).  
The composite diagram shows the \wise 3-color view of the galaxy.  The pinwheel diagram graphically displays the physical properties, with the center of the diagram representing average values for galaxies (see Table 4). The inner and outer rings demark the 50\% and 100\% departures from the canonical average values.
  In this example, NGC\,1365 exhibits `positive' attributes:  active SF and colors, massive and extended host. In contrast, the galaxy NGC\,4236 has `negative' properties:   minimal SF, low surface brightness, dwarf-like mass and extent.
\label{fig:ngc1365}}               
\end{figure*}

Inspired by the `starfish' diagrams created for the SAMI survey of galaxies (see e.g., \citealt{Kon14}),  we have constructed a polar-wedge schematic for each galaxy, which we will refer to as {\it pinwheel} diagrams.  For a given attribute, e.g., SFR, the goal is to convey the value relative to some average or typical value observed for a large sample.  In this way, the circular pinwheel can graphically show many attributes at once, conveying a signature pattern for that galaxy.  We can then use that pattern, for example, to match against other galaxies in order to associate similar kinds of galaxies or those in comparable phases of their stellar-population evolution.

\begin{table}[!ht]
\tablenum{4}
\caption{Pinwheel Diagram Reference Center}\label{table:pinwheel}
\begin{center}

\def\arraystretch{0.8}%
\footnotesize{
\begin{tabular}
{l | r | r | r | r|}
\hline
\hline
Attribute &  units  &  median value  &  floor  &  ceiling  \\
     --         &     --       &           --              &       (min)    &    (max)  \\
\hline
D   &  kpc   &  30  & 1  & 90 \\
Log SFR &  Log [M$_\odot$ yr$^{-1}$] & 0 & -2.0 & +2.0 \\
W2-W3 &  mag  &  2.75  & 0.45  &  5.05 \\
W1-W2 & mag  &   0.05  & -0.45 &  0.55 \\
SB$_e$ &  mag asec$^{-2}$ & 17.6 & 20.6 & 14.6\\
Log M$_\star$ & Log [M$_\odot$] & 10.0 & 8.0 & 12.0 \\
 \hline
\end{tabular}
\tablecomments{The center of the pinwheel diagram has the values presented here;  the median values are derived from the larger WXSC, several thousand galaxies. The exception is the central value for the Log\,SFR, adopted to be the approximate value of the Milky Way and its LG companion, M\,31. 
}
}
\hfill{}
\end{center}
\end{table}

An example of the pinwheel diagram, for the barred Fornax spiral galaxy, NGC\,1365, is shown in Fig.~\ref{fig:ngc1365}a.  The six attributes that make the pinwheel are the following:  mid-IR colors [W1-W2], [W2-W3], effective surface brightness (SBe), stellar mass (Log\,M$_\star$), diameter (D[kpc]), and star formation rate (Log\,SFR).  
Starting from the center, 0\% departure from average, circular rings demark the 50\% and 100\% departure value of the parameter from the typical value.
The center of the diagram represents median values for galaxies in the WXSC (defined in Table~\ref{table:pinwheel}), and so departures from the average value indicate either positive departure (upper half of diagram), or negative departure (lower half).  
For example, galaxies with a large SFR will have a red wedge pointing upwards, toward ``active SF"; whereas galaxies with low Log\,SFRs will have a blue wedge pointing downward, toward ``passive". In the case of NGC\,1365, it has large-valued colors indicating active star formation, somewhat high surface brightness, and is more massive and larger diameter than typical galaxies.  In the Fornax Cluster, NGC\,1365 stands out among its many cluster members.

In contrast, the faint galaxy, NGC\,4236 in Fig.~\ref{fig:ngc1365}b, has `negative' properties including low (dwarf) mass, low surface brightness, low SFR, and cool colors -- indeed, there is not much going on in this physically small galaxy, but it may be quite typical for its stellar mass and evolutionary SFH phase.
It is immediately apparent that NGC\,1365 and NGC\,4236 could not be more different galaxies in their measured and derived properties, visually (and easily) discerned with the pinwheel diagram. 

In Paper II, we will present the measurements and derived properties of the WXSC, and include pinwheel diagrams for each.  We will attempt to identify similar types of galaxies based on the pinwheel patterns, thereby linking their properties to their SF history and environment.

\section{The Brightest Globular Clusters}

\begin{figure}[!htbp]
  \begin{minipage}[b]{1.0\linewidth}
    \centering
     \includegraphics[width=\linewidth]{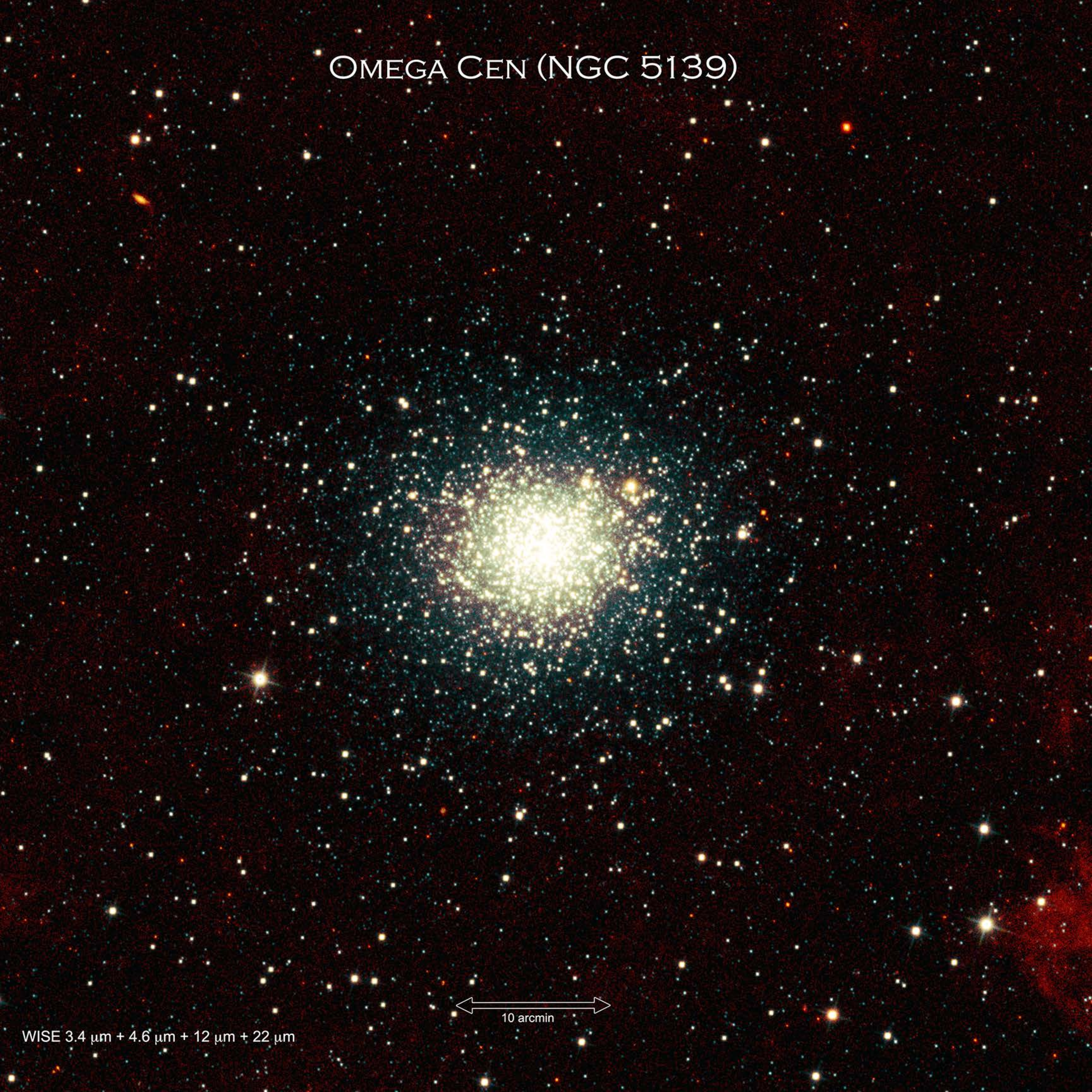}  
    \includegraphics[width=\linewidth]{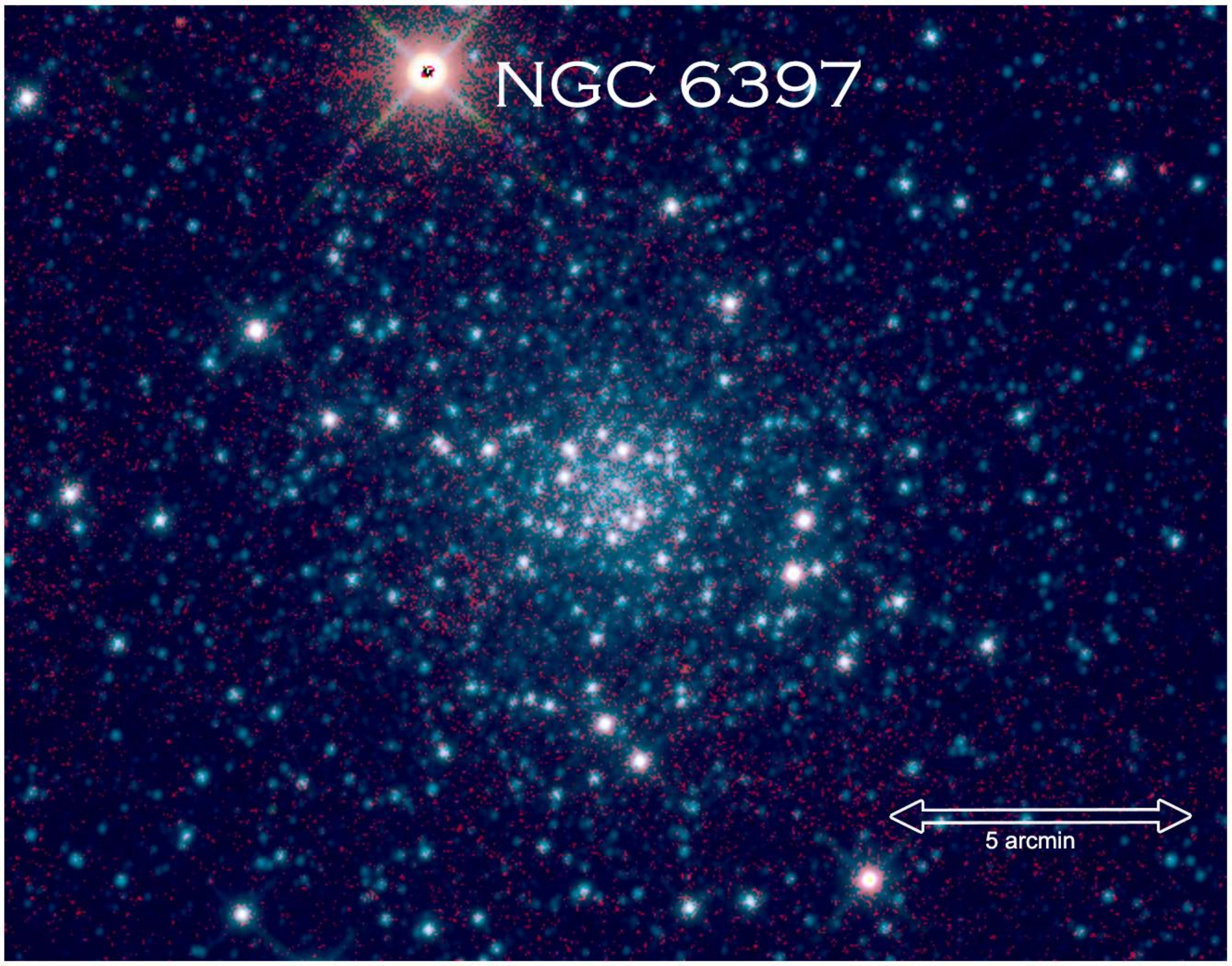}     
\caption{Nearby globular clusters Omega Centaurus (a) and  NGC\,6397 (b), as seen with \wise in all four bands.
Omega Cen is the brightest GC, and likely a dwarf galaxy.  NGC\,6397 has 
 unusually 'warm' W2-W3 colors for a GC, indicating the presence of 
 11.3\m\ PAH emission and significant dust mass, possibly arising from post-AGB shell ejection from a second generation of stars. 
\label{fig:GCs}}               
  \end{minipage}               
\end{figure}

For completeness we include here the brightest MW globular clusters (GCs), since many are also some of the largest and brightest objects in the sky; e.g., 47 Tucanae 
is a spectacular GC in the SMC field (see also Fig. 1), and Omega Centauri (NGC\,5139), Figure~\ref{fig:GCs}a, is so large that it is effectively a dwarf satellite galaxy.  Because we are unable to distinguish foreground (Milky Way) stars from those in the GCs, we measure their fluxes in a similar as was used for the Magellanic Clouds.  Specifically, we compute the mean flux per pixel of the ``sky" (annulus well outside of the GC), and remove that from each pixel value within the GC aperture. 
 

The resulting fluxes are shown in Table~\ref{table:globs}, which presents the 25 brightest GCs in the sky, all are part of the Milky Way system, and most of which are famed Messier objects.  For simplicity, we use the same circular aperture (column 4) for all four \wise bands, and report the integrated flux (sky-subtracted) in Vega magnitudes, and the enclosed surface brightness.  Omega Centauri (Figure~\ref{fig:GCs}a) is by far the brightest GC, with an integrated magnitude of +0.84 in W1 (143 Jy), while the last GC in the list, M\,72, is 6.2 magnitudes fainter (0.5 Jy).  A common property of GCs is their relatively uniform color, due to the dominate population -- evolved, low-mass stars -- the last survivors of these ancient star clusters; note how Omega\,Centauri and 47\,Tucanae (Fig. 1) appear homogenous in apparent properties, but with the occasional dusty (shell-enshrouded) red giant standing out.  Nevertheless, there is a slight spread in color across the 25 Brightest GCs, which would reflect accordingly the range in globular cluster ages and metallicities.

For this sample the mean colors are +0.002 $\pm$ 0.011 and $-$0.101 $\pm$ 0.161 for W1$-$W2 and W2$-$W3 respectively, which would land at the extreme ``blue" end of the galaxy color plot,  indicated in Fig.~\ref{fig:colors}.  As some of the oldest known objects in the Universe, GCs have low-mass, evolved member populations, and correspondingly similar colors to spheroidal galaxies.  None of the 25 brightest GCs have particularly deviant colors;  however, 
in terms of the morphology or SF activity indicator, W2$-$W3 color, the GC with the warmest (red) color is NGC\,6397 (Figure~\ref{fig:GCs}b, 
whose value is +0.56 $\pm$ 0.02 mag, about 3\sig\ warmer than the mean color for GCs.

One of the closest GCs relative to the Sun ($\sim$2.2 kpc; \citet{Alc97}),  NGC\,6397 has the typical GC appearance of a strongly contracted core; see 
\pagebreak
Figure~\ref{fig:GCs}b.  Upon closer inspection of the W3 (12\m) band, there is diffuse emission in the inner $\sim$3${\arcmin }$, consistent with the high(est) surface brightness (SBw3) seen in the sample (see Table 5).  The presence of 11.3\m\ PAH or mid-IR continuum emission indicates significant dust in the inter-GC medium, and consequently moving NGC\,6397 to the right of the GC mean in the color plot (Fig.~\ref{fig:colors}).
Possibly arising from post-AGB shell ejection from a second generation of stars \citep{Mil12}, this dust emission is diffuse and relatively weak (compared to W1 and W2), but the relatively close proximity allows clear detection with {\it WISE}.  Note that for M\,4, the closest GC system to the Sun (also $\sim$2 kpc), does not indicate any excess W3 emission -- 18.08 versus 16.08 mag/arcsec$^2$ in comparison to NGC\,6397 (see Table 5) -- and hence is unlikely to have a significant post-AGB population (see also Maclean et al. 2016).  A detailed analysis of the GC properties is beyond the scope of this present work, but this exercise of measuring the brightest GCs shows that interesting outliers, such as NGC6397, can be discerned and identified from the general photometric properties.

\nopagebreak

\vspace{1cm}

\begin{table}[!hp]
 \begin{minipage}[b]{1.0\linewidth}
\hspace*{-2.5cm}
\tablenum{5}

\caption{Brightest Globular Clusters}\label{table:globs}



\def\arraystretch{0.75}%
\hspace*{-12.5cm}
\footnotesize{
\begin{tabular}
{r | r | r | r | r | r | r | r | r | r | r | r |}
\hline
\hline
Cluster & R.A. &  Dec &  R$_{\rm W1}$ & m1 $\pm \Delta$m1 & SB1 & m2 $\pm \Delta$m2 & SB2 & m3 $\pm \Delta$m3 & SB3 & m4 $\pm \Delta$m4 & SB4 \\
  \, & deg & deg & arcmin & mag & mag $\cdot$ &  mag & mag $\cdot$ &  mag & mag $\cdot$ & mag & mag $\cdot$\\
  \, &    \,  &  \,   &   \,       &    \,   &   sec$^{-2}$        &    \,   &   sec$^{-2}$        &  \,   &   sec$^{-2}$        &  \,   &   sec$^{-2}$        \\
\hline
  NGC5139 & 201.69121 & -47.47686 &    27.50 & 0.84 0.01 & 18.17 & 0.84 0.01 & 18.17 & 1.02 0.01 & 17.03 & 0.96 0.02 & 16.71 \\
  NGC0104 & 6.02233 & -72.08144 &    25.00 & 1.17 0.01 & 18.29 & 1.20 0.01 & 18.33 & 1.13 0.01 & 16.89 & 0.99 0.01 & 15.73 \\
  MESSIER22 & 279.10086 & -23.90341 &    10.83 & 1.92 0.01 & 17.23 & 1.82 0.01 & 17.12 & 1.70 0.01 & 16.44 & -0.25 0.01 & 14.00 \\
  MESSIER04 & 245.89751 & -26.52552 &     9.83 & 2.72 0.01 & 17.81 & 2.61 0.01 & 17.71 & 2.99 0.04 & 18.08 & 2.22 0.05 & 16.47 \\
  NGC6397 & 265.17233 & -53.67369 &    11.67 & 3.11 0.01 & 18.58 & 3.03 0.01 & 18.50 & 2.47 0.01 & 16.43 & 3.42 0.06 & 17.38 \\
  NGC2808 & 138.01060 & -64.86282 &     9.17 & 3.13 0.01 & 18.08 & 3.13 0.01 & 18.07 & 2.73 0.01 & 17.47 & 1.95 0.01 & 16.46 \\
  NGC6752 & 287.71576 & -59.98185 &    10.00 & 3.16 0.01 & 18.29 & 3.13 0.01 & 18.26 & 3.26 0.01 & 18.20 & 3.67 0.10 & 18.18 \\
  MESSIER05 & 229.64064 & 2.08268 &    12.50 & 3.49 0.01 & 19.11 & 3.50 0.01 & 19.12 & 3.34 0.01 & 18.08 & 3.88 0.04 & 17.11 \\
  MESSIER13 & 250.42139 & 36.45853 &    10.83 & 3.57 0.01 & 18.88 & 3.58 0.01 & 18.89 & 3.66 0.01 & 17.91 & 3.67 0.07 & 17.63 \\
  MESSIER10 & 254.28746 & -4.09933 &    10.17 & 3.58 0.01 & 18.75 & 3.57 0.01 & 18.74 & 3.69 0.02 & 17.94 & 3.69 0.03 & 15.81 \\
  MESSIER14 & 264.40067 & -3.24592 &     8.75 & 3.65 0.01 & 18.50 & 3.65 0.01 & 18.49 & 3.73 0.02 & 17.36 & 3.05 0.03 & 15.80 \\
  NGC0362 & 15.80930 & -70.84821 &     8.67 & 3.98 0.01 & 18.80 & 3.98 0.01 & 18.80 & 3.88 0.01 & 18.39 & 3.83 0.05 & 17.06 \\
  MESSIER55 & 294.99750 & -30.96208 &     8.67 & 4.02 0.01 & 18.84 & 4.01 0.01 & 18.84 & 4.20 0.01 & 18.31 & 5.13 0.19 & 17.88 \\
  MESSIER03 & 205.54803 & 28.37712 &    10.83 & 4.06 0.01 & 19.37 & 4.07 0.01 & 19.38 & 3.98 0.01 & 17.61 & 3.95 0.04 & 16.70 \\
  MESSIER15 & 322.49323 & 12.16683 &     9.00 & 4.13 0.01 & 19.03 & 4.12 0.01 & 19.02 & 4.14 0.01 & 18.65 & 4.47 0.08 & 17.22 \\
  MESSIER12 & 251.81049 & -1.94782 &     8.33 & 4.14 0.01 & 18.87 & 4.14 0.01 & 18.87 & 4.43 0.03 & 18.55 & 4.27 0.11 & 17.89 \\
  MESSIER02 & 323.36255 & -0.82332 &     7.50 & 4.15 0.01 & 18.66 & 4.16 0.01 & 18.66 & 4.29 0.01 & 17.92 & 4.21 0.07 & 16.96 \\
  MESSIER92 & 259.28030 & 43.13652 &     9.58 & 4.34 0.01 & 19.38 & 4.33 0.01 & 19.37 & 4.58 0.02 & 17.04 & 4.54 0.04 & 16.67 \\
  MESSIER80 & 244.26045 & -22.97511 &     5.83 & 4.53 0.01 & 18.49 & 4.52 0.01 & 18.49 & 4.83 0.03 & 16.96 & 4.14 0.04 & 15.59 \\
  MESSIER54 & 283.76364 & -30.47850 &     5.42 & 4.63 0.01 & 18.43 & 4.55 0.01 & 18.35 & 4.27 0.01 & 17.50 & 3.94 0.04 & 16.06 \\
  MESSIER107 & 248.13298 & -13.05363 &     7.50 & 4.79 0.01 & 19.30 & 4.82 0.01 & 19.33 & 5.54 0.10 & 18.00 & 7.03 0.54 & 18.27 \\
  MESSIER53 & 198.23013 & 18.16912 &    10.42 & 5.43 0.02 & 20.66 & 5.43 0.04 & 20.66 & 5.35 0.02 & 19.32 & 5.15 0.08 & 16.87 \\
  MESSIER79 & 81.04414 & -24.52423 &     5.17 & 5.60 0.01 & 19.30 & 5.60 0.01 & 19.30 & 6.42 0.10 & 19.17 & 5.41 0.17 & 18.15 \\
  NGC0288 & 13.19106 & -26.58730 &     7.83 & 5.80 0.01 & 20.40 & 5.80 0.01 & 20.41 & 5.97 0.03 & 19.45 & 6.26 0.43 & 19.01 \\
  MESSIER72 & 313.36630 & -12.53706 &     2.92 & 7.02 0.01 & 19.48 & 7.05 0.01 & 19.51 & 7.35 0.03 & 18.36 & 7.47 0.35 & 17.21 \\
\hline
\end{tabular}
\tablecomments{Coordinates are in J2000;  photometry with circular apertures, radius R$_{\rm W1}$;  SB is the mean surface brightness for the photometric aperture.}

}
\hfill{}
 \end{minipage}
\end{table}

\clearpage

\section{Summary}


Building upon the early pilot study of \citet{Jar13}, this is the first paper in a series that encompasses the \wise Extended Source Catalogue (WXSC) data release, here showcasing the 100 largest angular-diameter galaxies.  These objects are unique in that we are able to discern parsec and sub-kpc scales with high signal-to-noise due to their apparent brightness and proximity, notably for the Local Group  (D $<$ 1 Mpc) and Local Volume ($<$ 10 Mpc), enabling high fidelity study of their internal composition and galactic `ecosystem'.  
In this study we measure their global properties, including flux, 
surface brightness and colors, as well as their two-component axi-symmetric radial distribution.

The largest of the large are the Magellanic Clouds and the Andromeda Galaxy, which are also the (integrated) brightest galaxies in the sky by several orders of magnitude.  From their apparent measurements and estimated distance, we derive physical properties such as luminosity, stellar mass, and star formation rate, and compare the largest galaxies with a statistically significant sample of bright nearby galaxies.  We have created a visual diagram for each galaxy that simultaneously depicts the key global properties, which we envision using to categorize and group galaxies with common evolutionary traits. 
Finally, we present the largest and brightest Milky Way globular clusters, integrated measurements of size, flux, color and surface brightness.  

\vspace{0.1cm}
To summarize the main results, we have investigated and presented the following:

\begin{enumerate} 

   \item We have constructed a whole-sky atlas of (chiefly) nearby, bright galaxies and globular clusters called the \wise Extended Source Catalogue (WXSC).   For each source, we have constructed new -- deep and wide --  mosaics from which to measure and characterize the target galaxy, as well as sources that are nearby in projected radial distance.  To date, we have measured over 70,000 sources.   
   For the first release, we choose the largest angular galaxies in the sky, starting from the Magellanic Clouds, to the major Local Group galaxies such as M\,31 and M\,33, to nearby galaxy groups and clusters (Virgo and Fornax). 
    The WXSC will allow us to compare the most well-studied galaxies with large and more distant galaxies samples from {\it WISE},  {\it Spitzer}, {\it Herschel}, {\it Euclid}, LSST, {\it JWST}, and the SKA pathfinders.
   
   \item We have measured the global and internal radial brightness properties of the 100 largest galaxies based on the \wise W1 (3.4\m) 1-\sig$_{\rm sky}$ isophotal radius (about 23 mag arcsec$^{-2}$), which typicially reaches axi-symmetric-averaged depths of 25 mag arcsec$^{-2}$, or approximately 28 mag arcsec$^{-2}$ in the AB system.  We present the results for angular diameters, integrated flux from isophotes and ``total" extractions from both asymptotic apertures and fitting to the radial profile to levels below the noise, surface brightness, and colors, W1$-$W2 and W2$-$W3.  The Magellanic Clouds, M\,31, and M\,33 are large enough that special methods were required to extract their light from the intervening foreground Galactic emission.
   
   \item Because of the high quality measurements from the largest and brightest galaxies in the sky (WXSC), 
   the \wise color-color diagram reveals a tight sequence that represents galaxy morphology and star-formation history, including present-day activity.   We use the rest-frame-corrected measurements from the largest and highest S/N galaxies to 
fit the functional form of this \wise color-color ``sequence", given by:
${\rm (W1-W2)} = [0.015 \times {\rm e}^{ \frac{{\rm (W2-W3)}}{1.38} }]  -  0.08$.
Offsets from the sequence arise from the usual photometric scatter, and more meaningfully, from excess infrared emission associated with nuclear activity from AGN and starbursts.
   
    \item  The sample is sub-divided by the \wise colors, which serve as proxies for four general types of galaxies:  bulge-dominated spheroidals, intermediate semi-quiescent disk, SF spirals, and AGN-dominated systems.   
Physical properties and attributes are computed based on their distance, notably the diameter, aggregate stellar mass and the dust-obscured star formation rate.   We use this classification scheme to study their global properties:  effective surface brightness, size, bulge-to-disk ratios,  luminosity.    We compare the half-light radii and surface brightnesses between \wise W1 (3.4\m) and 2MASS (2.2\m) measurements, showing that W1 radii are much larger and surface brightnesses much fainter than those extracted from the less sensitive 2MASS imaging, notably for dwarf and low-surface brightness galaxies.

  \item We find that the global properties are not remarkable compared to galaxies in the local universe, except in the sense that we can detect and discern the smallest and lowest-mass dwarf and satellite galaxies because of their close proximity.   Nevertheless, the 100 largest galaxies include bright cluster (Virgo and Fornax) galaxies (which have enormous diameters, $>$100 kpc, very high B/T ratios, and aggregate stellar masses; NGC\,1316 is the most massive), starbursts (such as NGC\,253 and M\,82 with SFRs ten times the rate of the Milky Way), and AGN (e.g. Circinus and NGC\,1068, with surface brightnesses that are so bright that they image-saturate in the mid-IR).    
    
    \item In terms of the star formation history, the 100 largest galaxies tend to have lower specific-SFRs compared to field galaxies, notably compared to a much larger sample of nearby galaxies belonging to the WXSC and compared to deeper redshift-selected samples, such as those from GAMA\,G12 \citep{Jar17}.
    
    \item The low mass end is dominated by dwarf spheroids (e.g., NGC\,0185), which have very low SFRs and hence, slowly building their bulge population.  Early-type disk galaxies, such as M\,81, are passively building with rates that fall below the sequence trend.    Late-type spirals, such as M\,83, are actively building with rates perfectly consistent with the SFH sequence observed in the GAMA\,G12 study. 
    
    \item  To efficiently display the attributes that we are capable of estimating with \wise measurements, we introduce a `pinwheel' diagram that depicts the physical properties with respect to the median value observed for galaxies in the WXSC.  These six attributes are the physical diameter, surface brightness, colors, SFR and stellar mass.   We show that with this diagram, it is possible to delineate between different kinds of galaxies, identifying those with similar SFHs, for example.  The Pinwheel Diagrams will be a featured product as part of the WXSC image, catalog and ancillary data archive.
    
    \item Finally, we present the 25 brightest globular clusters in the sky, for which many are also the largest and brightest objects outside of the Milky Way.  Most notably Omega\,Centauri, 47\,Tucanae and a number of famed night-sky targets (e.g., Hercules/M\,13).  GCs have mid-IR color properties that are similar to spheroidal galaxies, indeed Omega\,Centauri is essentially this type of object,  indicative of their dominant evolved-stellar populations.

\end{enumerate}

\acknowledgements

\section*{Acknowledgements}

THJ thanks Barry Madore (and the NED team, notably Joe Mazzarella) for the many wonderful and inspiring discussions of nearby galaxies over the years.  He would also like to thank the incredible \wise team for a job well done, and notably Ned Wright, Peter Eisenhardt and Roc Cutri for creating and shepherding \wise from idea to brilliant reality.
We thank the anonymous referee for helpful analysis suggestions.
THJ acknowledge support from the National Research Foundation (South Africa).
 MC is a recipient of an Australian Research Council Future Fellowship (project number FT170100273) funded by the Australian Government.
This research has made use of the NASA/IPAC Extragalactic Database (NED) and Wide-field Infrared Survey Explore (WISE), both of which are
operated by the Jet Propulsion Laboratory, California Institute of Technology, under contract with the National Aeronautics and Space Administration, and \wise is also 
a joint project with the University of California, Los Angeles.

\pagebreak

\begin{appendices}

\section{Total Fluxes:  Large Aperture and Radial-Profile Fitting Photometry}

This appendix is comprised of a table that lists the `total' fluxes for the 100 largest galaxies and LG galaxies detected by \wise
(except the Magellanic Clouds).  Two methods are used to estimate total fluxes:  (1) large asymptotic apertures
from curve-of-growth measurements, and double-Sersic Function fitting to the axi-symmetric radial profile.
The method and results are described in Section 4.1, with some statistical comparison results presented in
Table 2.

\setcounter{table}{0}
\renewcommand{\thetable}{A\arabic{table}}

\begin{table}[!hp]
\begin{minipage}{1.0\textwidth}
\caption{Asymptotic and  Total Integrated Brightness}\label{table:total}
\vspace*{0.5cm}


\def\arraystretch{0.70}%
\hspace*{-3.5cm}
\footnotesize{
\begin{tabular}
{r | r | r | r | r | r | r | r | r | r | r | r |}
\hline
\hline
iS &  Galaxy &
R$_{\rm W1}$ A$_{\rm W1} \pm \Delta$ & 
R$_{\rm W2}$ A$_{\rm W2} \pm \Delta$ & 
R$_{\rm W3}$ A$_{\rm W3} \pm \Delta$ & 
R$_{\rm W4}$ A$_{\rm W4} \pm \Delta$ & 
T$_{\rm W1} \pm \Delta$ & 
T$_{\rm W2} \pm \Delta$ & 
T$_{\rm W3} \pm \Delta$ & 
T$_{\rm W4} \pm \Delta$ \\
\, & \, & 
amin mag mag & 
amin mag mag & 
amin mag mag & 
amin mag mag & 
mag mag & 
mag mag & 
mag mag & 
mag mag \\
(1) & (2) & (3) & (4) & (5) & (6) & (7) & (8) & (9) & (10)\\
\hline
  3 &   M31                    &  149.39  0.08  0.01  &  132.78  0.13  0.01  &  148.80 -1.95  0.01  &   99.38 -3.11  0.01  &     0.06  0.01  &   0.10  0.01  &  -2.00  0.01  &  -3.13  0.01   \\
    4 &   M33                    &   65.32  3.03  0.01  &   38.83  3.00  0.01  &   37.28 -0.17  0.01  &   30.55 -1.94  0.01  &     3.08  0.01  &   3.02  0.01  &  -0.22  0.01  &  -2.02  0.01   \\
    5 &   NGC0253                &   25.27  3.46  0.01  &   24.39  3.26  0.01  &   25.13 -0.55  0.03  &   25.33 -2.94  0.03  &     3.46  0.01  &   3.25  0.01  &  -0.56  0.03  &  -2.95  0.03   \\
    6 &   NGC5128                &   23.45  2.99  0.01  &   23.36  2.98  0.01  &    9.08  0.41  0.01  &    8.32 -1.29  0.01  &     2.98  0.01  &   2.97  0.01  &   0.44  0.01  &  -1.26  0.01   \\
    7 &   NGC0055                &   23.36  5.42  0.01  &   23.31  5.33  0.01  &   18.76  2.80  0.01  &   18.66  0.29  0.01  &     5.43  0.01  &   5.30  0.02  &   2.70  0.03  &   0.22  0.02   \\
    8 &   MESSIER81              &   16.57  3.60  0.01  &   16.97  3.60  0.01  &   10.34  1.82  0.03  &   10.61  0.44  0.03  &     3.60  0.01  &   3.59  0.01  &   1.79  0.03  &   0.35  0.03   \\
    9 &   MESSIER101             &   13.01  5.22  0.01  &   13.62  5.10  0.01  &   10.38  1.61  0.01  &   10.28 -0.30  0.01  &     5.19  0.01  &   5.03  0.02  &   1.56  0.02  &  -0.38  0.02   \\
   10 &   NGC4945                &   15.49  4.06  0.01  &   14.72  3.87  0.01  &   12.76  0.31  0.01  &   13.70 -1.57  0.01  &     4.06  0.01  &   3.86  0.01  &   0.30  0.01  &  -1.58  0.01   \\
     11 &   IC0342                 &   15.48  3.95  0.01  &   15.42  3.88  0.01  &   11.40  0.22  0.01  &   10.96 -1.83  0.01  &     3.93  0.01  &   3.85  0.01  &   0.09  0.03  &  -1.84  0.01   \\
   12 &   NGC1316                &   15.21  5.06  0.01  &   12.80  5.11  0.01  &    8.32  4.27  0.01  &    4.05  3.18  0.02  &     5.06  0.01  &   5.10  0.01  &   4.36  0.03  &   3.13  0.05   \\
   13 &   MESSIER49              &   15.03  4.94  0.01  &   12.71  5.03  0.01  &    9.75  4.04  0.02  &    3.10  3.82  0.03  &     4.94  0.01  &   5.03  0.01  &   4.17  0.10  &   3.74  0.09   \\
   14 &   MESSIER86              &   14.23  5.45  0.01  &   11.85  5.52  0.01  &    5.87  5.43  0.02  &    2.08  4.99  0.06  &     5.45  0.01  &   5.51  0.02  &   5.70  0.05  &   4.92  0.10   \\
   15 &   MAFFEI1                &   13.82  3.91  0.01  &   12.32  3.93  0.02  &    7.05  3.22  0.02  &    2.54  2.68  0.01  &     3.90  0.01  &   3.92  0.02  &   3.26  0.08  &   2.50  0.13   \\
   16 &   MESSIER110             &   13.53  5.29  0.01  &   12.06  5.36  0.01  &   12.12  4.57  0.02  &   12.12  3.16  0.07  &     5.28  0.01  &   5.35  0.01  &   4.48  0.08  &   3.25  0.08   \\
   17 &   MESSIER104             &   13.05  4.60  0.01  &   11.05  4.64  0.01  &    9.30  3.54  0.01  &    4.36  2.53  0.02  &     4.60  0.01  &   4.62  0.01  &   3.63  0.02  &   2.42  0.03   \\
   18 &   NGC2403                &   13.11  5.67  0.01  &   12.95  5.57  0.01  &   12.77  2.25  0.01  &   10.92  0.28  0.01  &     5.66  0.01  &   5.53  0.02  &   2.21  0.02  &   0.19  0.02   \\
   19 &   NGC0247                &   13.02  6.45  0.01  &   12.88  6.49  0.01  &   12.91  4.04  0.01  &   12.16  2.27  0.03  &     6.44  0.01  &   6.42  0.02  &   3.85  0.06  &   2.08  0.08   \\
   20 &   MESSIER106             &   12.52  5.22  0.01  &   11.95  5.18  0.01  &    9.93  2.75  0.01  &   10.57  1.09  0.01  &     5.22  0.01  &   5.16  0.01  &   2.68  0.02  &   1.00  0.03   \\
   21 &   MESSIER87              &   12.39  5.29  0.01  &   10.38  5.41  0.01  &    6.75  4.75  0.02  &    2.22  3.79  0.02  &     5.28  0.01  &   5.41  0.02  &   4.96  0.06  &   3.73  0.08   \\
   22 &   NGC3628                &   12.02  5.75  0.01  &   10.21  5.62  0.01  &    9.79  2.42  0.01  &    9.78  0.50  0.01  &     5.74  0.01  &   5.61  0.01  &   2.39  0.02  &   0.46  0.02   \\
   23 &   NGC0300                &   11.91  5.70  0.01  &   11.43  5.72  0.01  &    9.93  3.13  0.01  &   12.00  1.27  0.02  &     5.69  0.01  &   5.65  0.02  &   null  null  &   null  null   \\
   24 &   NGC4736                &   11.37  4.75  0.01  &    9.67  4.74  0.01  &    7.34  2.03  0.01  &    7.47  0.32  0.01  &     4.75  0.01  &   4.74  0.02  &   1.99  0.01  &   0.28  0.02   \\
   25 &   NGC6822                &   11.41  5.73  0.01  &   11.38  5.76  0.01  &   11.40  3.41  0.01  &   11.07  1.13  0.02  &     5.71  0.02  &   5.66  0.03  &   3.36  0.09  &   1.13  0.11   \\
   26 &   NGC1532                &   11.28  6.42  0.01  &    9.38  6.38  0.01  &    8.85  3.58  0.01  &    8.85  1.97  0.02  &     6.42  0.01  &   6.37  0.01  &   3.52  0.04  &   1.90  0.04   \\
   27 &   NGC5236                &   10.53  4.24  0.01  &    9.52  4.12  0.01  &   10.94  0.32  0.01  &    9.27 -1.86  0.01  &     4.24  0.01  &   4.11  0.01  &   0.31  0.01  &  -1.88  0.01   \\
   28 &   NGC0147                &   11.12  6.50  0.01  &    9.30  6.57  0.01  &    9.02  4.55  0.01  &    null  null  null  &     6.47  0.02  &   6.52  0.05  &   null  null  &   null  null   \\
   29 &   NGC6744                &   10.77  5.47  0.01  &   10.88  5.45  0.01  &    8.94  2.27  0.01  &   10.02  0.68  0.01  &     5.47  0.01  &   5.44  0.01  &   2.12  0.04  &   0.58  0.03   \\
   30 &   MESSIER63              &   10.36  5.25  0.01  &    9.94  5.19  0.01  &   10.17  1.87  0.01  &    9.89  0.24  0.01  &     5.24  0.01  &   5.16  0.01  &   1.85  0.01  &   0.20  0.02   \\
   31 &   NGC1553                &   10.41  5.85  0.01  &    8.69  5.95  0.01  &    5.72  5.19  0.01  &    2.69  4.24  0.02  &     5.87  0.01  &   5.96  0.01  &   5.26  0.02  &   4.13  0.05   \\
   32 &   NGC1399                &   10.23  5.80  0.01  &    8.68  5.90  0.01  &    6.00  5.28  0.02  &    2.45  4.77  0.04  &     5.80  0.01  &   5.88  0.01  &   5.51  0.04  &   4.69  0.06   \\
   33 &   NGC4236                &   10.36  7.84  0.01  &   10.32  7.94  0.01  &   10.38  5.93  0.03  &   10.35  3.06  0.04  &     7.81  0.02  &   7.80  0.05  &   5.67  0.09  &   null  null   \\
   34 &   NGC4565                &   10.01  5.75  0.01  &    9.73  5.71  0.01  &   10.19  3.19  0.01  &   10.23  1.63  0.01  &     5.75  0.01  &   5.70  0.01  &   3.15  0.02  &   1.58  0.02   \\
   35 &   Maffei2                &   10.12  4.74  0.01  &    9.99  4.57  0.01  &   10.04  1.05  0.01  &   10.15 -1.06  0.01  &     4.74  0.01  &   4.57  0.01  &   1.01  0.02  &  -1.07  0.01   \\
   36 &   NGC4631                &    9.88  5.99  0.01  &    9.41  5.80  0.01  &    9.92  1.98  0.01  &    9.72 -0.07  0.01  &     5.99  0.01  &   5.78  0.01  &   1.96  0.01  &  -0.10  0.01   \\
   37 &   MESSIER60              &    9.69  5.37  0.01  &    8.18  5.42  0.01  &    6.31  4.33  0.02  &    2.59  3.85  0.04  &     5.36  0.01  &   5.41  0.02  &   4.76  0.11  &   3.77  0.12   \\
   38 &   NGC4636                &    9.59  6.05  0.01  &    8.04  6.13  0.01  &    6.47  5.00  0.02  &    1.98  4.69  0.05  &     6.04  0.01  &   6.11  0.02  &   5.11  0.13  &   4.78  0.17   \\
   39 &   NGC2768                &    9.52  6.74  0.01  &    7.97  6.77  0.01  &    5.90  6.20  0.03  &    2.15  5.50  0.06  &     6.72  0.01  &   6.76  0.02  &   6.45  0.07  &   5.41  0.09   \\
   40 &   NGC3585                &    9.43  6.26  0.01  &    7.97  6.35  0.01  &    5.70  5.74  0.03  &    1.61  5.28  0.07  &     6.25  0.01  &   6.34  0.02  &   5.90  0.07  &   5.10  0.08   \\
   41 &   ESO270-G017            &    9.52  8.41  0.01  &    7.96  8.51  0.01  &    7.26  6.10  0.02  &    7.26  3.91  0.04  &     8.36  0.03  &   8.45  0.08  &   6.09  0.10  &   3.85  0.07   \\
   42 &   MESSIER51a             &    9.22  5.07  0.01  &    9.09  4.96  0.01  &    9.09  1.18  0.01  &    8.98 -0.62  0.01  &     5.08  0.01  &   4.94  0.01  &   1.16  0.01  &  -0.65  0.01   \\
   43 &   NGC3115                &    9.06  5.58  0.01  &    7.67  5.60  0.01  &    5.68  4.55  0.01  &    1.88  4.52  0.03  &     5.58  0.01  &   5.59  0.02  &   4.68  0.08  &   4.49  0.04   \\
   44 &   NGC3923                &    9.15  5.95  0.01  &    7.66  6.07  0.01  &    5.91  5.43  0.03  &    1.58  5.29  0.07  &     5.95  0.01  &   6.06  0.02  &   5.93  0.09  &   5.16  0.09   \\
   45 &   NGC4365                &    8.92  6.25  0.01  &    7.54  6.38  0.01  &    1.25  6.75  0.01  &    1.02  5.77  0.05  &     6.24  0.01  &   6.38  0.02  &   6.75  0.03  &   5.56  0.12   \\
   46 &   NGC1313                &    8.91  6.68  0.01  &    8.93  6.62  0.01  &    8.96  3.44  0.01  &    8.96  0.98  0.01  &     6.66  0.01  &   6.57  0.02  &   3.35  0.05  &   0.90  0.04   \\
   47 &   MESSIER84              &    8.93  5.71  0.01  &    7.45  5.76  0.01  &    4.68  5.42  0.02  &    1.56  4.77  0.05  &     5.70  0.01  &   5.79  0.01  &   5.64  0.04  &   4.72  0.09   \\
   48 &   NGC0185                &    8.92  6.19  0.01  &    7.44  6.27  0.01  &    5.84  5.25  0.02  &    2.47  4.55  0.04  &     6.16  0.02  &   6.26  0.02  &   5.36  0.03  &   4.38  0.08   \\
   49 &   NGC6946                &    8.84  5.00  0.01  &    8.88  4.82  0.01  &    7.96  0.94  0.01  &    8.71 -1.04  0.01  &     5.00  0.01  &   4.80  0.01  &   0.93  0.01  &  -1.07  0.01   \\
   50 &   NGC1395                &    8.88  6.43  0.01  &    7.43  6.51  0.01  &    5.67  5.85  0.02  &    1.52  5.37  0.07  &     6.42  0.01  &   6.51  0.02  &   6.24  0.06  &   5.28  0.10   \\
\hline
\end{tabular}

\tablecomments{columns:   (1) order of W1 3.4\m\ isophotal angular size, see Table 1;  (2) galaxy name;
(3-6) W1, W2, W3, and W4 (respectively), asymptotic measurements:  Radius (arcmin), magnitude and its uncertainty;
(7-10) total magnitude and its uncertainty (W1, W2, W3, W4, respectively) estimated using a double-Sersic fitting to the radial surface brightness profile.}

}
\hfill{}
\end{minipage}
\end{table}

\clearpage

\setcounter{table}{0}

\begin{table}[!ht]
\caption{... continued:  Asymptotic and  Total Integrated Brightness}\label{table:total2}
\vspace*{0.5cm}


\def\arraystretch{0.70}%
\hspace*{-3.5cm}
\footnotesize{
\begin{tabular}
{r | r | r | r | r | r | r | r | r | r | r | r |}
\hline
\hline
iS &  Galaxy &
R$_{\rm W1}$ A$_{\rm W1} \pm \Delta$ & 
R$_{\rm W2}$ A$_{\rm W2} \pm \Delta$ & 
R$_{\rm W3}$ A$_{\rm W3} \pm \Delta$ & 
R$_{\rm W4}$ A$_{\rm W4} \pm \Delta$ & 
T$_{\rm W1} \pm \Delta$ & 
T$_{\rm W2} \pm \Delta$ & 
T$_{\rm W3} \pm \Delta$ & 
T$_{\rm W4} \pm \Delta$ \\
\, & \, & 
amin mag mag & 
amin mag mag & 
amin mag mag & 
amin mag mag & 
mag mag & 
mag mag & 
mag mag & 
mag mag \\
(1) & (2) & (3) & (4) & (5) & (6) & (7) & (8) & (9) & (10)\\
\hline
   51 &   IC0010                 &    8.73  5.08  0.01  &    8.38  5.01  0.01  &    5.53  2.28  0.01  &    4.62 -0.35  0.01  &     5.07  0.01  &   5.01  0.01  &   2.30  0.01  &  -0.35  0.01   \\
   52 &   NGC4517                &    8.57  7.07  0.01  &    8.55  6.98  0.01  &    8.64  3.89  0.01  &    8.61  2.10  0.02  &     7.07  0.01  &   6.96  0.01  &   3.84  0.03  &   2.02  0.02   \\
   53 &   NGC1291                &    8.62  5.34  0.01  &    8.65  5.39  0.01  &    6.22  4.29  0.01  &    8.09  3.09  0.05  &     5.33  0.01  &   5.37  0.01  &   4.20  0.06  &   3.02  0.10   \\
   54 &   NGC2683                &    8.40  6.13  0.01  &    7.46  6.08  0.01  &    7.47  3.77  0.01  &    3.88  2.56  0.01  &     6.13  0.01  &   6.06  0.01  &   3.73  0.02  &   2.56  0.03   \\
   55 &   NGC4697                &    8.51  5.98  0.01  &    7.12  6.05  0.01  &    4.98  5.50  0.02  &    1.58  4.72  0.04  &     5.96  0.01  &   6.03  0.02  &   5.83  0.06  &   4.61  0.14   \\
   56 &   NGC3521                &    8.39  5.43  0.01  &    7.13  5.35  0.01  &    5.87  1.96  0.01  &    5.90  0.31  0.01  &     5.42  0.01  &   5.35  0.01  &   1.94  0.01  &   0.27  0.02   \\
   57 &   NGC3109                &    8.50  7.93  0.01  &    8.49  7.95  0.01  &    8.26  6.63  0.05  &    null  null  null  &     7.89  0.03  &   7.80  0.05  &   null  null  &   null  null   \\
   58 &   NGC0891                &    8.27  5.50  0.01  &    8.43  5.29  0.01  &    8.49  1.95  0.01  &    8.47  0.22  0.01  &     5.50  0.01  &   5.29  0.01  &   1.93  0.01  &   0.19  0.01   \\
   59 &   MESSIER85              &    8.43  5.78  0.01  &    7.08  5.82  0.01  &    5.33  5.29  0.02  &    2.07  4.41  0.04  &     5.77  0.01  &   5.81  0.01  &   5.45  0.05  &   4.36  0.10   \\
   60 &   NGC4244                &    8.47  7.48  0.01  &    8.45  7.43  0.01  &    8.46  5.32  0.01  &    7.52  3.26  0.03  &     7.47  0.01  &   7.39  0.03  &   5.18  0.05  &   3.11  0.07   \\
   61 &   NGC4762                &    8.41  7.09  0.01  &    7.09  7.14  0.01  &    5.18  6.43  0.03  &    1.92  5.98  0.10  &     7.09  0.01  &   7.12  0.02  &   6.73  0.07  &   5.88  0.09   \\
   62 &   NGC5084                &    8.13  6.81  0.01  &    8.15  6.83  0.01  &    8.22  5.41  0.02  &    8.15  3.93  0.08  &     6.80  0.01  &   6.79  0.02  &   5.44  0.09  &   4.08  0.08   \\
   63 &   NGC5907                &    7.91  6.46  0.01  &    8.04  6.34  0.01  &    8.05  3.22  0.01  &    8.03  1.55  0.01  &     6.46  0.01  &   6.34  0.01  &   3.20  0.01  &   1.52  0.02   \\
   64 &   NGC4395                &    8.04  7.70  0.01  &    8.02  7.73  0.01  &    7.98  5.43  0.03  &    7.77  2.99  0.05  &     7.67  0.02  &   7.58  0.07  &   5.34  0.09  &   2.97  0.11   \\
   65 &   NGC1407                &    7.92  6.19  0.01  &    6.61  6.28  0.01  &    5.28  5.55  0.03  &    1.53  5.38  0.08  &     6.18  0.01  &   6.27  0.02  &   5.95  0.09  &   5.25  0.10   \\
   66 &   NGC3627                &    7.83  5.55  0.01  &    6.56  5.45  0.01  &    6.14  2.02  0.01  &    6.15  0.01  0.01  &     5.54  0.01  &   5.44  0.01  &   2.00  0.01  &  -0.02  0.02   \\
   67 &   NGC4438                &    7.76  6.77  0.01  &    6.58  6.77  0.01  &    5.53  5.03  0.01  &    2.23  3.78  0.02  &     6.76  0.01  &   6.71  0.02  &   5.20  0.05  &   3.73  0.05   \\
   68 &   NGC1365                &    7.79  6.03  0.01  &    7.43  5.74  0.01  &    7.65  2.11  0.01  &    7.40 -0.46  0.01  &     6.02  0.01  &   5.73  0.01  &   2.09  0.01  &  -0.48  0.01   \\
   69 &   NGC2903                &    7.57  5.66  0.01  &    7.75  5.57  0.01  &    7.70  2.06  0.01  &    7.68  0.09  0.01  &     5.66  0.01  &   5.56  0.01  &   2.03  0.01  &   0.05  0.02   \\
   70 &   NGC5846                &    7.68  6.38  0.01  &    6.45  6.47  0.01  &    4.59  6.08  0.03  &    1.41  5.69  0.07  &     6.36  0.01  &   6.47  0.03  &   6.37  0.08  &   5.52  0.10   \\
   71 &   NGC4725                &    7.67  6.07  0.01  &    6.46  6.12  0.01  &    5.22  3.99  0.01  &    4.09  2.60  0.02  &     6.06  0.01  &   6.11  0.01  &   3.90  0.02  &   2.50  0.03   \\
   72 &   NGC1549                &    7.66  6.33  0.01  &    6.80  6.43  0.01  &    5.63  5.64  0.01  &    2.27  5.16  0.04  &     6.33  0.01  &   6.41  0.01  &   5.90  0.08  &   5.04  0.07   \\
   73 &   WLM                    &    7.70  8.85  0.01  &    7.68  8.93  0.02  &    7.69  8.17  0.22  &    4.97  5.36  0.17  &     8.79  0.03  &   8.75  0.07  &   null  null  &   null  null   \\
   74 &   NGC2841                &    7.68  5.88  0.01  &    6.33  5.89  0.01  &    4.68  3.78  0.01  &    5.90  2.17  0.02  &     5.87  0.01  &   5.88  0.01  &   3.74  0.01  &   2.10  0.03   \\
   75 &   CircinusGalaxy         &    7.62  4.39  0.01  &    7.61  3.74  0.01  &   11.02 -0.31  0.04  &   10.28 -2.59  0.04  &     4.37  0.02  &   3.73  0.01  &  -0.31  0.04  &  -2.60  0.04   \\
   76 &   NGC3621                &    7.58  6.32  0.02  &    6.32  6.18  0.02  &    6.24  2.55  0.01  &    6.24  0.83  0.01  &     6.31  0.02  &   6.15  0.02  &   2.52  0.01  &   0.78  0.02   \\
   77 &   NGC5078                &    7.56  6.69  0.01  &    6.32  6.65  0.01  &    4.59  4.00  0.01  &    3.26  2.44  0.01  &     6.68  0.01  &   6.64  0.01  &   4.01  0.01  &   2.40  0.02   \\
   78 &   NGC1023                &    7.53  6.03  0.01  &    6.27  6.09  0.01  &    3.69  5.62  0.02  &    1.92  4.93  0.05  &     6.03  0.01  &   6.09  0.01  &   5.72  0.07  &   4.78  0.08   \\
   79 &   NGC7331                &    7.45  5.69  0.01  &    7.21  5.60  0.01  &    7.25  2.35  0.01  &    7.18  0.66  0.01  &     5.69  0.01  &   5.59  0.01  &   2.33  0.01  &   0.62  0.01   \\
   80 &   MESSIER64              &    7.18  5.18  0.01  &    6.14  5.17  0.01  &    5.95  2.95  0.01  &    5.96  1.19  0.01  &     5.18  0.01  &   5.17  0.01  &   2.91  0.02  &   1.15  0.02   \\
   81 &   MESSIER59              &    7.21  6.43  0.01  &    6.11  6.51  0.01  &    4.53  5.31  0.02  &    1.73  5.09  0.07  &     6.42  0.01  &   6.51  0.02  &   5.41  0.12  &   5.06  0.07   \\
   82 &   NGC4696                &    7.18  6.64  0.01  &    6.05  6.70  0.01  &    5.09  5.92  0.02  &    1.94  5.37  0.06  &     6.63  0.01  &   6.69  0.01  &   6.35  0.09  &   5.32  0.14   \\
   83 &   MESSIER82              &    6.92  4.06  0.01  &    6.82  3.61  0.01  &    9.00 -0.96  0.01  &    8.81 -4.14  0.01  &     4.06  0.01  &   3.61  0.01  &  -0.97  0.01  &  -4.14  0.01   \\
   84 &   ESO274-001             &    7.07  7.82  0.01  &    7.07  7.85  0.01  &    3.29  6.36  0.03  &    3.38  3.01  0.01  &     7.77  0.33  &   7.77  0.09  &   6.65  0.09  &   2.98  0.02   \\
   85 &   MESSIER77              &    6.60  4.59  0.01  &    5.01  2.89  0.01  &    6.94 -0.68  0.04  &    7.33 -2.45  0.04  &     4.59  0.01  &   2.89  0.01  &  -0.68  0.04  &  -2.45  0.04   \\
   86 &   NGC3077                &    6.95  6.77  0.01  &    5.84  6.71  0.01  &    4.53  3.98  0.01  &    3.16  1.61  0.01  &     6.76  0.01  &   6.69  0.02  &   4.00  0.02  &   1.58  0.02   \\
   87 &   MESSIER65              &    6.95  5.89  0.01  &    5.88  5.91  0.01  &    5.85  4.14  0.01  &    5.88  2.74  0.03  &     5.89  0.01  &   5.90  0.01  &   4.09  0.02  &   2.64  0.04   \\
   88 &   NGC7213                &    6.91  6.46  0.01  &    5.82  6.41  0.01  &    4.02  4.23  0.01  &    2.64  2.42  0.01  &     6.45  0.01  &   6.41  0.02  &   4.26  0.02  &   2.38  0.07   \\
   89 &   IC0356                 &    6.74  5.76  0.01  &    5.92  5.76  0.01  &    3.09  3.74  0.01  &    5.96  1.91  0.02  &     5.74  0.01  &   5.75  0.01  &   3.56  0.02  &   1.81  0.04   \\
   90 &   NGC1560                &    6.90  8.56  0.01  &    5.82  8.69  0.01  &    3.03  7.15  0.05  &    1.88  5.35  0.04  &     8.55  0.01  &   8.67  0.02  &   null  null  &   null  null   \\
   91 &   NGC2663                &    6.90  6.22  0.01  &    6.00  6.32  0.01  &    5.06  5.86  0.03  &    1.43  4.90  0.04  &     6.21  0.01  &   6.29  0.02  &   6.30  0.04  &   4.83  0.04   \\
   92 &   NGC4216                &    6.92  6.36  0.01  &    5.73  6.37  0.01  &    5.62  4.22  0.01  &    5.30  2.89  0.02  &     6.35  0.01  &   6.36  0.01  &   4.16  0.02  &   2.78  0.03   \\
   93 &   NGC1055                &    6.79  6.74  0.01  &    6.49  6.63  0.01  &    6.82  3.06  0.01  &    6.77  1.30  0.01  &     6.74  0.01  &   6.60  0.01  &   3.04  0.01  &   1.27  0.02   \\
   94 &   NGC5170                &    6.73  7.42  0.01  &    5.67  7.40  0.01  &    5.68  5.21  0.01  &    5.19  3.79  0.03  &     7.41  0.01  &   7.39  0.01  &   5.14  0.03  &   3.61  0.04   \\
   95 &   MESSIER98              &    6.62  6.70  0.01  &    6.69  6.66  0.01  &    6.77  3.81  0.01  &    6.73  2.19  0.02  &     6.70  0.01  &   6.64  0.01  &   3.77  0.02  &   2.11  0.03   \\
   96 &   NGC2997                &    6.73  6.13  0.01  &    6.54  6.02  0.01  &    6.57  2.34  0.01  &    6.74  0.53  0.01  &     6.13  0.01  &   6.00  0.01  &   2.30  0.01  &   0.48  0.02   \\
   97 &   NGC4125                &    6.68  6.48  0.01  &    5.62  6.55  0.01  &    3.89  5.98  0.02  &    1.42  5.20  0.04  &     6.48  0.01  &   6.54  0.01  &   6.10  0.05  &   5.00  0.07   \\
   98 &   NGC7793                &    6.72  6.50  0.01  &    6.62  6.41  0.01  &    5.99  3.21  0.01  &    4.68  1.54  0.01  &     6.50  0.01  &   6.38  0.01  &   3.16  0.01  &   1.47  0.02   \\
   99 &   NGC5363                &    6.65  6.55  0.01  &    5.60  6.58  0.01  &    3.55  5.46  0.01  &    1.79  4.30  0.03  &     6.54  0.01  &   6.56  0.02  &   5.57  0.03  &   4.23  0.07   \\
  100 &   NGC4217                &    6.69  7.21  0.01  &    5.60  7.05  0.01  &    5.01  3.72  0.01  &    5.02  2.05  0.01  &     7.20  0.01  &   7.05  0.01  &   3.71  0.01  &   2.04  0.02   \\
  101 &   IC1613                 &    5.62  8.35  0.01  &    5.56  8.59  0.02  &    2.53  8.31  0.12  &    null  null  null  &     8.26  0.06  &   8.41  0.09  &   null  null  &   null  null   \\
  102 &   MESSIER32              &    4.32  5.01  0.01  &    3.56  5.07  0.01  &    3.49  4.42  0.01  &    1.90  3.57  0.03  &     5.01  0.01  &   5.06  0.01  &   4.37  0.02  &   3.49  0.04   \\
  103 &   UGC05373               &    3.46  9.65  0.01  &    2.92  9.66  0.02  &    null  null  null  &    2.11  6.39  0.33  &     9.57  0.04  &   9.56  0.09  &   null  null  &   null  null   \\
  104 &   ESO245-007             &    2.61 10.88  0.02  &    2.61 11.12  0.04  &    null  null  null  &    null  null  null  &    10.53  0.10  &   null  null  &   null  null  &   null  null   \\
\hline
\end{tabular}

\tablecomments{columns:   (1) order of W1 3.4\m\ isophotal angular size, see Table 1;  (2) galaxy name;
(3-6) W1, W2, W3, and W4 (respectively), asymptotic measurements:  Radius (arcmin), magnitude and its uncertainty;
(7-10) total magnitude and its uncertainty (W1, W2, W3, W4, respectively) estimated using a double-Sersic fitting to the radial surface brightness profile.}

}
\hfill{}
\end{table}

\clearpage

\section{SEDs of a Classic Galaxy Types}

Four examples of galaxies that represents the \wise color sets (Fig. 10), comprising spheroids, intermediate-SF, active-SF disks, and AGN-dominated.  The SEDs are constructed from 2MASS XSC and \wise global measurements, and templates that best fit the data.  Since these Brown et al. (2014a) templates have real mid-infrared spectral data from {\it Spitzer}-IRS, they are accurate representations of the emission and continuum features.
Also indicated are the \wise relative system response curves (RSRs), here normalized to unity for easy comparison.  Note that the actual throughput quantum efficiency is significantly different, with W1 the most sensitive band, followed closely by W2, and the long wave-bands considerably less sensitive (Jarrett et al. 11).  The 12\m\ W3 detector, however, makes up for less sensitivity by having a very wide band, enclosing molecular emission and dust continuum components. 
\vspace{-0.5cm}
\begin{figure*}[ht!]
\gridline{\leftfig{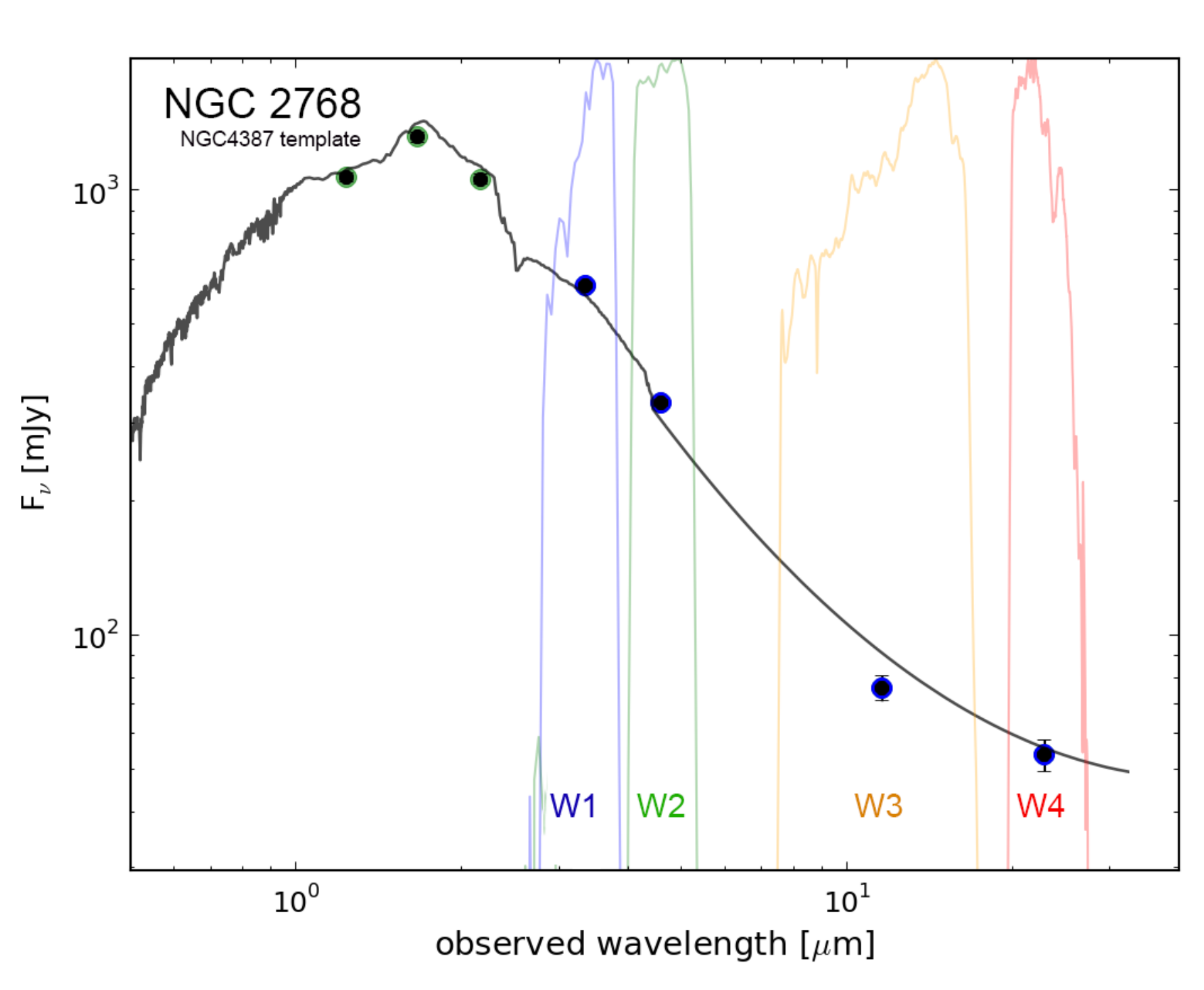}{0.48\textwidth}{(a)} 
              \rightfig{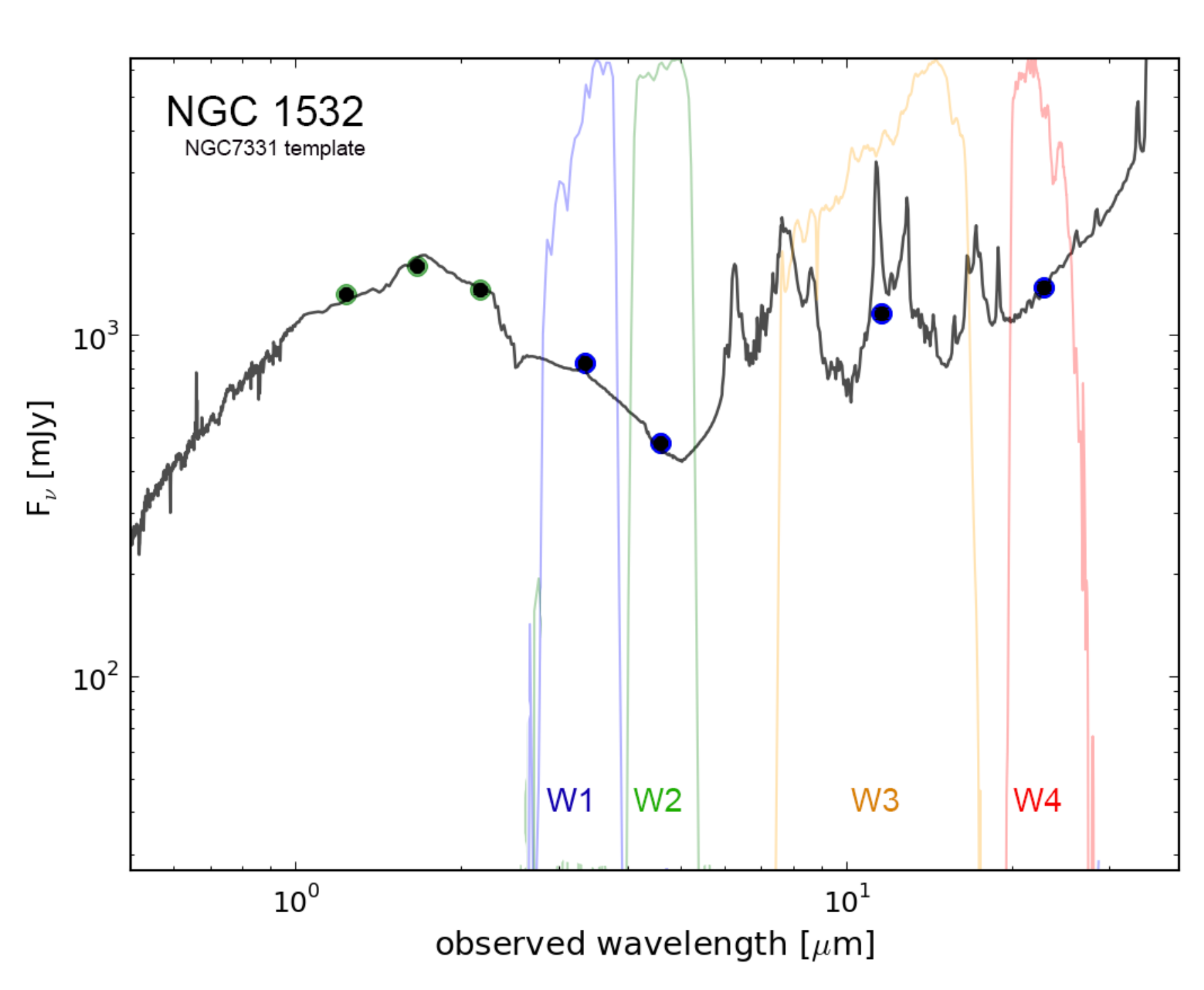}{0.48\textwidth}{(b)}}  
\gridline{\leftfig{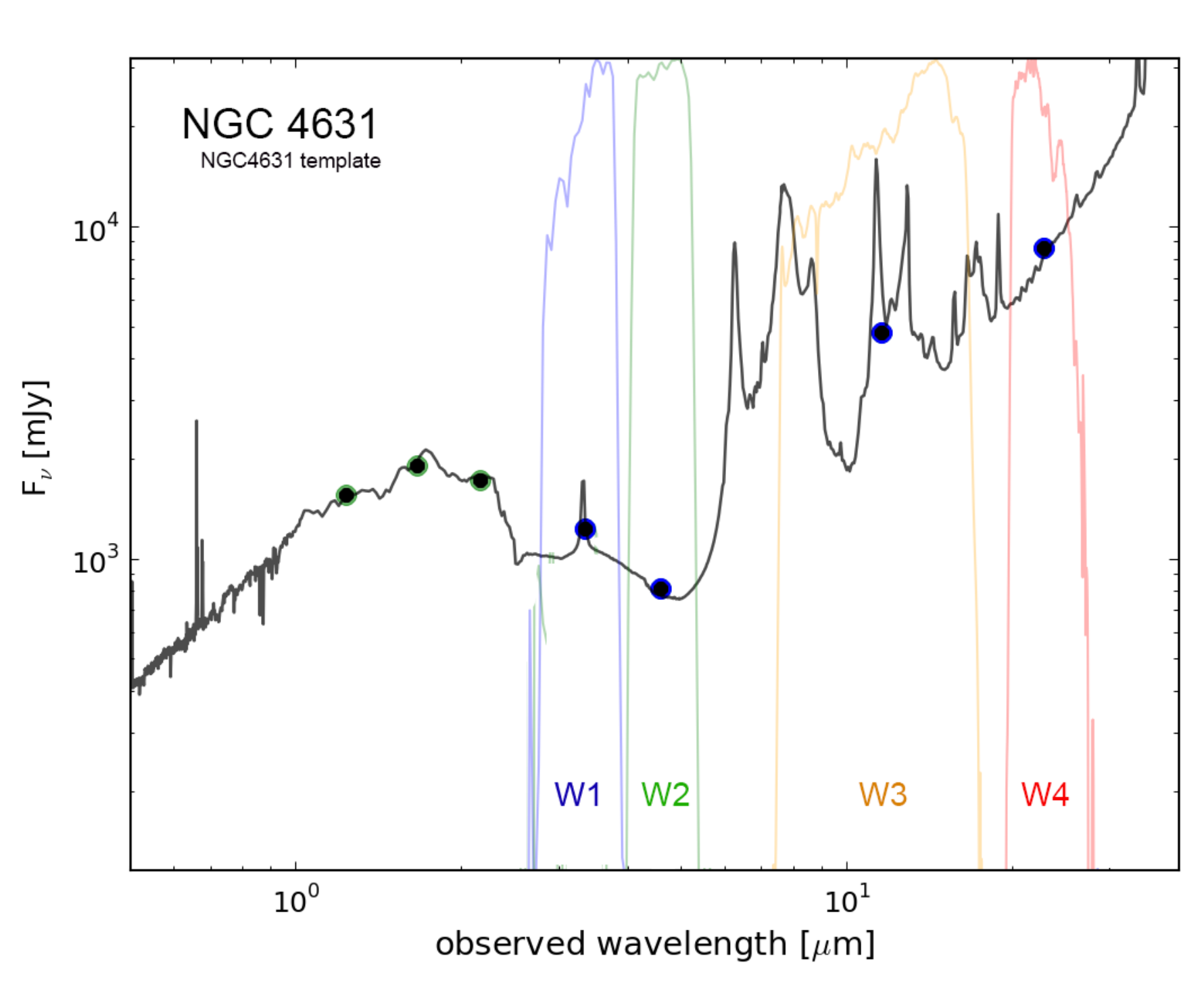}{0.48\textwidth}{(c)} 
              \rightfig{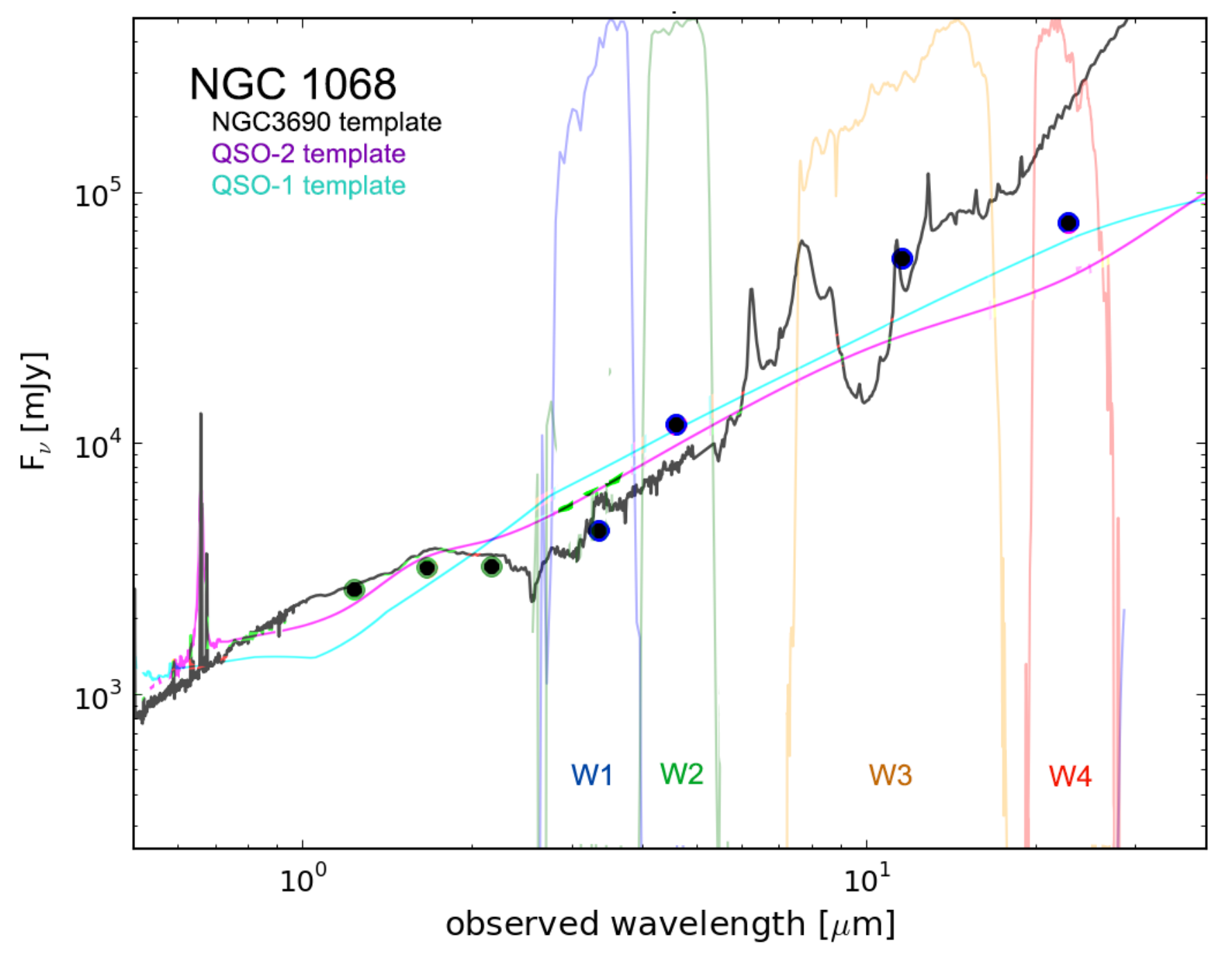}{0.48\textwidth}{(d)}}  
\caption{Infrared spectral energy distributions of four galaxies, demonstrating the broad classification revealed by \wise colors.  (a) The first is a typical early-type elliptical galaxy, stellar-bulge dominated;  (b) the second is an intermediate (Sb) galaxy; (c) the third is a late-type (Sd), SF-dominated spiral, and (d) the last  demonstrates an AGN-dominated host galaxy.  The data points are global measurements from 2MASS XSC and \wise (this study).  The "template" is a best fit to the data, from the Brown et al. (2014a) and GRASIL (Silva et al. 1998) suite of spectral-templates.   The \wise filter bands (normalized to unity) are indicated for each.
 \label{fig:SED}}               
\end{figure*}

\section{Data Products of the WXSC}

{\bf Images:} The foundation of the WXSC are the native-resolution mosaics, constructed specifically for resolved galaxies.   For each galaxy, the images range in size from 0.25 degrees to many degrees (e.g., LG galaxies),  large enough to encompass the galaxy and its local environment.  The pixel scale is 1 arcsec, except for the Magellanic Clouds (8 arcsec) and M\,31 (1.5 arcsec).  The flux calibration is in the Vega system, and the zero point magnitude is in the FITS header.  The orientation is standard, and the headers have full WCS information.

There are four bands of \wise, for each band we construct three kinds of images:  integrated signal (INT), the corresponding uncertainty (UNC) and the frame coverage (COV).  

Processing includes identification and removal of foreground stars, background galaxies and neighboring (satellite) galaxies, as well as the (rare) artifact or image glitch.  These `cleaned' images are also part of the available data products, and are used to measure and study the target galaxy.

{\bf Photometry and Catalogued Values:} Full source characterization is carried out on the cleaned images, featuring size, shape, integrated fluxes, surface brightness, and a number of other measurements.  A README detailing the catalog columns is part of the data release.   For those galaxies with a distance (either redshift-independent, or a redshift-based luminosity distance), physical attributes (luminosities, SFRs, stellar mass) are derived from k-corrected fluxes.  The k-correction is carried out using templates and SED fitting (see \citet{Jar17} for more details).   Likewise, there is a README file that details the physical values in a separate `derived' catalog.

{\bf Ancillary Products:} The images and catalogs are the primary data products, but there are a number of other useful products.  3-color (RGB) images of the WXSC galaxies that show the before and after star removal, SEDs where the flux data are plotted and compared with the best-fit SED template, and the pinwheel diagrams (see below), that graphically show the physical attributes of the galaxy with respect to the mean values of the greater WXSC sample (Table 4).

\vspace{-0.5cm}
\begin{figure*}[ht!]   
\hspace{5mm}
  \includegraphics[width=143mm]{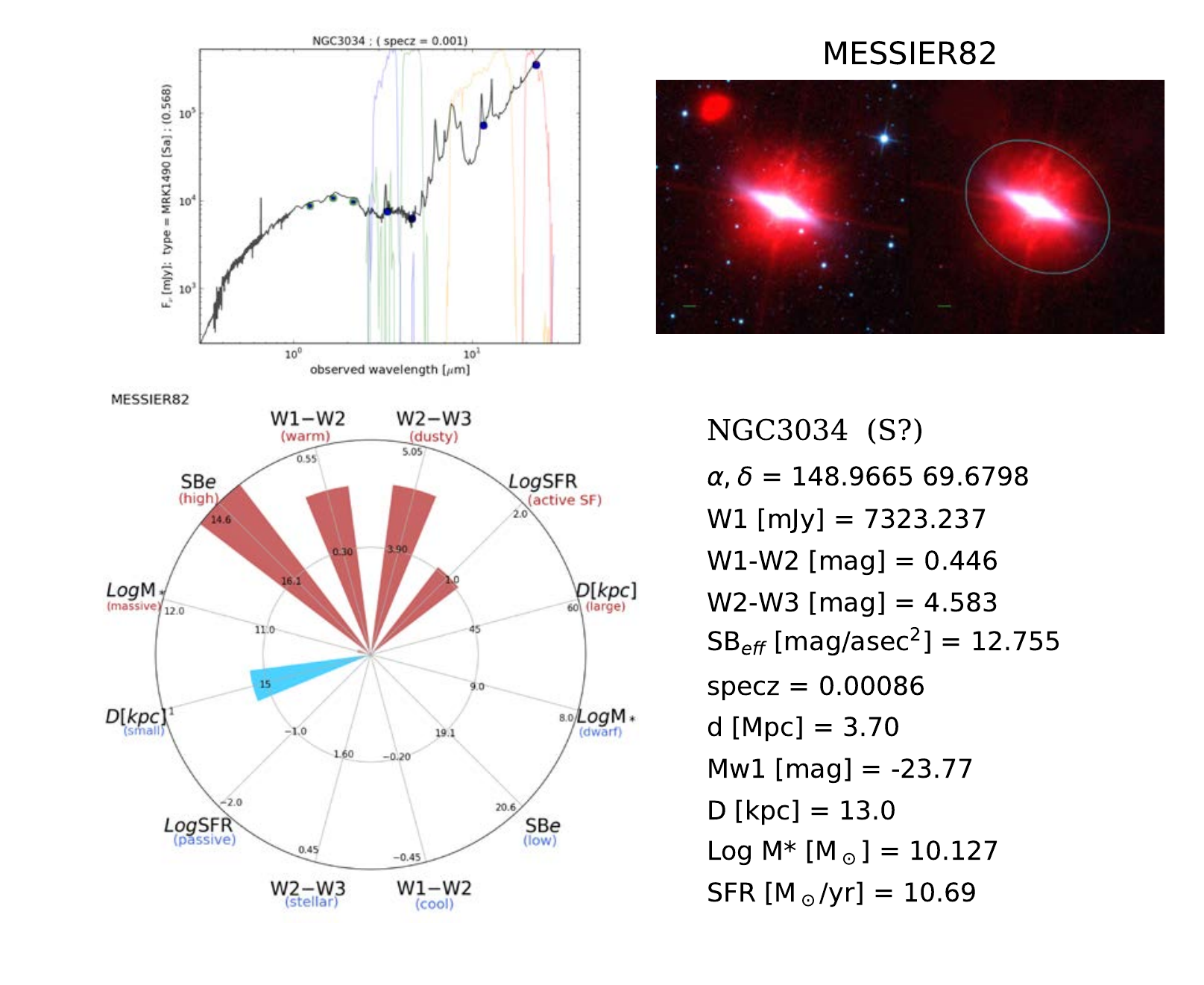}
  \caption{Physical properties of starburst NGC\,3034 (M\,82). The image shows the \wise view before/after star subtraction.  The ellipse denotes the 1-\sig$_{\rm sky}$ isophote aperture.  The spectral energy distribution includes the four wise band measurements,  three 2MASS measurements and the best-fit galaxy template.
\label{fig:ngc3034pw}}               
\end{figure*}

\pagebreak

\section{3-Color Images of the Largest Galaxies}


\begin{figure}[!htbp]
 \begin{minipage}[b]{0.92\linewidth}
 \vspace{-3.5mm}
    \centering
    \includegraphics[width=\linewidth]{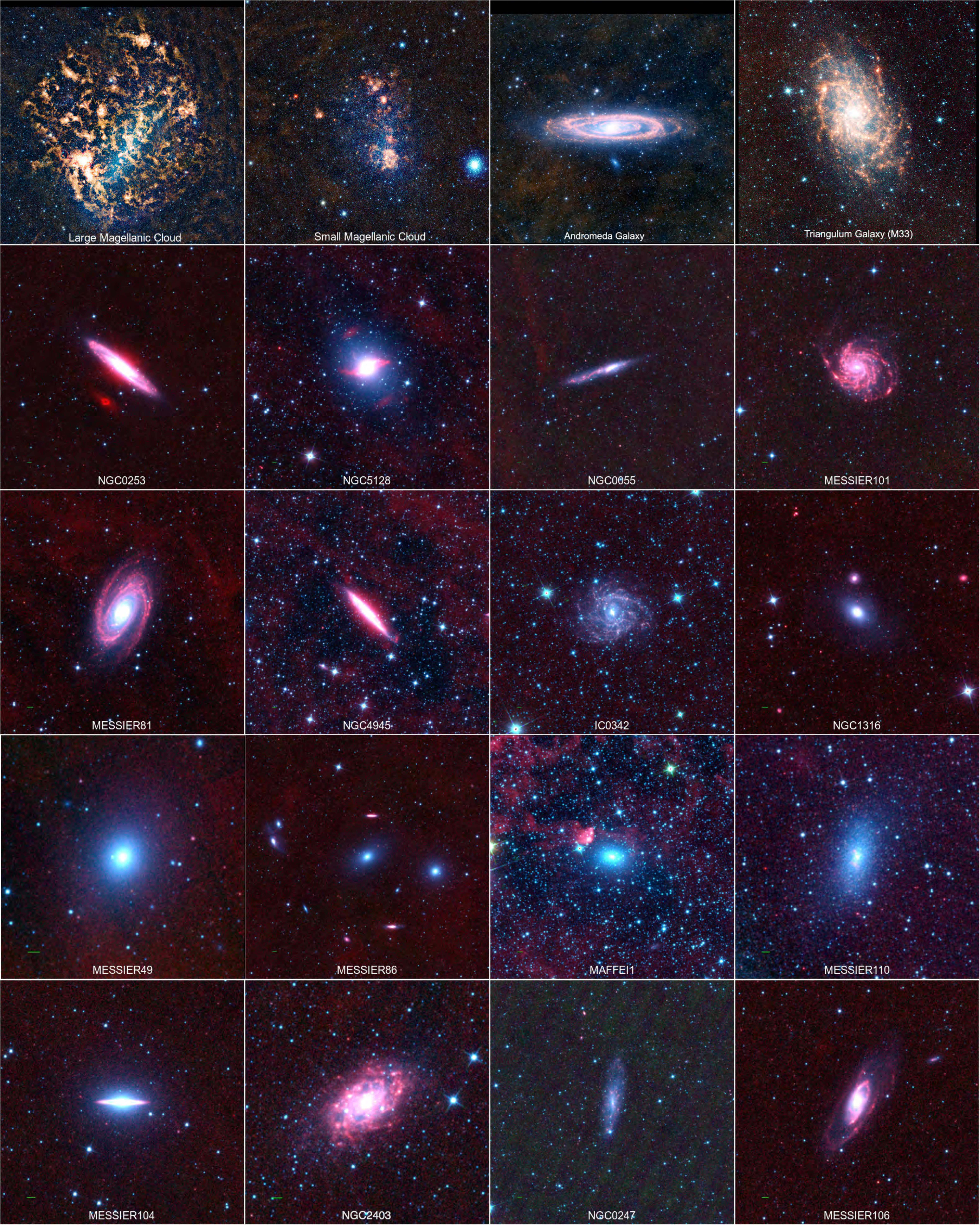}
    \caption{Color combination of the W1, W2 and W3 bands.  Assigning the blue (W1) and green (W2) renders the stellar-dominated light to have a blue-cyan hue.  Assigning red to W3 highlights star formation sites, appearing yellow-orange-red.  Sources that have strong AGN emission will have strong W2 relative to W1, and hence appear yellow/green. In the lower left corner, the green dash specifies 1 arcmin in scale.
     \label{fig:mon_0}}   
     \end{minipage}         
\end{figure}

\begin{figure}[!htbp] 
    \centering
    \includegraphics[width=\linewidth]{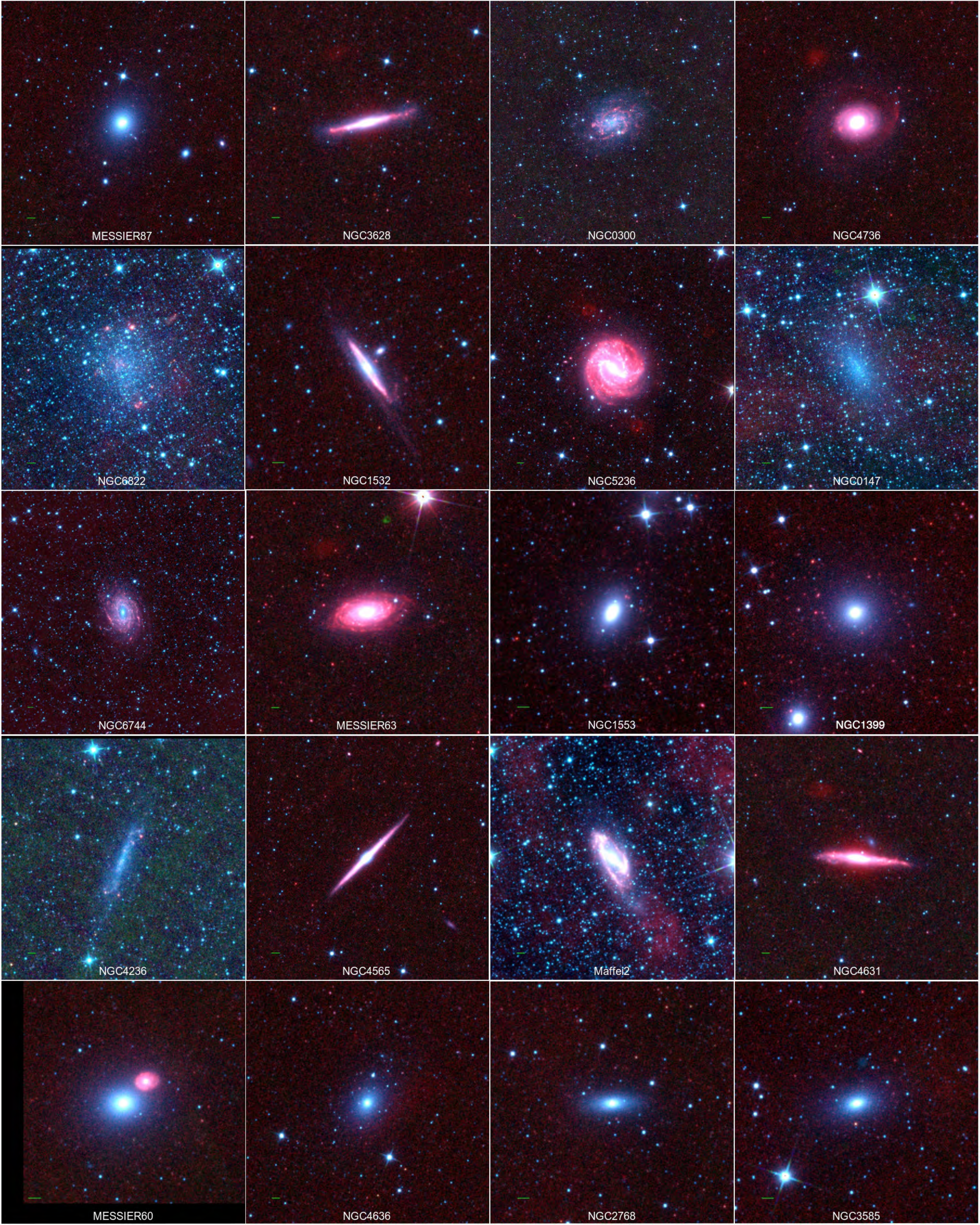}
     \caption{See Fig 21 for details.}{\label{fig:mon_1}}              
\end{figure}

\begin{figure}[!htbp] 
    \centering
    \includegraphics[width=\linewidth]{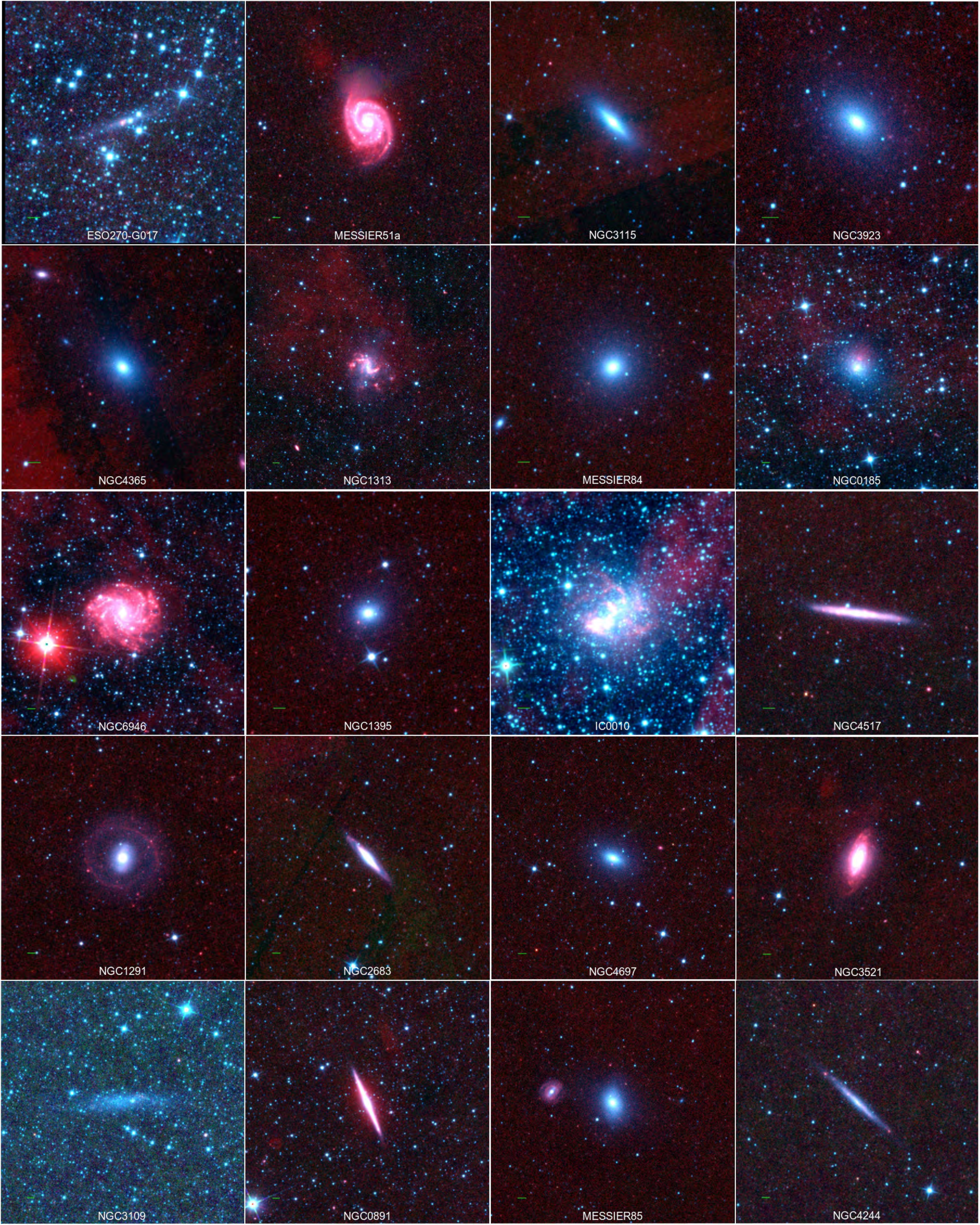}
     \caption{See Fig 21 for details.}{\label{fig:mon_2}}               
\end{figure}

\begin{figure}[!htbp] 
    \centering
    \includegraphics[width=\linewidth]{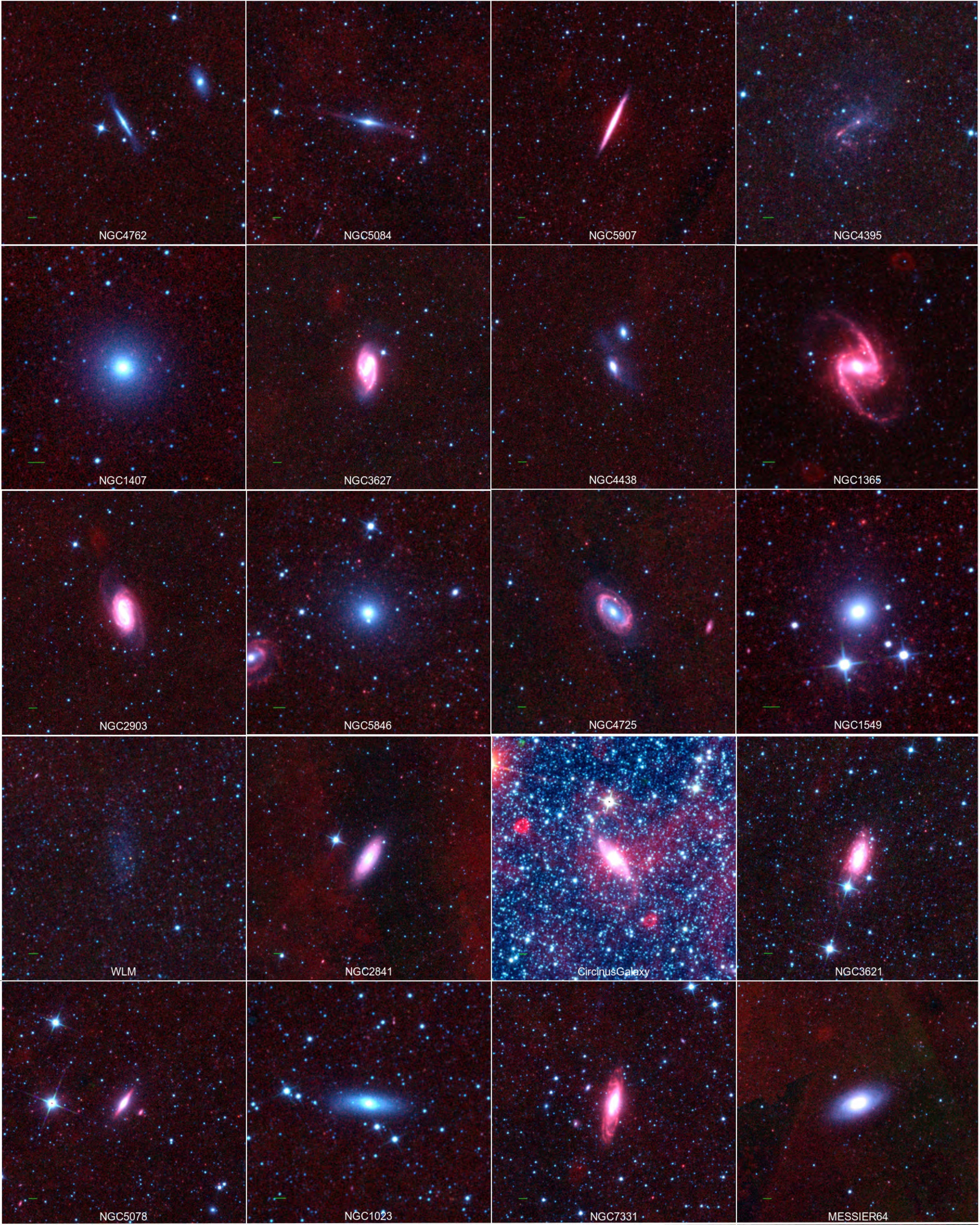}
     \caption{See Fig 21 for details.}{\label{fig:mon_3}}               
\end{figure}

\begin{figure}[!htbp] 
    \centering
    \includegraphics[width=\linewidth]{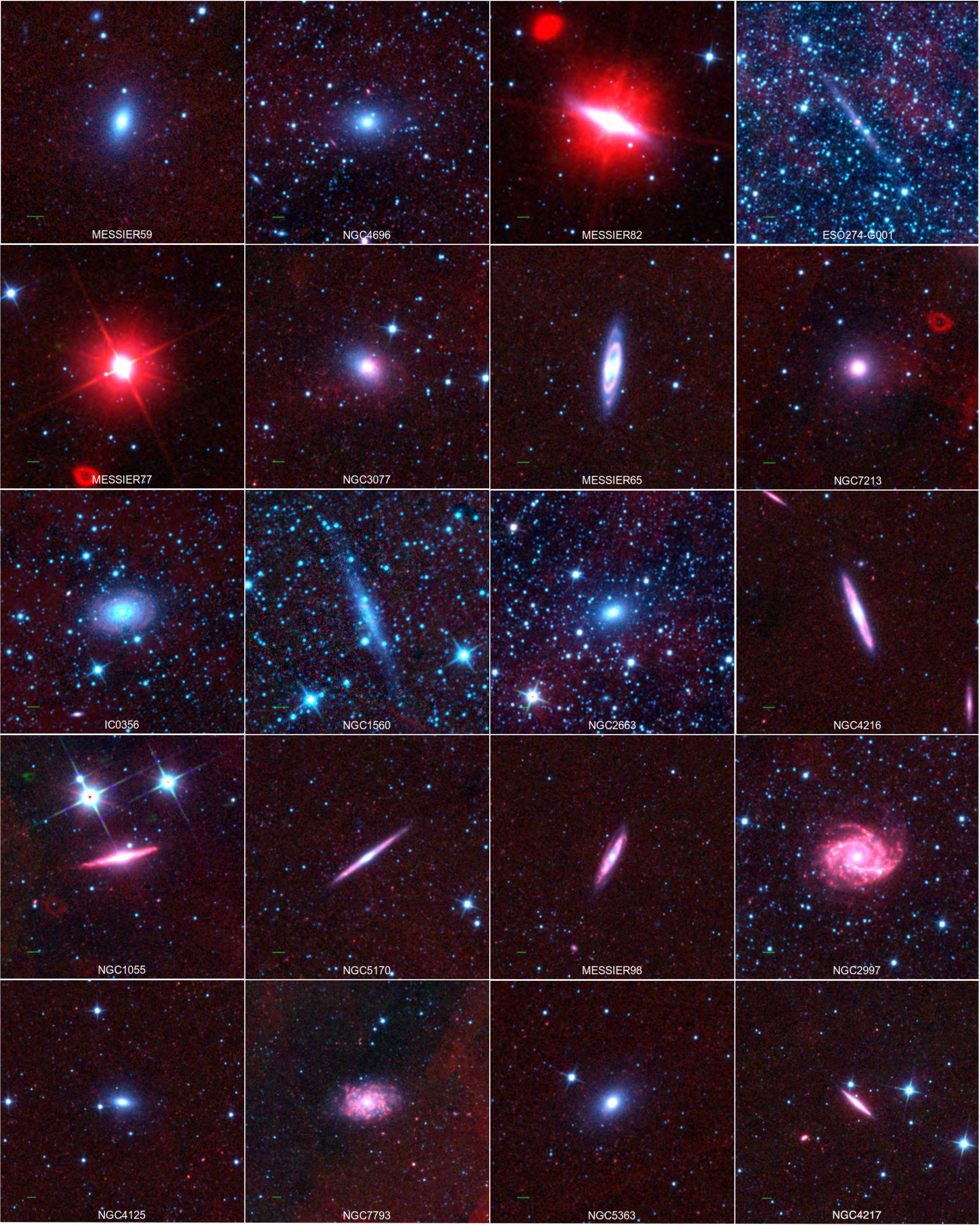}
     \caption{See Fig 21 for details.}{\label{fig:mon_4}}               
\end{figure}

\end{appendices}

\end{document}